\documentclass[12pt,onecolumn,draftcls]{IEEEtran}
  \usepackage{amssymb}
  \usepackage{amsmath}
  \usepackage{rotate}
  \usepackage{color}

\newtheorem{theorem}{Theorem}
\newtheorem{lemma}{Lemma}
\newtheorem{proposition}{Proposition}

\newcommand{\beq}   {\begin{equation}}
\newcommand{\eeq}   {\end{equation}}
\newcommand\bydef{\triangleq}
\newcommand{\Bydef} {~\bydef~}
\newcommand{\mem}[1]{\label{eq:#1}}
\newcommand{\rec}[1]{(\ref{eq:#1})}
\newcommand{\nono}  {\nonumber}
\newcommand{\bm}    {\boldmath}

\newcommand  {\rr}    {\mbox{\boldmath $r$}}
\newcommand  {\phiphi} {\mbox{\boldmath $\phi$}}
\newcommand  {\xx}    {\mbox{\boldmath $x$}}
\newcommand  {\xxb}   {\mbox{\scriptsize\boldmath $x$}}
\newcommand  {\0}     {{\bf 0}}
\newcommand  {\C}     {\mbox{${\textstyle \,{\cal C}\!\!\!\!\!\!\!\sim}$}}
\newcommand  {\dis}   {\displaystyle}

\newcounter{step}
\begin{document}

\title{Analysis of Sequential Decoding Complexity Using
       the Berry-Esseen Inequality$^\ast$\thanks{$^\ast$This work is supported by
       the {\it National Science Council} of Taiwan, R.O.C., under
       the projects of NSC 88-2219-E-009-004 and NSC 88-2213-E-260-006.}}
\author{Po-Ning Chen$^\dag$ ({\it Senior Member}, {\it IEEE}),
       Yunghsiang S.~Han$^\ddag$ ({\it Member}, {\it IEEE}),\\
       Carlos, R.~P.~Hartmann$^\S$ ({\it Fellow}, {\it IEEE}) and
       Hong-Bin Wu$^\dag$\\
       {\normalsize~\\
       $\dag$Department of Communications Engineering\\
       National Chiao Tung University, Hsin Chu\\
       Taiwan 30050, R.O.C.\\[5mm]
       $\ddag$Graduate Institute of Communication Engineering\\
       National Taipei University, Taipei\\
       Taiwan, R.O.C.\\[5mm]
       $\S$Department of Electrical Engineering and Computer\\
       Science, Syracuse University, Syracuse, NY 13244-4100, USA\\[5mm]
       ~}\\
       \parbox[l]{6in}{
       {\bf Subject Area}: Coding Theory\\
       {\bf Corresponding Author}: Prof. Yunghsiang S. Han\\
       \hspace*{10mm}Graduate Institute of Communication Engineering\\
       \hspace*{10mm}National Taipei University\\
       \hspace*{10mm}Taipei, Taiwan, R.O.C.\\
       \hspace*{10mm}E-mail: yshan@mail.ntpu.edu.tw\\
       \hspace*{10mm}Tel:+1-886-2-86746401\quad\quad\quad Fax:+1-886-2-26744448}}
\maketitle

\begin{abstract}
This study presents a novel technique to estimate the computational complexity
of sequential decoding using the Berry-Esseen theorem. Unlike the theoretical
bounds determined by the conventional central limit theorem argument, which
often holds only for sufficiently large codeword length, the new bound obtained
from the Berry-Esseen theorem is valid for any blocklength. The accuracy of the
new bound is then examined for two sequential decoding algorithms, an
ordering-free variant of the generalized Dijkstra's algorithm (GDA)(or
simplified GDA) and the maximum-likelihood sequential decoding algorithm
(MLSDA). Empirically investigating codes of small blocklength reveals that the
theoretical upper bound for the simplified GDA almost matches the simulation
results as the signal-to-noise ratio (SNR) per information bit ($\gamma_b$) is
greater than or equal to $8$ dB. However, the theoretical bound may become
markedly higher than the simulated average complexity when $\gamma_b$ is small.
 For the MLSDA,
the theoretical upper bound is quite close to the simulation results for both
high SNR ($\gamma_b\geq 6$ dB) and low SNR ($\gamma_b\leq 2$ dB). Even for
moderate SNR, the simulation results and the theoretical bound differ by at
most $0.8$ on a $\log_{10}$ scale.
\end{abstract}


\section{Introduction}
\pagestyle{plain}

The Berry-Esseen theorem \cite[sec.XVI.~5]{FEL66} states that the
distribution of the sum of independent zero-mean random variables
$\{X_i\}_{i=1}^{n}$, normalized by the standard deviation of the
sum, differs from the unit Gaussian distribution by no more than
$C\,r_n/s_n^3$, where $s_n^2$ and $r_n$ are, respectively, the sums of
the marginal variances and the marginal absolute third moments,
and the Berry-Esseen coefficient, $C$, is an absolute constant.
Specifically, for every $a\in\Re$,
\begin{equation}
\mem{oribe} \left|\Pr\left\{\frac
1{s_n}\left(X_1+\cdots+X_{n}\right)\leq
a\right\}-\Phi(a)\right|\leq C\ \frac{r_n}{s_n^3},
\end{equation}
where $\Phi(\cdot)$ represents the unit Gaussian cumulative
distribution function (cdf). The remarkable aspect of this theorem
is that the upper bound depends only on the variance and the
absolute third moment, and therefore, can provide a good probability
estimate through the first three moments. A typical estimate of the
absolute constant is six \cite[sec.XVI.~5, Thm.~2]{FEL66}. When
$\{X_n\}_{i=1}^n$ are identically distributed, in addition to
independent, the absolute constant can be reduced to three, and
has been reported to be improved down to $2.05$ \cite[sec.XVI.~5,
Thm.~1]{FEL66}. In 1972, Beek sharpened the constant to 0.7975
\cite{Bee72}. Later, Shiganov further improved the constant down
to 0.7915 for an independent sample sum, and, 0.7655, if these
samples are also identically distributed \cite{SHI86}. Shiganov's
result is generally considered to be the best result yet obtained thus far
\cite{VLA98}.

In applying this inequality to analyze the computational
complexity of sequential decoding algorithms, the original
analytical problem is first transformed into one that concerns
the asymptotic probability mass of the sum of independent random
samples. Inequality \rec{oribe} can therefore be applied.
The complexities of two sequential maximum-likelihood decoding algorithms
are then analyzed. One is an
ordering-free variant of the generalized Dijkstra's algorithm
(GDA) \cite{HAN93} operated over a code tree of linear block
codes, and the other is the maximum-likelihood sequential decoding
algorithm (MLSDA) \cite{HANCHEN982} that searches for the codeword
over a trellis of binary convolutional codes.

The computational effort required by sequential decoding is conventionally
determined using a random coding technique, which averages the
computational effort over the ensemble of random tree codes
\cite{JAC67,JEL691,SAV66}. Branching process analysis
on sequential decoding complexity has been recently proposed
\cite{HAC84,JOH79,JOH85}; the results, however, were still derived
by averaging over semi-random tree codes. Chevillat
and Costello proposed to analyze the computational
effort of sequential decoding in terms of the column distance
function of a specific time-invariant code \cite{CHE78}; but, the
analysis only applied to a situation in which the code was
transmitted via binary symmetric channels.

In light of the Berry-Esseen inequality and the large deviations technique,
this work presents an alternative approach to derive the theoretical
upper bounds on the computational effort of the simplified GDA and
the MLSDA for binary codes antipodally transmitted through an
additive white Gaussian noise (AWGN) channel.
Unlike the bounds established in terms of the conventional
central limit theorem argument, which often holds only for
sufficiently large codeword length, the new bound is valid for any
blocklength. Empirically investigating codes of small
blocklength shows that for the trellis-based MLSDA, the
theoretical upper bound is quite close to the simulation results
for both high SNR and low SNR; even for moderate SNR, the
theoretical upper bound and the simulation results differ by no more than
$0.579966$ on a $\log_{10}$ scale.
 For the tree-based ordering-free GDA,
the theoretical bound coincides with the simulation results at high
SNR; however, the bound tends to be substantially larger than the
simulation results at very low SNR. The possible cause of the
inaccuracy of the bound at low SNR for the tree-based ordering-free GDA is
addressed at the end of this study.

The rest of this paper is organized as follows.
Section~\ref{SEC:BET} derives a probability bound for use of analyzing
the sequential decoding complexity
due to the Berry-Esseen inequality. Section \ref{SEC:GDA} presents
an analysis of the average computational complexity of the GDA.
Section \ref{SEC:ANA} briefly introduces the MLSDA, and then
analyzes its complexity upper bound. Conclusions are finally drawn
in Section~\ref{SEC:CON}.

Throughout this article, $\Phi(\cdot)$ denotes the unit Gaussian
cdf.

\section{Berry-Esseen Theorem and probability bound}
\label{SEC:BET}

This section derives an upper probability bound for the sum of
independent random samples using the Berry-Esseen inequality.
This bound is essential to the analysis of the
computational effort of sequential decoding algorithms.

The approach used here is the {\it large deviations}
technique, which is generally applied to compute
the {\it exponent} of an exponentially decaying
probability mass. The
Berry-Esseen inequality is also applied to evaluate the {\it
subexponential} detail of the concerned probability. With these two
techniques, an upper bound of the
concerned probability can be established.

\begin{lemma}\label{lm:ubound}
Let $Y_n=\sum_{i=1}^nX_i$ be the sum of i.i.d.~random variables whose marginal
distribution is $F(\cdot)$. Define the {\it twisted} distribution with
parameter $\theta$ corresponding to $F(\cdot)$ as:
$$
dF^{(\theta)}(x) \triangleq \frac{\exp\{\theta x\}\,dF(x)}{M(\theta)},
$$
where $M(\theta)\triangleq E[e^{\theta X_1}]$. Let the random variable with
probability distribution $F^{(\theta)}(\cdot)$ be $X^{(\theta)}$. Then, for
every $\theta<0$,
$$\Pr\left\{Y_n\leq -n\alpha\right\}\leq
\displaystyle A_n(\theta,\alpha)e^{\theta\alpha n} M^n(\theta),$$
where $A_n(\theta,\alpha)=\min\{B_n(\theta,\alpha),1\}$,
$$
B_n(\theta,\alpha)\Bydef\left\{\begin{array}{ll}
\dis\frac{\sigma(\theta)}{\sqrt{2\pi n}[(\mu(\theta)+\alpha)-\theta\sigma^2(\theta)]}
e^{-(\mu(\theta)+\alpha)^2n/[2\sigma^2(\theta)]}+2C\frac{\rho(\theta)}{\sigma^3(\theta)\sqrt{n}},&\mbox{if }\alpha>\theta\sigma^2(\theta)-\mu(\theta);\\
e^{\theta[\theta\sigma^2(\theta)-2(\mu(\theta)+\alpha)]n/2}
+2C\dis\frac{\rho(\theta)}{\sigma^3(\theta)\sqrt{n}},
&\mbox{otherwise},
\end{array}\right.
$$
$$\mu(\theta)=E[X^{(\theta)}],\quad\sigma^2(\theta)=E[|X^{(\theta)}-\mu(\theta)|^2],\quad
\rho(\theta)=E[|X^{(\theta)}-\mu(\theta)|^3]$$ and $C=0.7655$.
\end{lemma}

\begin{proof}
Define
$F_n^{(\theta)}(y)=\Pr[X_1^{(\theta)}+X_2^{(\theta)}+\cdots+X_n^{(\theta)}\leq
y]$, and let the distribution of
$[(X_1^{(\theta)}-\mu(\theta))+\cdots+(X_n^{(\theta)}-\mu(\theta))]/[\sigma(\theta)\sqrt{n}]$
be $H_n(\cdot)$, where in the evaluation of the above two statistics,
$\{X_i^{(\theta)}\}_{i=1}^n$ are assumed independent with common marginal
distribution $F^{(\theta)}(\cdot)$. Then, by denoting
$Y_n^{(\theta)}=X_1^{(\theta)}+X_2^{(\theta)}+\cdots+X_n^{(\theta)}$, we
obtain:
\begin{eqnarray}
\Pr\left(Y_n\leq -n\alpha\right)&=&\int_{[x_1+\cdots+x_n\leq -n\alpha]} dF(x_1)dF(x_2)\cdots dF(x_n)\nono\\
&=&M^n(\theta)\int_{[x_1+\cdots+x_n\leq -n\alpha]}
e^{-\theta(x_1+\cdots+x_n)}dF^{(\theta)}(x_1)dF^{(\theta)}(x_2)\cdots dF^{(\theta)}(x_n)\nono\\
&=&M^n(\theta)E\left[e^{-\theta(X_1^{(\theta)}+\cdots+X_n^{(\theta)})}
{\bf 1}\{X_1^{(\theta)}+\cdots+X_n^{(\theta)}\leq -n\alpha\}\right]\nono\\
&=&M^n(\theta)E\left[e^{-\theta Y_n^{(\theta)}}
{\bf 1}\{Y_n^{(\theta)}\leq -n\alpha\}\right]\nono\\
&=&M^n(\theta)\int_{-\infty}^{-n\alpha} e^{-\theta y}dF_n^{(\theta)}(y)\nono\quad\quad(y\rightarrow\sigma(\theta)\sqrt{n} y'+\mu(\theta)n)\\
&=&M^n(\theta)\int_{-\infty}^{-(\mu(\theta)+\alpha)\sqrt{n}/\sigma(\theta)}
e^{-\theta\sigma(\theta)\sqrt{n}y'-\theta\mu(\theta)n}dH_n(y')\mem{pre-final}\\
&=&e^{\theta\alpha n}M^n(\theta)\int_{-\infty}^{-(\mu(\theta)+\alpha)\sqrt{n}/\sigma(\theta)}
e^{-\theta\sigma(\theta)\sqrt{n}\,[y'+(\mu(\theta)+\alpha)\sqrt{n}/\sigma(\theta)]}dH_n(y'),\mem{final}
\end{eqnarray}
where ${\bf 1}\{\cdot\}$ is the set indicator function, and \rec{pre-final}
follows from $H_n(y)=F_n^{(\theta)}(\sigma(\theta)\sqrt{n}y+\mu(\theta)n)$.

Integrating by parts on \rec{final} with
$\lambda(dy)\Bydef-\theta\sigma(\theta)\sqrt{n}\exp\{-\theta\sigma(\theta)\sqrt{n}
[y+(\mu(\theta)+\alpha)\sqrt{n}/\sigma(\theta)]\}dy$ defined over
$(-\infty,-(\mu(\theta)+\alpha)\sqrt{n}/\sigma(\theta)]$, and then applying
equation \rec{oribe} yields
\begin{eqnarray}
\lefteqn{\int_{-\infty}^{-(\mu(\theta)+\alpha)\sqrt{n}/\sigma(\theta)}
e^{-\theta\sigma(\theta)\sqrt{n}\,[y+(\mu(\theta)+\alpha)\sqrt{n}/\sigma(\theta)]}dH_n(y)}\\
&=&\int_{-\infty}^{-(\mu(\theta)+\alpha)\sqrt{n}/\sigma(\theta)}
\left[H_n\left(-\frac{(\mu(\theta)+\alpha)\sqrt{n}}{\sigma(\theta)}\right)-H_n(y)\right]\lambda(dy)\nono\\
&\leq&\int_{-\infty}^{-(\mu(\theta)+\alpha)\sqrt{n}/\sigma(\theta)}
\left[\Phi\left(-\frac{(\mu(\theta)+\alpha)\sqrt{n}}{\sigma(\theta)}\right)-\Phi(y)
+2C\frac{\rho(\theta)}{\sigma^3(\theta)\sqrt{n}}\right]
\lambda(dy)\nono\\
&=&\int_{-\infty}^{-(\mu(\theta)+\alpha)\sqrt{n}/\sigma(\theta)}
\left[\Phi\left(-\frac{(\mu(\theta)+\alpha)\sqrt{n}}{\sigma(\theta)}\right)-\Phi(y)\right]\lambda(dy)
+2C\frac{\rho(\theta)}{\sigma^3(\theta)\sqrt{n}}\nono\\
&=&\int_{-\infty}^{-(\mu(\theta)+\alpha)\sqrt{n}/\sigma(\theta)}
e^{-\theta\sigma(\theta)\sqrt{n}\,[y+(\mu(\theta)+\alpha)\sqrt{n}/\sigma(\theta)]}\frac 1{\sqrt{2\pi}}e^{-y^2/2}dy
+2C\frac{\rho(\theta)}{\sigma^3(\theta)\sqrt{n}}\mem{ibp}\\
&=&e^{\theta^2\sigma^2(\theta)n/2}e^{-\theta(\mu(\theta)+\alpha)n}
\Phi\left(\theta\sigma(\theta)\sqrt{n}-\frac{(\mu(\theta)+\alpha)\sqrt{n}}{\sigma(\theta)}\right)+2C\frac{\rho(\theta)}{\sigma^3(\theta)\sqrt{n}}\nono\\
&\leq&
\left\{\begin{array}{ll}
\dis\frac{\sigma(\theta)}{\sqrt{2\pi n}[(\mu(\theta)+\alpha)-\theta\sigma^2(\theta)]}
e^{-(\mu(\theta)+\alpha)^2n/[2\sigma^2(\theta)]}+2C\frac{\rho(\theta)}{\sigma^3(\theta)\sqrt{n}},&\mbox{if }\alpha>\theta\sigma^2(\theta)-\mu(\theta);\\
e^{\theta^2\sigma^2(\theta)n/2}e^{-\theta(\mu(\theta)+\alpha)n}
+2C\dis\frac{\rho(\theta)}{\sigma^3(\theta)\sqrt{n}},
&\mbox{otherwise},
\end{array}\right.
\mem{iii}
\end{eqnarray}
where \rec{ibp} holds by, again, applying integration by part,
and \rec{iii} follows from
$$\Phi(-u)\leq
\frac{1}{\sqrt{2\pi}u}e^{-u^2/2}\quad{\rm and}\quad
\Phi(u)\leq 1\quad\mbox{ for }u>0.$$
It remains to show that
$$\int_{-\infty}^{-(\mu(\theta)+\alpha)\sqrt{n}/\sigma(\theta)}
e^{-\theta\sigma(\theta)\sqrt{n}\,[y+(\mu(\theta)+\alpha)\sqrt{n}/\sigma(\theta)]}dH_n(y)\leq 1,$$
which be established by observing that
\begin{eqnarray}
e^{\theta\alpha n}M^n(\theta)\int_{-\infty}^{-(\mu(\theta)+\alpha)\sqrt{n}/\sigma(\theta)}
e^{-\theta\sigma(\theta)\sqrt{n}\,[y+(\mu(\theta)+\alpha)\sqrt{n}/\sigma(\theta)]}dH_n(y)
&=&\Pr\{Y_n\leq -n\alpha\}\label{chernoff1}\\
&=&\Pr\left\{e^{\theta (Y_n+n\alpha)}\geq 1\right\}\nono\\
&\leq&E[e^{\theta (Y_n+n\alpha)}]\nono\\
&=&e^{\theta\alpha n}M^n(\theta).\label{chernoff2}
\end{eqnarray}
\end{proof}

Some remarks are made following Lemma~\ref{lm:ubound} as follows. First, the
upper probability bound in Lemma~\ref{lm:ubound} consists of two parts, the
exponentially decaying $e^{\theta\alpha n}M^n(\theta)$ and the subexponentially
bounded $A_n(\theta,\alpha)$. When $\alpha>\theta\sigma^2(\theta)-\mu(\theta)$
and $\alpha\neq-\mu(\theta)$,
$$B_n(\theta,\alpha)=\dis\frac{\sigma(\theta)}{\sqrt{2\pi n}[(\mu(\theta)+\alpha)-\theta\sigma^2(\theta)]}
e^{-(\mu(\theta)+\alpha)^2n/[2\sigma^2(\theta)]}+2C\frac{\rho(\theta)}{\sigma^3(\theta)\sqrt{n}}
\approx 2C\frac{\rho(\theta)}{\sigma^3(\theta)\sqrt{n}}$$
since the first term decays exponentially fast, and $B_n(\theta,\alpha)$ reduces to the Berry-Esseen
probability bound. However, when $\theta$ is taken to satisfy $\mu(\theta)=-\alpha$,
$$B_n(\theta,\alpha)=\dis\frac{1}{\sqrt{2\pi n}|\theta|\sigma(\theta)}
+2C\frac{\rho(\theta)}{\sigma^3(\theta)\sqrt{n}},$$
and a larger bound (than the Berry-Esseen one) is resulted.
In either case, $B_n(\theta,\alpha)$ vanishes exactly at the speed of $1/\sqrt{n}$.
Secondly,
when  $A_n(\theta,\alpha)=1$, the upper probability bound reduces to
the simple Chernoff bound $e^{\theta\alpha n}M^n(\theta)$
for which a four-line proof from
\eqref{chernoff1} to \eqref{chernoff2} is sufficient \cite[Eq.~(5.4.9)]{Gal68},
and is always valid for every $\theta<0$, regardless of whether
$\alpha>\theta\sigma^2(\theta)-\mu(\theta)$ or not.

The independent samples $\{X_i\}_{i=1}^n$ with which our decoding problems are
concerned actually consist of two i.i.d.~sequences, one of which is Gaussian
distributed and the other is non-Gaussian distributed. One way to bound the
desired probability of $\Pr[\sum_{i=1}^nX_i\leq 0]$ is to directly use the
Berry-Esseen inequality for independent but non-identical samples (which can be
done following similar proof of Lemma~\ref{lm:ubound}). However, in order to
manage a better bound, we will apply Lemma~\ref{lm:ubound} only to those
non-Gaussian i.i.d.~samples, and manipulate the remaining Gaussian samples
directly by way of their known probability densities in the below lemma
(cf.~The derivation in \eqref{cc}).

\begin{lemma}\label{coro}
Let $Y_n=\sum_{i=1}^nX_i$ be the sum of independent random
variables $\{X_i\}_{i=1}^n$, among which $\{X_i\}_{i=1}^d$ are
identically Gaussian distributed with positive mean $\mu$ and
non-zero variance $\sigma^2$, and $\{X_i\}_{i=d+1}^n$ have common marginal
distribution as $\min\{X_1,0\}$. Let
$\gamma\Bydef(1/2)(\mu^2/\sigma^2)$. Then
$$\Pr\left\{Y_n\leq 0\right\}\leq
{\cal B}\left(d,n-d,\gamma\right),$$ where
$${\cal B}\left(d,n-d,\gamma\right)=
\left\{\begin{array}{ll}
\Phi(-\sqrt{2\gamma n}),&\mbox{if }d=n;\\[5mm]
\Phi\left(-\frac{(n-d)\hat\mu+d\sqrt{2\gamma}}{\sqrt{d}}\right)+\tilde A_{n-d}(\lambda)&\\
\ \times
\left[\Phi(-\lambda)e^{-\gamma}e^{\lambda^2/2}+\Phi(\sqrt{2\gamma})\right]^{n-d}&\\
\
\times e^{d(-\gamma+\lambda^2/2)}\Phi\left(\frac{(n-d)\hat\mu+\lambda d}{\sqrt{d}}\right),&
\mbox{if }1>\frac dn\geq 1-\frac{\sqrt{4\pi\gamma}e^\gamma}{1
+\sqrt{4\pi\gamma}e^\gamma\Phi(\sqrt{2\gamma})};\\[5mm]
1,&\mbox{otherwise},
\end{array}\right.$$
\begin{eqnarray*}
a&\triangleq&-\hat{\mu}+(\sqrt{2\gamma}-\lambda)\tilde\sigma^2(\lambda)+\tilde\mu(\lambda),\\
\tilde A_{n-d}(\lambda)&\bydef& \min\left({\bf
1}\{a>0\}\left[\frac{\tilde\sigma(\lambda)}{a\sqrt{2\pi(n-d)}}
+2C\frac{\tilde\rho(\lambda)}{\tilde\sigma^3(\lambda)\sqrt{n-d}}\right]+{\bf 1}\{a\leq 0\},1\right),\\
\hat\mu&\bydef&E[X_{d+1}]=-(1/\sqrt{2\pi})e^{-\gamma}+\sqrt{2\gamma}\Phi(-\sqrt{2\gamma}),\\
\tilde\mu(\lambda)&=&-\frac d{n-d}\lambda,\\
\tilde\sigma^2(\lambda)
&\bydef&-\frac{d}{n-d}-\frac{nd}{(n-d)^2}\lambda^2
+\frac n{n-d}\frac{1}
{1+\sqrt{2\pi}\lambda e^\gamma\Phi(\sqrt{2\gamma})},\\
\tilde\rho(\lambda)&\bydef&\frac n{(n-d)}\frac\lambda
{[1+\sqrt{2\pi}\lambda e^\gamma\Phi(\sqrt{2\gamma})]}
\left\{1-\frac{d(n+d)}{(n-d)^2}\lambda^2\right.\nono\\
&&+2\left[\frac{n^2}{(n-d)^2}\lambda^2+2\right]
e^{-d(2n-d)\lambda^2/[2(n-d)^2]}\nono\\
&&-\frac d{n-d}\left[\frac{n+d}{n-d}\lambda^2+3\right]
\sqrt{2\pi}\lambda e^\gamma\Phi(\sqrt{2\gamma})\nono\\
&&\left.-\frac{2n}{n-d}\left[\frac{n^2}{(n-d)^2}\lambda^2+3\right]
\sqrt{2\pi}\lambda e^{\lambda^2/2}\Phi\left(-\frac n{n-d}\lambda\right)\right\},
\end{eqnarray*}
and $\lambda$ is the unique solution (in $[0,\sqrt{2\gamma})$) of
$$\lambda e^{(1/2)\lambda^2}\Phi(-\lambda)=
\frac 1{\sqrt{2\pi}}\left(1-\frac dn\right) -\frac
dne^{\gamma}\Phi(\sqrt{2\gamma})\lambda.$$\end{lemma}

\begin{proof} Only the bound for $d<n$ is proved since
the case of $d=n$ can be easily substantiated.

Let
$$\tilde\mu(\theta)=\frac{E[X_{d+1}^{(\theta)}]}{\sigma},\quad
\tilde\sigma(\theta)=\frac{\mbox{Var}[X_{d+1}^{(\theta)}]}{\sigma^2},\quad
{\rm and}\quad\tilde\rho(\theta)=\frac{E[|X_{d+1}^{(\theta)}-E[X_{d+1}^{(\theta)}]|^3]}{\sigma^3},$$
and let $\hat\mu=E[X_{d+1}]/\sigma$.
 By noting that
$(\mu/\sigma)=\sqrt{2\gamma}$, and for any $\theta<0$ satisfying that
$$a\triangleq -\hat{\mu}-\sigma\theta\tilde\sigma^2(\theta)+\tilde\mu(\theta)>0,$$
$\Pr(Y_n\leq 0)$ can be bounded by
\begin{eqnarray}
\lefteqn{\Pr(Y_n\leq 0)}\nonumber\\
&=&\Pr\left\{X_1+\cdots+X_d+X_{d+1}+\cdots+X_n\leq 0\right\}\nonumber\\
&=&\int_{-\infty}^\infty
\Pr\left\{X_{d+1}+\cdots+X_n\leq -x\right\}\frac 1{\sqrt{2\pi d\sigma^2}}
e^{-\frac{(x-d\mu)^2}{2d\sigma^2}}dx,\quad(x\rightarrow\sigma x')\nonumber\\
&=&\int_{-\infty}^\infty
\Pr\left\{X_{d+1}+\cdots+X_n\leq -\sigma x'\right\}\frac 1{\sqrt{2\pi d}}
e^{-\frac{(x'-d\sqrt{2\gamma})^2}{2d}}dx',\quad(x'\rightarrow(n-d)x'')\nonumber\\
&=&\int_{-\infty}^\infty
\Pr\left\{X_{d+1}+\cdots+X_n\leq -\sigma (n-d)x''\right\}\frac 1{\sqrt{2\pi d/(n-d)^2}}
e^{-\frac{(x''-d\sqrt{2\gamma}/(n-d))^2}{2d/(n-d)^2}}dx''\nonumber\\
&=&\int_{-\infty}^{\sigma\theta\tilde\sigma^2(\theta)-\tilde\mu(\theta)+a}
\Pr\left\{X_{d+1}+\cdots+X_n\leq -\sigma (n-d)x''\right\}\frac 1{\sqrt{2\pi
d/(n-d)^2}}
e^{-\frac{(x''-d\sqrt{2\gamma}/(n-d))^2}{2d/(n-d)^2}}dx''\nonumber\\
&+&\int_{\sigma\theta\tilde\sigma^2(\theta)-\tilde\mu(\theta)+a}^\infty
\Pr\left\{X_{d+1}+\cdots+X_n\leq -\sigma (n-d)x''\right\}\frac 1{\sqrt{2\pi
d/(n-d)^2}}
e^{-\frac{(x''-d\sqrt{2\gamma}/(n-d))^2}{2d/(n-d)^2}}dx''\nonumber\\
&\leq&\int_{-\infty}^{\sigma\theta\tilde\sigma^2(\theta)-\tilde\mu(\theta)+a}
\frac 1{\sqrt{2\pi d/(n-d)^2}}
e^{-\frac{(x''-d\sqrt{2\gamma}/(n-d))^2}{2d/(n-d)^2}}dx''\nonumber\\
&+&\int_{\sigma\theta\tilde\sigma^2(\theta)-\tilde\mu(\theta)+a}^\infty
\min\left(\frac{\tilde\sigma(\theta)}{a\sqrt{2\pi(n-d)}}
e^{-(\tilde\mu(\theta)+x'')^2(n-d)/[2\tilde\sigma^2(\theta)]}
+2C\frac{\tilde\rho(\theta)}{\tilde\sigma^3(\theta)\sqrt{n-d}},1\right)\nonumber\\
&&\times e^{\theta\sigma(n-d)x''}M^{n-d}(\theta)\frac 1{\sqrt{2\pi d/(n-d)^2}}
e^{-\frac{(x''-d\sqrt{2\gamma}/(n-d))^2}{2d/(n-d)^2}}dx'',\label{cc}
\end{eqnarray}
where $M(\theta)=E[e^{\theta X_{d+1}}]$, and the last inequality follows from
Lemma \ref{lm:ubound}. Observe that
$$e^{-(\tilde\mu(\theta)+x'')^2(n-d)/[2\tilde\sigma^2(\theta)]}\leq 1.$$
Thus,
\begin{eqnarray}
\Pr(Y_n\leq 0)&\leq&\int_{-\infty}^{-\hat\mu}
\frac 1{\sqrt{2\pi d/(n-d)^2}}
e^{-\frac{(x''-d\sqrt{2\gamma}/(n-d))^2}{2d/(n-d)^2}}dx''\nonumber\\
&+&\int_{-\hat\mu}^\infty
\min\left(\frac{\tilde\sigma(\theta)}{a\sqrt{2\pi(n-d)}}
+2C\frac{\tilde\rho(\theta)}{\tilde\sigma^3(\theta)\sqrt{n-d}},1\right)\nonumber\\
&&\times e^{\theta\sigma(n-d)x''}M^{n-d}(\theta)\frac 1{\sqrt{2\pi d/(n-d)^2}}
e^{-\frac{(x''-d\sqrt{2\gamma}/(n-d))^2}{2d/(n-d)^2}}dx''\nonumber\\
&=&\Phi\left(-\frac{(n-d)\hat\mu+d\sqrt{2\gamma}}{\sqrt{d}}\right)\nonumber\\&&+
\tilde A_{n-d}(\theta)M^{n-d}(\theta)e^{d(\theta\sigma\sqrt{2\gamma}+\theta^2\sigma^2/2)}
\Phi\left(\frac{(n-d)\hat\mu+d\sqrt{2\gamma}}{\sqrt{d}}+\theta\sigma\sqrt{d}\right),\label{cc1}
\end{eqnarray}
where for $a>0$,
$$\tilde A_{n-d}(\theta)=\min\left(\frac{\tilde\sigma(\theta)}{a\sqrt{2\pi(n-d)}}
+2C\frac{\tilde\rho(\theta)}{\tilde\sigma^3(\theta)\sqrt{n-d}},1\right).$$ Now
for $\theta<0$ and $a\leq 0$, we can use Chernoff bound in \eqref{cc} instead,
in which case the derivation up to \eqref{cc1} similarly follows with $\tilde
A_{n-d}(\theta)=1$.

We then note that
$$M^{n-d}(\theta)e^{d(\theta\sigma\sqrt{2\gamma}+\theta^2\sigma^2/2)}$$
is exactly the moment
generating function of $Y_n=\sum_{i=1}^nX_i$; hence, if $E[Y_n]=d\mu+(n-d)\sigma\hat\mu>0$,
then the solution $\theta$ of $\partial E[e^{\theta Y_n}]/\partial\theta=0$ is
definitely negative.

For notational convenience, we
let $\lambda=(\mu/\sigma)+\sigma\theta=\sqrt{2\gamma}+\sigma\theta$,
and yield that
$$
M(\theta)
=\Phi\left(-\lambda\right)e^{-\gamma}e^{\lambda^2/2}
+\Phi(\sqrt{2\gamma})\quad{\rm and}\quad
e^{\theta\sigma\sqrt{2\gamma}+\theta^2\sigma^2/2}=
e^{-\gamma}e^{\lambda^2/2}.$$
Accordingly, the chosen $\lambda=\sqrt{2\gamma}+\sigma\theta$ should satisfy
$$
\frac{\partial\left(\left[\Phi(-\lambda)e^{-\gamma}e^{\lambda^2/2}+\Phi(\sqrt{2\gamma})\right]^{n-d}
e^{d(-\gamma+\lambda^2/2)}\right)}{\partial\lambda}=0,
$$
or equivalently,
\begin{equation}
\mem{thetahata} e^{(1/2)\lambda^2}\Phi(-\lambda)= \frac
1{\sqrt{2\pi}\lambda}\left(1-\frac dn\right) -\frac
dne^{\gamma}\Phi(\sqrt{2\gamma}).
\end{equation}
As it turns out, the solution $\lambda=\lambda(\gamma)$ of the above equation
depends only on $\gamma$. Now, by replacing $e^{(1/2)\lambda^2}\Phi(-\lambda)$
with $\left(1-d/n\right)/(\sqrt{2\pi}\lambda)
-(d/n)e^{\gamma}\Phi(\sqrt{2\gamma})$, we obtain \begin{eqnarray*}
\tilde\mu(\lambda)&=&\left.\frac{E\left[X_{d+1}^{(\theta)}\right]}{\sigma}\right|_{\theta=
(\lambda-\sqrt{2\gamma})/\sigma}=-\frac{d}{n-d}\lambda
\end{eqnarray*}
\begin{eqnarray*}
\tilde\sigma^2(\lambda)
&\bydef&\left.\frac{\mbox{Var}\left[X_{d+1}^{(\theta)}\right]}{\sigma^2}\right|_{\theta=
(\lambda-\sqrt{2\gamma})/\sigma}\\
&=&-\frac{d}{n-d}-\frac{nd}{(n-d)^2}\lambda^2
+\frac n{n-d}\frac{1}
{1+\sqrt{2\pi}\lambda e^\gamma\Phi(\sqrt{2\gamma})},
\end{eqnarray*}
and
\begin{eqnarray*}
\tilde\rho(\lambda)&\bydef&\left.\frac{
E\left[\left|X_{d+1}^{(\theta)}-\hat\mu\right|^3\right]}{\sigma^3}\right|_{\theta=
(\lambda-\sqrt{2\gamma})/\sigma}\\
&=&\frac n{(n-d)}\frac\lambda
{[1+\sqrt{2\pi}\lambda e^\gamma\Phi(\sqrt{2\gamma})]}
\left\{1-\frac{d(n+d)}{(n-d)^2}\lambda^2\right.\nono\\
&&+2\left[\frac{n^2}{(n-d)^2}\lambda^2+2\right]
e^{-d(2n-d)\lambda^2/[2(n-d)^2]}\nono\\
&&-\frac d{n-d}\left[\frac{n+d}{n-d}\lambda^2+3\right]
\sqrt{2\pi}\lambda e^\gamma\Phi(\sqrt{2\gamma})\nono\\
&&\left.-\frac{2n}{n-d}\left[\frac{n^2}{(n-d)^2}\lambda^2+3\right]
\sqrt{2\pi}\lambda e^{\lambda^2/2}\Phi\left(-\frac n{n-d}\lambda\right)\right\}
\end{eqnarray*}
Hence, the previously obtained upper bound for $\Pr(Y_n\leq 0)$ can be reformulated as
\begin{eqnarray*}
\lefteqn{\Phi\left(-\frac{(n-d)\hat\mu+d\sqrt{2\gamma}}{\sqrt{d}}\right)}\\
&&+\tilde A_{n-d}(\lambda)
\left[\Phi(-\lambda)e^{-\gamma}e^{\lambda^2/2}+\Phi(\sqrt{2\gamma})\right]^{n-d}
e^{d(-\gamma+\lambda^2/2)}\Phi\left(\frac{(n-d)\hat\mu+\lambda
d}{\sqrt{d}}\right),
\end{eqnarray*}
where
$$\tilde A_{n-d}(\lambda)=\min\left({\bf 1}\{a>0\}\left[\frac{\tilde\sigma(\lambda)}{a\sqrt{2\pi(n-d)}}
+2C\frac{\tilde\rho(\lambda)}{\tilde\sigma^3(\lambda)\sqrt{n-d}}\right]+{\bf
1}\{a\leq 0\},1\right).$$
 Finally, a simple derivation yields
\begin{eqnarray*}
E[Y_n]&=&dE[X_1]+(n-d)E[X_{d+1}]\\
&=&\sigma\left(d\sqrt{2\gamma}+(n-d)\left[-(1/\sqrt{2\pi})e^{-\gamma}+\sqrt{2\gamma}
\Phi(-\sqrt{2\gamma})\right]\right),
\end{eqnarray*}
and hence, the condition of $E[Y_n]>0$ can be equivalently
replaced by
$$\frac dn\geq 1-\frac{\sqrt{4\pi\gamma}e^\gamma}{1
+\sqrt{4\pi\gamma}e^\gamma\Phi(\sqrt{2\gamma})}.$$
\end{proof}

Again, if the simple Chernoff inequality is used instead in the derivation of
\eqref{cc}, the bound remains of the same form in Lemma~\ref{coro} except that
$\tilde A_{n-d}(\lambda)$ is always equal to one.

Empirical evaluations of $\tilde{A}_{n-d}(\lambda)$ in Figs.~\ref{and1} and
\ref{and2} indicates that when the sample number $n\leq 50$,
$\tilde{A}_{n-d}(\lambda)$ will be close to $1$, and the subexponential
analysis based on the Berry-Esseen inequality does not help improving the upper
probability bound. However, for a slightly larger $n$ such as $n=200$, a
visible reduction in the probability bound can be obtained through the
introduction of the Berry-Esseen inequality.

One of the main studied subjects in this paper is to examine whether the
introduction of the subexponential analysis can help improving the complexity
bound at practical code length. The observation from Figs.~\ref{and1} and
\ref{and2} does coincide with what we obtained in later applications. That is,
some visible improvement in complexity bound can really be obtained for a
little larger codeword length in the MLSDA (specifically, $N=2(60+6)$ or
$2(100+6)$). However, since the simulated codes are only of lengths $24$ and
$48$, no improvement can be observed for the GDA algorithm.

\begin{figure}[hbt]
\begin{center}
\setlength{\unitlength}{0.240900pt}
\ifx\plotpoint\undefined\newsavebox{\plotpoint}\fi
\sbox{\plotpoint}{\rule[-0.200pt]{0.400pt}{0.400pt}}
\begin{picture}(1500,900)(0,0)
\font\gnuplot=cmr10 at 10pt
\gnuplot
\sbox{\plotpoint}{\rule[-0.200pt]{0.400pt}{0.400pt}}
\put(140.0,123.0){\rule[-0.200pt]{4.818pt}{0.400pt}}
\put(120,123){\makebox(0,0)[r]{0}}
\put(1419.0,123.0){\rule[-0.200pt]{4.818pt}{0.400pt}}
\put(140.0,287.0){\rule[-0.200pt]{4.818pt}{0.400pt}}
\put(120,287){\makebox(0,0)[r]{1(0)}}
\put(1419.0,287.0){\rule[-0.200pt]{4.818pt}{0.400pt}}
\put(140.0,450.0){\rule[-0.200pt]{4.818pt}{0.400pt}}
\put(120,450){\makebox(0,0)[r]{1(0)}}
\put(1419.0,450.0){\rule[-0.200pt]{4.818pt}{0.400pt}}
\put(140.0,614.0){\rule[-0.200pt]{4.818pt}{0.400pt}}
\put(120,614){\makebox(0,0)[r]{1(0)}}
\put(1419.0,614.0){\rule[-0.200pt]{4.818pt}{0.400pt}}
\put(140.0,777.0){\rule[-0.200pt]{4.818pt}{0.400pt}}
\put(120,777){\makebox(0,0)[r]{1}}
\put(1419.0,777.0){\rule[-0.200pt]{4.818pt}{0.400pt}}
\put(273.0,123.0){\rule[-0.200pt]{0.400pt}{4.818pt}}
\put(273,82){\makebox(0,0){ 50}}
\put(273.0,757.0){\rule[-0.200pt]{0.400pt}{4.818pt}}
\put(440.0,123.0){\rule[-0.200pt]{0.400pt}{4.818pt}}
\put(440,82){\makebox(0,0){ 100}}
\put(440.0,757.0){\rule[-0.200pt]{0.400pt}{4.818pt}}
\put(606.0,123.0){\rule[-0.200pt]{0.400pt}{4.818pt}}
\put(606,82){\makebox(0,0){ 150}}
\put(606.0,757.0){\rule[-0.200pt]{0.400pt}{4.818pt}}
\put(773.0,123.0){\rule[-0.200pt]{0.400pt}{4.818pt}}
\put(773,82){\makebox(0,0){ 200}}
\put(773.0,757.0){\rule[-0.200pt]{0.400pt}{4.818pt}}
\put(939.0,123.0){\rule[-0.200pt]{0.400pt}{4.818pt}}
\put(939,82){\makebox(0,0){ 250}}
\put(939.0,757.0){\rule[-0.200pt]{0.400pt}{4.818pt}}
\put(1106.0,123.0){\rule[-0.200pt]{0.400pt}{4.818pt}}
\put(1106,82){\makebox(0,0){ 300}}
\put(1106.0,757.0){\rule[-0.200pt]{0.400pt}{4.818pt}}
\put(1272.0,123.0){\rule[-0.200pt]{0.400pt}{4.818pt}}
\put(1272,82){\makebox(0,0){ 350}}
\put(1272.0,757.0){\rule[-0.200pt]{0.400pt}{4.818pt}}
\put(1439.0,123.0){\rule[-0.200pt]{0.400pt}{4.818pt}}
\put(1439,82){\makebox(0,0){ 400}}
\put(1439.0,757.0){\rule[-0.200pt]{0.400pt}{4.818pt}}
\put(140.0,123.0){\rule[-0.200pt]{312.929pt}{0.400pt}}
\put(1439.0,123.0){\rule[-0.200pt]{0.400pt}{157.549pt}}
\put(140.0,777.0){\rule[-0.200pt]{312.929pt}{0.400pt}}
\put(789,21){\makebox(0,0){$n$}}
\put(789,839){\makebox(0,0){$d/n=0.2$}}
\put(1173,254){\makebox(0,0)[l]{$\gamma=-5$dB}}
\put(1173,417){\makebox(0,0)[l]{$\gamma=-3$dB}}
\put(1173,581){\makebox(0,0)[l]{$\gamma=-1$dB}}
\put(1173,744){\makebox(0,0)[l]{$\gamma=1$dB}}
\put(-59,515){\makebox(0,0)[l]{$\tilde{A}_{n-d}(\lambda)$}}
\put(140.0,123.0){\rule[-0.200pt]{0.400pt}{157.549pt}}
\put(140,287){\rule{1pt}{1pt}}
\put(173,287){\rule{1pt}{1pt}}
\put(207,287){\rule{1pt}{1pt}}
\put(240,287){\rule{1pt}{1pt}}
\put(273,274){\rule{1pt}{1pt}}
\put(307,260){\rule{1pt}{1pt}}
\put(340,251){\rule{1pt}{1pt}}
\put(373,242){\rule{1pt}{1pt}}
\put(406,235){\rule{1pt}{1pt}}
\put(440,230){\rule{1pt}{1pt}}
\put(473,225){\rule{1pt}{1pt}}
\put(506,220){\rule{1pt}{1pt}}
\put(540,217){\rule{1pt}{1pt}}
\put(573,213){\rule{1pt}{1pt}}
\put(606,210){\rule{1pt}{1pt}}
\put(640,207){\rule{1pt}{1pt}}
\put(673,205){\rule{1pt}{1pt}}
\put(706,202){\rule{1pt}{1pt}}
\put(740,200){\rule{1pt}{1pt}}
\put(773,198){\rule{1pt}{1pt}}
\put(806,196){\rule{1pt}{1pt}}
\put(839,195){\rule{1pt}{1pt}}
\put(873,193){\rule{1pt}{1pt}}
\put(906,192){\rule{1pt}{1pt}}
\put(939,190){\rule{1pt}{1pt}}
\put(973,189){\rule{1pt}{1pt}}
\put(1006,188){\rule{1pt}{1pt}}
\put(1039,187){\rule{1pt}{1pt}}
\put(1073,186){\rule{1pt}{1pt}}
\put(1106,184){\rule{1pt}{1pt}}
\put(1139,184){\rule{1pt}{1pt}}
\put(1173,183){\rule{1pt}{1pt}}
\put(1206,182){\rule{1pt}{1pt}}
\put(1239,181){\rule{1pt}{1pt}}
\put(1272,180){\rule{1pt}{1pt}}
\put(1306,179){\rule{1pt}{1pt}}
\put(1339,178){\rule{1pt}{1pt}}
\put(1372,178){\rule{1pt}{1pt}}
\put(1406,177){\rule{1pt}{1pt}}
\put(1439,176){\rule{1pt}{1pt}}
\put(140,450){\rule{1pt}{1pt}}
\put(173,450){\rule{1pt}{1pt}}
\put(207,450){\rule{1pt}{1pt}}
\put(240,450){\rule{1pt}{1pt}}
\put(273,444){\rule{1pt}{1pt}}
\put(307,431){\rule{1pt}{1pt}}
\put(340,420){\rule{1pt}{1pt}}
\put(373,412){\rule{1pt}{1pt}}
\put(406,404){\rule{1pt}{1pt}}
\put(440,398){\rule{1pt}{1pt}}
\put(473,393){\rule{1pt}{1pt}}
\put(506,388){\rule{1pt}{1pt}}
\put(540,384){\rule{1pt}{1pt}}
\put(573,381){\rule{1pt}{1pt}}
\put(606,378){\rule{1pt}{1pt}}
\put(640,375){\rule{1pt}{1pt}}
\put(673,372){\rule{1pt}{1pt}}
\put(706,370){\rule{1pt}{1pt}}
\put(740,367){\rule{1pt}{1pt}}
\put(773,365){\rule{1pt}{1pt}}
\put(806,364){\rule{1pt}{1pt}}
\put(839,362){\rule{1pt}{1pt}}
\put(873,360){\rule{1pt}{1pt}}
\put(906,359){\rule{1pt}{1pt}}
\put(939,357){\rule{1pt}{1pt}}
\put(973,356){\rule{1pt}{1pt}}
\put(1006,354){\rule{1pt}{1pt}}
\put(1039,353){\rule{1pt}{1pt}}
\put(1073,352){\rule{1pt}{1pt}}
\put(1106,351){\rule{1pt}{1pt}}
\put(1139,350){\rule{1pt}{1pt}}
\put(1173,349){\rule{1pt}{1pt}}
\put(1206,348){\rule{1pt}{1pt}}
\put(1239,347){\rule{1pt}{1pt}}
\put(1272,346){\rule{1pt}{1pt}}
\put(1306,345){\rule{1pt}{1pt}}
\put(1339,344){\rule{1pt}{1pt}}
\put(1372,344){\rule{1pt}{1pt}}
\put(1406,343){\rule{1pt}{1pt}}
\put(1439,342){\rule{1pt}{1pt}}
\sbox{\plotpoint}{\rule[-0.400pt]{0.800pt}{0.800pt}}
\put(140,614){\rule{1pt}{1pt}}
\put(173,614){\rule{1pt}{1pt}}
\put(207,614){\rule{1pt}{1pt}}
\put(240,614){\rule{1pt}{1pt}}
\put(273,614){\rule{1pt}{1pt}}
\put(307,609){\rule{1pt}{1pt}}
\put(340,597){\rule{1pt}{1pt}}
\put(373,587){\rule{1pt}{1pt}}
\put(406,579){\rule{1pt}{1pt}}
\put(440,573){\rule{1pt}{1pt}}
\put(473,567){\rule{1pt}{1pt}}
\put(506,562){\rule{1pt}{1pt}}
\put(540,558){\rule{1pt}{1pt}}
\put(573,554){\rule{1pt}{1pt}}
\put(606,550){\rule{1pt}{1pt}}
\put(640,547){\rule{1pt}{1pt}}
\put(673,544){\rule{1pt}{1pt}}
\put(706,541){\rule{1pt}{1pt}}
\put(740,539){\rule{1pt}{1pt}}
\put(773,537){\rule{1pt}{1pt}}
\put(806,535){\rule{1pt}{1pt}}
\put(839,533){\rule{1pt}{1pt}}
\put(873,531){\rule{1pt}{1pt}}
\put(906,529){\rule{1pt}{1pt}}
\put(939,528){\rule{1pt}{1pt}}
\put(973,526){\rule{1pt}{1pt}}
\put(1006,525){\rule{1pt}{1pt}}
\put(1039,523){\rule{1pt}{1pt}}
\put(1073,522){\rule{1pt}{1pt}}
\put(1106,521){\rule{1pt}{1pt}}
\put(1139,520){\rule{1pt}{1pt}}
\put(1173,519){\rule{1pt}{1pt}}
\put(1206,518){\rule{1pt}{1pt}}
\put(1239,517){\rule{1pt}{1pt}}
\put(1272,516){\rule{1pt}{1pt}}
\put(1306,515){\rule{1pt}{1pt}}
\put(1339,514){\rule{1pt}{1pt}}
\put(1372,513){\rule{1pt}{1pt}}
\put(1406,512){\rule{1pt}{1pt}}
\put(1439,511){\rule{1pt}{1pt}}
\sbox{\plotpoint}{\rule[-0.500pt]{1.000pt}{1.000pt}}
\put(140,777){\rule{1pt}{1pt}}
\put(173,777){\rule{1pt}{1pt}}
\put(207,777){\rule{1pt}{1pt}}
\put(240,777){\rule{1pt}{1pt}}
\put(273,777){\rule{1pt}{1pt}}
\put(307,777){\rule{1pt}{1pt}}
\put(340,777){\rule{1pt}{1pt}}
\put(373,776){\rule{1pt}{1pt}}
\put(406,766){\rule{1pt}{1pt}}
\put(440,758){\rule{1pt}{1pt}}
\put(473,752){\rule{1pt}{1pt}}
\put(506,746){\rule{1pt}{1pt}}
\put(540,740){\rule{1pt}{1pt}}
\put(573,736){\rule{1pt}{1pt}}
\put(606,732){\rule{1pt}{1pt}}
\put(640,728){\rule{1pt}{1pt}}
\put(673,725){\rule{1pt}{1pt}}
\put(706,721){\rule{1pt}{1pt}}
\put(740,718){\rule{1pt}{1pt}}
\put(773,716){\rule{1pt}{1pt}}
\put(806,713){\rule{1pt}{1pt}}
\put(839,711){\rule{1pt}{1pt}}
\put(873,709){\rule{1pt}{1pt}}
\put(906,707){\rule{1pt}{1pt}}
\put(939,705){\rule{1pt}{1pt}}
\put(973,703){\rule{1pt}{1pt}}
\put(1006,702){\rule{1pt}{1pt}}
\put(1039,700){\rule{1pt}{1pt}}
\put(1073,698){\rule{1pt}{1pt}}
\put(1106,697){\rule{1pt}{1pt}}
\put(1139,696){\rule{1pt}{1pt}}
\put(1173,694){\rule{1pt}{1pt}}
\put(1206,693){\rule{1pt}{1pt}}
\put(1239,692){\rule{1pt}{1pt}}
\put(1272,691){\rule{1pt}{1pt}}
\put(1306,690){\rule{1pt}{1pt}}
\put(1339,689){\rule{1pt}{1pt}}
\put(1372,688){\rule{1pt}{1pt}}
\put(1406,687){\rule{1pt}{1pt}}
\put(1439,686){\rule{1pt}{1pt}}
\sbox{\plotpoint}{\rule[-0.200pt]{0.400pt}{0.400pt}}
\put(140,287){\usebox{\plotpoint}}
\put(140.0,287.0){\rule[-0.200pt]{312.929pt}{0.400pt}}
\put(140,450){\usebox{\plotpoint}}
\put(140.0,450.0){\rule[-0.200pt]{312.929pt}{0.400pt}}
\put(140,614){\usebox{\plotpoint}}
\put(140.0,614.0){\rule[-0.200pt]{312.929pt}{0.400pt}}
\end{picture}\\
{\caption{$\tilde{A}_{n-d}(\lambda)$ for fixed $d/n=0.2$ with respect to different $\gamma$.
Notation ``$1(0)$'' represents that the $y$-tic is either $1$ (for the curve below) or $0$ (for the curve above).}\label{and1}}
\end{center}
\end{figure}
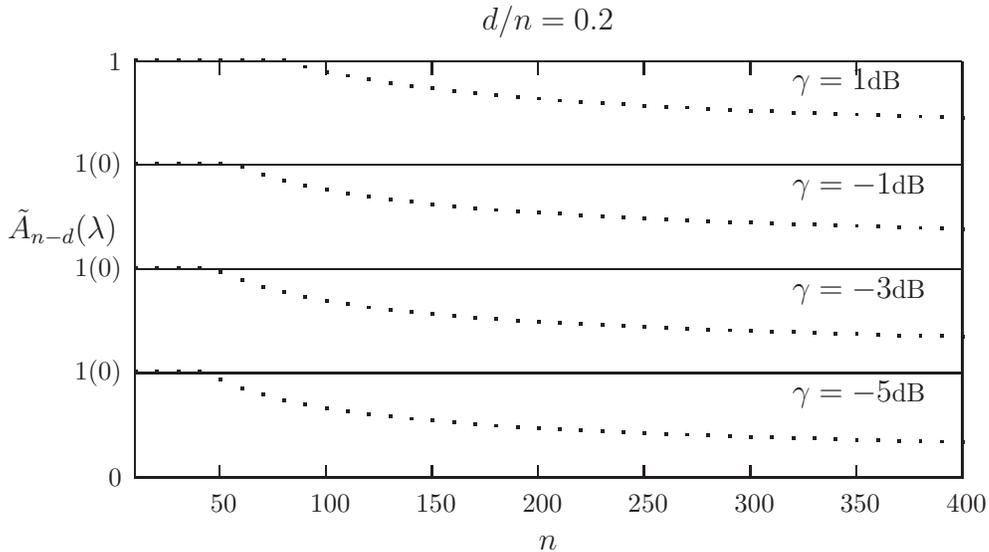

\begin{figure}[t]
\begin{center}
\setlength{\unitlength}{0.240900pt}
\ifx\plotpoint\undefined\newsavebox{\plotpoint}\fi
\sbox{\plotpoint}{\rule[-0.200pt]{0.400pt}{0.400pt}}
\begin{picture}(1500,900)(0,0)
\font\gnuplot=cmr10 at 10pt
\gnuplot
\sbox{\plotpoint}{\rule[-0.200pt]{0.400pt}{0.400pt}}
\put(140.0,123.0){\rule[-0.200pt]{4.818pt}{0.400pt}}
\put(120,123){\makebox(0,0)[r]{0}}
\put(1419.0,123.0){\rule[-0.200pt]{4.818pt}{0.400pt}}
\put(140.0,287.0){\rule[-0.200pt]{4.818pt}{0.400pt}}
\put(120,287){\makebox(0,0)[r]{1(0)}}
\put(1419.0,287.0){\rule[-0.200pt]{4.818pt}{0.400pt}}
\put(140.0,450.0){\rule[-0.200pt]{4.818pt}{0.400pt}}
\put(120,450){\makebox(0,0)[r]{1(0)}}
\put(1419.0,450.0){\rule[-0.200pt]{4.818pt}{0.400pt}}
\put(140.0,614.0){\rule[-0.200pt]{4.818pt}{0.400pt}}
\put(120,614){\makebox(0,0)[r]{1(0)}}
\put(1419.0,614.0){\rule[-0.200pt]{4.818pt}{0.400pt}}
\put(140.0,777.0){\rule[-0.200pt]{4.818pt}{0.400pt}}
\put(120,777){\makebox(0,0)[r]{1}}
\put(1419.0,777.0){\rule[-0.200pt]{4.818pt}{0.400pt}}
\put(273.0,123.0){\rule[-0.200pt]{0.400pt}{4.818pt}}
\put(273,82){\makebox(0,0){ 50}}
\put(273.0,757.0){\rule[-0.200pt]{0.400pt}{4.818pt}}
\put(440.0,123.0){\rule[-0.200pt]{0.400pt}{4.818pt}}
\put(440,82){\makebox(0,0){ 100}}
\put(440.0,757.0){\rule[-0.200pt]{0.400pt}{4.818pt}}
\put(606.0,123.0){\rule[-0.200pt]{0.400pt}{4.818pt}}
\put(606,82){\makebox(0,0){ 150}}
\put(606.0,757.0){\rule[-0.200pt]{0.400pt}{4.818pt}}
\put(773.0,123.0){\rule[-0.200pt]{0.400pt}{4.818pt}}
\put(773,82){\makebox(0,0){ 200}}
\put(773.0,757.0){\rule[-0.200pt]{0.400pt}{4.818pt}}
\put(939.0,123.0){\rule[-0.200pt]{0.400pt}{4.818pt}}
\put(939,82){\makebox(0,0){ 250}}
\put(939.0,757.0){\rule[-0.200pt]{0.400pt}{4.818pt}}
\put(1106.0,123.0){\rule[-0.200pt]{0.400pt}{4.818pt}}
\put(1106,82){\makebox(0,0){ 300}}
\put(1106.0,757.0){\rule[-0.200pt]{0.400pt}{4.818pt}}
\put(1272.0,123.0){\rule[-0.200pt]{0.400pt}{4.818pt}}
\put(1272,82){\makebox(0,0){ 350}}
\put(1272.0,757.0){\rule[-0.200pt]{0.400pt}{4.818pt}}
\put(1439.0,123.0){\rule[-0.200pt]{0.400pt}{4.818pt}}
\put(1439,82){\makebox(0,0){ 400}}
\put(1439.0,757.0){\rule[-0.200pt]{0.400pt}{4.818pt}}
\put(140.0,123.0){\rule[-0.200pt]{312.929pt}{0.400pt}}
\put(1439.0,123.0){\rule[-0.200pt]{0.400pt}{157.549pt}}
\put(140.0,777.0){\rule[-0.200pt]{312.929pt}{0.400pt}}
\put(789,21){\makebox(0,0){$n$}}
\put(789,839){\makebox(0,0){$\gamma=-3$dB}}
\put(1173,254){\makebox(0,0)[l]{$d/n=0.1$}}
\put(1173,417){\makebox(0,0)[l]{$d/n=0.2$}}
\put(1173,581){\makebox(0,0)[l]{$d/n=0.3$}}
\put(1173,744){\makebox(0,0)[l]{$d/n=0.4$}}
\put(-59,515){\makebox(0,0)[l]{$\tilde{A}_{n-d}(\lambda)$}}
\put(140.0,123.0){\rule[-0.200pt]{0.400pt}{157.549pt}}
\put(140,287){\rule{1pt}{1pt}}
\put(173,287){\rule{1pt}{1pt}}
\put(207,287){\rule{1pt}{1pt}}
\put(240,287){\rule{1pt}{1pt}}
\put(273,287){\rule{1pt}{1pt}}
\put(307,281){\rule{1pt}{1pt}}
\put(340,269){\rule{1pt}{1pt}}
\put(373,260){\rule{1pt}{1pt}}
\put(406,252){\rule{1pt}{1pt}}
\put(440,245){\rule{1pt}{1pt}}
\put(473,240){\rule{1pt}{1pt}}
\put(506,235){\rule{1pt}{1pt}}
\put(540,230){\rule{1pt}{1pt}}
\put(573,226){\rule{1pt}{1pt}}
\put(606,223){\rule{1pt}{1pt}}
\put(640,220){\rule{1pt}{1pt}}
\put(673,217){\rule{1pt}{1pt}}
\put(706,214){\rule{1pt}{1pt}}
\put(740,212){\rule{1pt}{1pt}}
\put(773,210){\rule{1pt}{1pt}}
\put(806,207){\rule{1pt}{1pt}}
\put(839,205){\rule{1pt}{1pt}}
\put(873,204){\rule{1pt}{1pt}}
\put(906,202){\rule{1pt}{1pt}}
\put(939,200){\rule{1pt}{1pt}}
\put(973,199){\rule{1pt}{1pt}}
\put(1006,197){\rule{1pt}{1pt}}
\put(1039,196){\rule{1pt}{1pt}}
\put(1073,195){\rule{1pt}{1pt}}
\put(1106,194){\rule{1pt}{1pt}}
\put(1139,192){\rule{1pt}{1pt}}
\put(1173,191){\rule{1pt}{1pt}}
\put(1206,190){\rule{1pt}{1pt}}
\put(1239,189){\rule{1pt}{1pt}}
\put(1272,188){\rule{1pt}{1pt}}
\put(1306,187){\rule{1pt}{1pt}}
\put(1339,187){\rule{1pt}{1pt}}
\put(1372,186){\rule{1pt}{1pt}}
\put(1406,185){\rule{1pt}{1pt}}
\put(1439,184){\rule{1pt}{1pt}}
\put(140,450){\rule{1pt}{1pt}}
\put(173,450){\rule{1pt}{1pt}}
\put(207,450){\rule{1pt}{1pt}}
\put(240,450){\rule{1pt}{1pt}}
\put(273,444){\rule{1pt}{1pt}}
\put(307,431){\rule{1pt}{1pt}}
\put(340,420){\rule{1pt}{1pt}}
\put(373,412){\rule{1pt}{1pt}}
\put(406,404){\rule{1pt}{1pt}}
\put(440,398){\rule{1pt}{1pt}}
\put(473,393){\rule{1pt}{1pt}}
\put(506,388){\rule{1pt}{1pt}}
\put(540,384){\rule{1pt}{1pt}}
\put(573,381){\rule{1pt}{1pt}}
\put(606,378){\rule{1pt}{1pt}}
\put(640,375){\rule{1pt}{1pt}}
\put(673,372){\rule{1pt}{1pt}}
\put(706,370){\rule{1pt}{1pt}}
\put(740,367){\rule{1pt}{1pt}}
\put(773,365){\rule{1pt}{1pt}}
\put(806,364){\rule{1pt}{1pt}}
\put(839,362){\rule{1pt}{1pt}}
\put(873,360){\rule{1pt}{1pt}}
\put(906,359){\rule{1pt}{1pt}}
\put(939,357){\rule{1pt}{1pt}}
\put(973,356){\rule{1pt}{1pt}}
\put(1006,354){\rule{1pt}{1pt}}
\put(1039,353){\rule{1pt}{1pt}}
\put(1073,352){\rule{1pt}{1pt}}
\put(1106,351){\rule{1pt}{1pt}}
\put(1139,350){\rule{1pt}{1pt}}
\put(1173,349){\rule{1pt}{1pt}}
\put(1206,348){\rule{1pt}{1pt}}
\put(1239,347){\rule{1pt}{1pt}}
\put(1272,346){\rule{1pt}{1pt}}
\put(1306,345){\rule{1pt}{1pt}}
\put(1339,344){\rule{1pt}{1pt}}
\put(1372,344){\rule{1pt}{1pt}}
\put(1406,343){\rule{1pt}{1pt}}
\put(1439,342){\rule{1pt}{1pt}}
\sbox{\plotpoint}{\rule[-0.400pt]{0.800pt}{0.800pt}}
\put(140,614){\rule{1pt}{1pt}}
\put(173,614){\rule{1pt}{1pt}}
\put(207,614){\rule{1pt}{1pt}}
\put(240,614){\rule{1pt}{1pt}}
\put(273,607){\rule{1pt}{1pt}}
\put(307,594){\rule{1pt}{1pt}}
\put(340,583){\rule{1pt}{1pt}}
\put(373,575){\rule{1pt}{1pt}}
\put(406,567){\rule{1pt}{1pt}}
\put(440,561){\rule{1pt}{1pt}}
\put(473,556){\rule{1pt}{1pt}}
\put(506,551){\rule{1pt}{1pt}}
\put(540,547){\rule{1pt}{1pt}}
\put(573,544){\rule{1pt}{1pt}}
\put(606,541){\rule{1pt}{1pt}}
\put(640,538){\rule{1pt}{1pt}}
\put(673,535){\rule{1pt}{1pt}}
\put(706,533){\rule{1pt}{1pt}}
\put(740,530){\rule{1pt}{1pt}}
\put(773,528){\rule{1pt}{1pt}}
\put(806,527){\rule{1pt}{1pt}}
\put(839,525){\rule{1pt}{1pt}}
\put(873,523){\rule{1pt}{1pt}}
\put(906,522){\rule{1pt}{1pt}}
\put(939,520){\rule{1pt}{1pt}}
\put(973,519){\rule{1pt}{1pt}}
\put(1006,517){\rule{1pt}{1pt}}
\put(1039,516){\rule{1pt}{1pt}}
\put(1073,515){\rule{1pt}{1pt}}
\put(1106,514){\rule{1pt}{1pt}}
\put(1139,513){\rule{1pt}{1pt}}
\put(1173,512){\rule{1pt}{1pt}}
\put(1206,511){\rule{1pt}{1pt}}
\put(1239,510){\rule{1pt}{1pt}}
\put(1272,509){\rule{1pt}{1pt}}
\put(1306,508){\rule{1pt}{1pt}}
\put(1339,508){\rule{1pt}{1pt}}
\put(1372,507){\rule{1pt}{1pt}}
\put(1406,506){\rule{1pt}{1pt}}
\put(1439,505){\rule{1pt}{1pt}}
\sbox{\plotpoint}{\rule[-0.500pt]{1.000pt}{1.000pt}}
\put(140,777){\rule{1pt}{1pt}}
\put(173,777){\rule{1pt}{1pt}}
\put(207,777){\rule{1pt}{1pt}}
\put(240,777){\rule{1pt}{1pt}}
\put(273,776){\rule{1pt}{1pt}}
\put(307,762){\rule{1pt}{1pt}}
\put(340,752){\rule{1pt}{1pt}}
\put(373,743){\rule{1pt}{1pt}}
\put(406,735){\rule{1pt}{1pt}}
\put(440,729){\rule{1pt}{1pt}}
\put(473,723){\rule{1pt}{1pt}}
\put(506,719){\rule{1pt}{1pt}}
\put(540,714){\rule{1pt}{1pt}}
\put(573,711){\rule{1pt}{1pt}}
\put(606,707){\rule{1pt}{1pt}}
\put(640,705){\rule{1pt}{1pt}}
\put(673,702){\rule{1pt}{1pt}}
\put(706,699){\rule{1pt}{1pt}}
\put(740,697){\rule{1pt}{1pt}}
\put(773,695){\rule{1pt}{1pt}}
\put(806,693){\rule{1pt}{1pt}}
\put(839,691){\rule{1pt}{1pt}}
\put(873,689){\rule{1pt}{1pt}}
\put(906,688){\rule{1pt}{1pt}}
\put(939,686){\rule{1pt}{1pt}}
\put(973,685){\rule{1pt}{1pt}}
\put(1006,683){\rule{1pt}{1pt}}
\put(1039,682){\rule{1pt}{1pt}}
\put(1073,681){\rule{1pt}{1pt}}
\put(1106,680){\rule{1pt}{1pt}}
\put(1139,679){\rule{1pt}{1pt}}
\put(1173,678){\rule{1pt}{1pt}}
\put(1206,677){\rule{1pt}{1pt}}
\put(1239,676){\rule{1pt}{1pt}}
\put(1272,675){\rule{1pt}{1pt}}
\put(1306,674){\rule{1pt}{1pt}}
\put(1339,673){\rule{1pt}{1pt}}
\put(1372,672){\rule{1pt}{1pt}}
\put(1406,672){\rule{1pt}{1pt}}
\put(1439,671){\rule{1pt}{1pt}}
\sbox{\plotpoint}{\rule[-0.200pt]{0.400pt}{0.400pt}}
\put(140,287){\usebox{\plotpoint}}
\put(140.0,287.0){\rule[-0.200pt]{312.929pt}{0.400pt}}
\put(140,450){\usebox{\plotpoint}}
\put(140.0,450.0){\rule[-0.200pt]{312.929pt}{0.400pt}}
\put(140,614){\usebox{\plotpoint}}
\put(140.0,614.0){\rule[-0.200pt]{312.929pt}{0.400pt}}
\end{picture}
{\caption{$\tilde{A}_{n-d}(\lambda)$ for fixed $\gamma=-3dB$ with respect to different $d/n$ ratios.
Notation ``$1(0)$'' represents that the $y$-tic is either $1$ (for the curve below) or $0$ (for the curve above).}\label{and2}}
\end{center}
\end{figure}
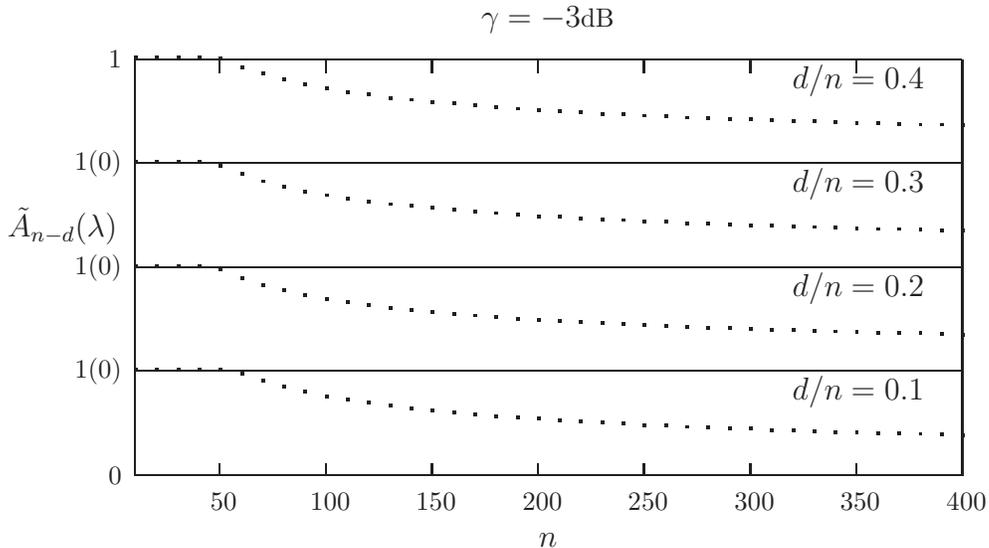

We end this section by presenting the operational meanings of the three
arguments in function ${\cal B}(\cdot,\cdot,\cdot)$ before their practice in
subsequent sections. When in use for sequential-type decoding complexity
analysis, the first integer argument is the Hamming distance between the
transmitted codeword and the examined codeword up to the level of the currently
visited tree node. The second integer argument represents a prediction of the
future route, which is not yet occurred, and hence in our complexity analysis,
is always equal to the maximum length of the codewords (resp.~$n$ for GDA
algorithm and $N$ for MLSDA algorithm) minus the length of the codeword portion
of the current visited node (resp.~$\ell$ for GDA algorithm and $\ell n$ for
MLSDA algorithm).\footnote{ The metric for use of sequential-type decoding can
be generally divided into two parts, where the first part is determined by the
{\em past} branches traversed thus far, while the second part helps predicting
the {\em future} route to speed up the code search process \cite{Han}. For
example, by adding a constant term $\sum_{i=1}^N\log_2\Pr(y_i)$ to the
accumulant Fano metric
$\sum_{i=1}^q\left(\log_2[\Pr(y_j|b_j)/\Pr(y_j)]-R\right)$ up to level $q$, it
can be seen that $\sum_{i=1}^q\left(\log_2(\Pr(y_j|b_j)-R\right)$ weights the
history, and $\sum_{i=q+1}^N\log_2\Pr(y_j)$ is the expectation of branch
metrics to be added for possible future routes. Based on the intuition, the
first argument and the second argument respectively realize the {\em
historical} known part and the {\em future} predictive part of the decoding
metric.}  The third argument is exactly the signal-to-noise ratio for the
decoding environment, and is reasonably assumed to be always positive.

\section{Analysis of the Computational Effort of the Simplified Generalized
Dijkstra's Algorithm} \label{SEC:GDA}

In 1993, a novel and fast maximum-likelihood soft-decision
decoding algorithm for linear block codes was proposed in
\cite{HAN93}, and was called the generalized Dijkstra's
algorithm (GDA). Computer simulations have shown that
the algorithm is highly
efficient (that is, with small average computational
effort) for certain number of linear block codes \cite{EKR96,HAN93}.
Improvements of the GDA have been subsequently reported
\cite{AGU98,EKR96,GAZ97,HAN981,HAN98,KSC98,SHI98}.

The authors of \cite{HAN98} proposed an upper bound on the average
computational effort of an ordering-free variant of the GDA for linear block
codes antipodally transmitted via the AWGN channel; however, the bound is valid
only for codes with sufficiently large codeword length. In terms of the large
deviations technique and Berry-Esseen inequality, an alternative upper bound
that holds for {\it any} (thus including, {\em small}) codeword length can be
given.

\subsection{Notations and definitions}
\label{SEC:GDAA}

Let \C\ be an $(n,k)$ binary linear block code with codeword length $n$
and dimension $k$, and let $R\Bydef k/n$
 be  the {\it code rate} of \C. Denote the codeword of \C\ by
$\xx\Bydef(x_0, x_1, . . . , x_{n-1})$. Also, denote by
$\rr=(r_0,r_1,\ldots,r_{n-1})$ the received vector due to a
codeword of \C\ is transmitted via a time-discrete memoryless
channel.

 From \cite{BEE86} (also \cite{VAR91a,VAR91}),
the {\em maximum-likelihood \em{(ML)} estimate} $\hat{\xx}$=$(\hat
x_0$, $\hat x_1$, $\ldots$, $\hat x_{n-1})$ for a time-discrete
memoryless channel, upon the receipt of $\rr$, satisfies
\begin{equation}
\mem{ml}
\sum^{n-1}_{j=0}\ \left(\phi_j-(-1)^{\hat x_{j}}\right)^2\le
\sum^{n-1}_{j=0} \ \left( \phi_j-(-1)^{x_{j}}\right)^2\ \mbox{ for
all } \xx\in\C,
\end{equation}
 where
 $ \phi_j~\bydef~\ln[\Pr(r_j|0)/\Pr(r_j|1)]$.
An immediate implication of equation \rec{ml} is
that using the log-likelihood ratio vector $\mbox{\bm
$\phi$}=(\phi_0,$ $\phi_1,$ $\ldots,$ $\phi_{n-1})$ rather than the
received vector $\rr$ is sufficient in ML decoding.

When the linear block code is antipodally transmitted through the
AWGN channel, the relationship between the binary codeword $\xx$ and
the received vector {\boldmath $r$} can be characterized by
\beq
\mem{awgn}
r_j=(-1)^{x_j}\sqrt{E}+e_j\quad\mbox{for}\ 0\leq j\leq n-1,
\eeq
where $E$ is the signal energy per channel bit, and $e_j$
represents a noise sample of a Gaussian process with single-sided
noise power per hertz $N_0$.
The signal-to-noise ratio for the
channel is therefore $\gamma\Bydef E/N_0$. In order to account for
the code redundancy for different code rates, the SNR
per information bit $\gamma_b=\gamma/R$ is used instead of $\gamma$ in the following
discussions.

A {\it code tree} of an $(n,k)$ binary linear block code is formed
by representing every codeword as a {\em code path} on a binary
tree of $(n+1)$ levels. A {\em code path} is a particular {\em
path} that begins at the {\em start node} at level 0, and ends at
one of the {\em leaf nodes} at level $n$. There are two branches,
respectively labelled by 0 and 1, that leave each node at the first
$k$ levels. The remaining nodes at levels $k$ through $(n-1)$
consist of only a single leaving branch. The $2^k$ rightmost nodes
at level $n$ are referred to as {\it goal nodes}. In notation,
$\xx_{[\ell]}$ is used to denote a path labelled by
$(x_0,x_1,\ldots,x_{\ell-1})$. For notational convenience, the
subscript ``$[n]$'' is dropped for the label sequence of a code
path, namely $\xx_{[n]}$ is briefed by $\xx$. The same notational
convention is adopted for other notation including the received
vector \rr\ and the log-likelihood ratio vector $\phiphi$.

\subsection{Brief description of the GDA}
\label{SEC:GDAA2}

For completeness,
we brief the GDA decoding algorithm in \cite{HAN93} in this
subsection.

After obtaining the log-likelihood ratio vector $\mbox{\bm
$\phi$}=(\phi_0,$ $\phi_1,$ $\ldots,$ $\phi_{n-1})$, the GDA
algorithm first permutes the positions of codeword components such
that the codeword component that corresponds to larger absolute
value of log-likelihood ratio appears earlier in its position
whenever possible, and still the first $k$ positions uniquely
determine a code path. The post-permutation codewords thereby
result in a new code tree \C$^{\ast}$. Let {\bm
$\phi$}$^\ast~\bydef~(\phi^\ast_0,\phi^\ast_1,\ldots,\phi^\ast_{n-1})$
be the new log-likelihood ratio vector after permutation, and
define the path metric of a path $\xx_{[\ell]}$ (over the new code
tree \C$^{\ast}$) as
$\sum_{j=0}^{\ell-1}(\phi^\ast_{j}-(-1)^{x_{j}})^2.$ The path
metric of a code path $\xx$ is thus given by
$\sum_{j=0}^{n-1}(\phi^\ast_{j}-(-1)^{x_{j}})^2$. The algorithm
then searches for the code path with the minimum path metric over
$\C^{\ast}$, which, from equation \rec{ml}, is exactly the code
path labelled by the permuted ML codeword. As expected, the final
step of the algorithm is to output the de-permuted version of the
labels of the minimum-metric code path.

The search process of the GDA algorithm is guided by an evaluation function
$f(\cdot)$, defined for all paths of a code tree. A simple evaluation function
\cite{HAN981} that guarantees the ultimate finding of the minimum-metric code
path is \begin{eqnarray}
f(\xx_{[\ell]}|\phiphi^\ast)=\sum_{j=0}^{\ell-1}\left(\phi_j^\ast-(-1)^{x_j}\right)^2+
\sum_{j=\ell}^{n-1}\left(|\phi_j^\ast|-1\right)^2.\label{evaluation_function}
\end{eqnarray}
Hence, when a path $\xx_{[\ell]}$ is extended to its immediate successor path
$\xx_{[\ell+1]}$, the evaluation function value is updated by adding the branch
metric, $(\phi_\ell^\ast-(-1)^{x_\ell})^2-\left(|\phi_\ell^\ast|-1\right)^2$,
to its original value. The algorithm begins the search from the path that
contains only the start node. It then extends, among the paths that have been
visited, the path with the smallest $f$-function value. Once the algorithm
chooses to extend a path that ends at a goal node, the search process
terminates. Notably, any path that ends at level $k$ has already uniquely
determined a code path. Hence, once a length-$k$ path is visited and the
$f$-function value associated with its respective code path does not exceed the
associated $f$-function value of any of the later top paths in the stack, the
algorithm can ensure that this code path is the targeted one with the minimum
code path metric. This indicates that the computational complexity of the GDA
is dominantly contributed by those paths up to level $k$. This justifies our
later analysis of the decoding complexity of the GDA, where only the
computations due to those paths up to level $k$ are considered.

The {\em simplified GDA algorithm} is an unpermuted variant of the GDA
algorithm. In other words, its codeword search is operated over the unpermuted
original code tree \C. Although both algorithms yield the same output, the
simplified one was demonstrated to involve a larger branch metric computational
load \cite{HAN98}. We quote the algorithm below.

\begin{list}{Step~\arabic{step}.}
    {\usecounter{step}
    \setlength{\labelwidth}{1cm}
    \setlength{\leftmargin}{1.6cm}\slshape}
\item Put the path that contains only the start node of the code tree
      into the Stack, and assign its evaluation function value as zero.
\item Compute the evaluation function value (as in \eqref{evaluation_function}) for each of the successor paths of the top
      path $\xx_{[\ell]}$ in the Stack by adding the branch metric of
      the extended branch to the evaluation function value of the top path.
      Delete the top path from the Stack.
\item Insert the  successor paths into the Stack in order of
      ascending evaluation function value.
\item If the top path in the Stack ends
      at a goal node, output the codeword corresponding to the top
      path, and the algorithm stops;
       otherwise go to Step 2.
\end{list}

It can be seen from the above algorithm that the simplified GDA algorithm
resembles the stack algorithm except that it uses the evaluation function in
\eqref{evaluation_function} instead of the Fano metric to guide the search on
the code tree, and is designed to decode the block codes rather than the
convolutional codes. In addition, the simplified GDA algorithm is
maximum-likelihood in performance as contrary to the sub-optimality of the
stack algorithm.

\subsection{Analysis of the computational effort of the simplified GDA}
\label{SEC:ANA3}

The computational effort of the
simplified GDA can now be analyzed.

\begin{theorem}[Complexity of the simplified GDA]
\label{THM:MAINGDA}Consider an $(n,k)$ binary linear block code antipodally transmitted
via an AWGN channel.
The average number of branch metric computations evaluated by
the simplified GDA, denoted by ${L_{\rm SGDA}}(\gamma_b)$, is upper-bounded by
\beq
\mem{newbound}
{L_{\rm SGDA}}(\gamma_b)\leq 2\sum_{\ell=0}^{k-1}\sum_{d=0}^\ell
{\ell\choose d}{\cal B}\left(d,n-\ell,k\gamma_b/n\right),
\eeq
where function ${\cal
B}(\cdot,\cdot,\cdot)$ is defined in Lemma~\ref{coro}.
\end{theorem}
\begin{proof}
Assume without loss of generality that the
all-zero codeword \0 is transmitted.

Let $\xx^\ast$ label the minimum-metric code path for a given
log-likelihood ratio vector $\phiphi$. Then we quote from
\cite{HAN98} that for any path $\xx_{[\ell]}$ selected for
extension by the simplified GDA,
$$f(\xx_{[\ell]}|\phiphi)
\le \sum_{j=0}^{n-1}\left(\phi_j-(-1)^{x^\ast_j}\right)^2,$$
which implies that for $\ell<k$,
\begin{eqnarray}
&&\Pr\left[{\rm path}\ \xx_{[\ell]}\ {\rm is\ extended\ by\ the\
simplified\ GDA}\right]\nono\\
&\leq& \Pr\left[f(\xx_{[\ell]}|\phiphi)\leq
\sum_{j=0}^{n-1}\left(\phi_j-(-1)^{x^\ast_j}\right)^2\right]\mem{last0}\\
&\leq&\Pr\left[f(\xx_{[\ell]}|\phiphi)\leq
\sum_{j=0}^{n-1}\left(\phi_j-(-1)^{0}\right)^2\right],\mem{last1}\\
&=&\Pr\left[\sum_{j=0}^{\ell-1}
\left(\phi_j-(-1)^{x_j}\right)^2+\sum_{j=\ell}^{n-1}\left(|\phi_j|-1\right)^2
\leq \sum_{j=0}^{n-1}(\phi_j-1)^2\right],\mem{last2}
\end{eqnarray}
where \rec{last1} follows from the assumption
that the path metric of the $\xx^\ast$-labelled code path
is the smallest with respect to $\phiphi$, and hence, does not
exceed that of the $\0$-labelled code path.

Now denote by ${\cal J}={\cal J}(\xx_{[\ell]})$ the set of index
$j$, where $0\leq j\leq\ell -1$, for which $x_j=1$. Then
\rec{last2} can be rewritten as
\begin{eqnarray}
&&\Pr\left[{\rm path}\ \xx_{[\ell]}\ {\rm is\ extended\ by\ the\
simplified\ GDA}\right]\nono\\
&\leq&\Pr\left[\sum_{j\in{\cal J}}\phi_j
+\sum_{j=\ell}^{n-1}\min(\phi_j,0)\leq 0\right]\nono\\
&=&\Pr\left[\sum_{j\in{\cal J}}r_j
+\sum_{j=\ell}^{n-1}\min(r_j,0)\leq 0\right]\mem{newrequest}
\end{eqnarray}
where \rec{newrequest} holds since for the AWGN channel specified
in \rec{awgn}, $\phi_j=4\sqrt{E}r_j/N_0$.
As the all-zero codeword is assumed to be transmitted, $r_j$ is
Gaussian distributed with mean $\sqrt{E}$ and variance $N_0/2$.
Hence, Lemma~\ref{coro} can be applied to obtain
$$
\Pr\left[{\rm path}\ \xx_{[\ell]}\ {\rm is\ extended\ by\ the\
simplified\ GDA}\right]\leq{\cal
B}\left(d,n-\ell,R\gamma_b\right),
$$
where $d=|{\cal J}|$ is the Hamming weight of $\xx_{[\ell]}$.

Observe that the extension of each path that ends at level $\ell$,
where $\ell<k$, causes two branch metric computations.
Therefore, the expectation value of the number of branch metric
evaluations satisfies
$$
{L_{\rm SGDA}}(\gamma_b)\leq 2\sum_{\ell=0}^{k-1}\sum_{d=0}^\ell
{\ell\choose d}{\cal B}\left(d,n-\ell,R\gamma_b\right).
$$
\end{proof}

\subsection{Numerical and simulation results}
\label{NSRGDA}

The accuracy of the previously derived
theoretical upper bound for the average
computational effort of the simplified GDA is now empirically studied.
Two linear block codes are considered ---
one is a $(24,12)$ binary extended Golay code,
and the other is
a $(48,24)$ binary extended quadratic residue code.

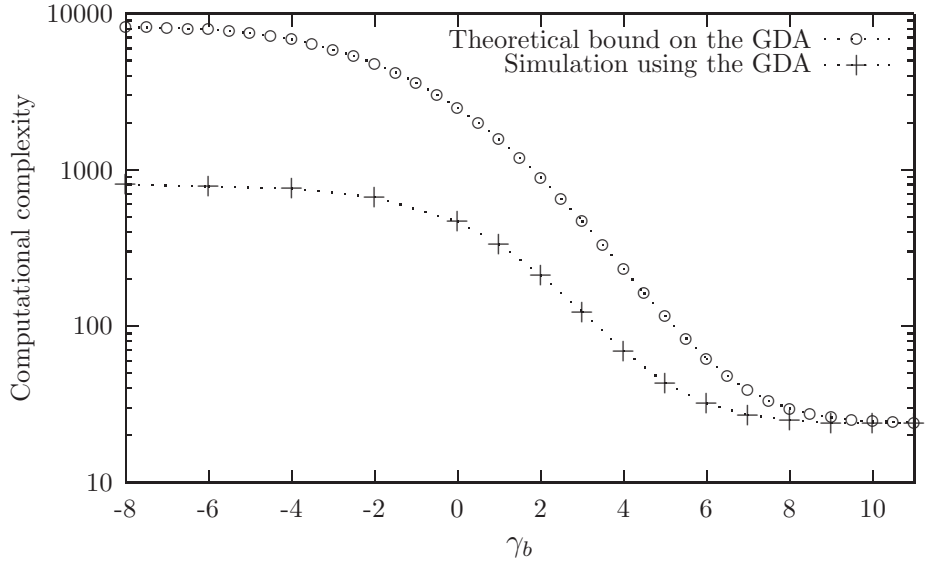
\begin{figure}[hbt]
\begin{center}
\setlength{\unitlength}{0.240900pt}
\ifx\plotpoint\undefined\newsavebox{\plotpoint}\fi
\sbox{\plotpoint}{\rule[-0.200pt]{0.400pt}{0.400pt}}
\begin{picture}(1500,900)(0,0)
\font\gnuplot=cmr10 at 10pt \gnuplot
\sbox{\plotpoint}{\rule[-0.200pt]{0.400pt}{0.400pt}}
\put(201.0,123.0){\rule[-0.200pt]{4.818pt}{0.400pt}}
\put(181,123){\makebox(0,0)[r]{10}}
\put(1419.0,123.0){\rule[-0.200pt]{4.818pt}{0.400pt}}
\put(201.0,197.0){\rule[-0.200pt]{2.409pt}{0.400pt}}
\put(1429.0,197.0){\rule[-0.200pt]{2.409pt}{0.400pt}}
\put(201.0,240.0){\rule[-0.200pt]{2.409pt}{0.400pt}}
\put(1429.0,240.0){\rule[-0.200pt]{2.409pt}{0.400pt}}
\put(201.0,271.0){\rule[-0.200pt]{2.409pt}{0.400pt}}
\put(1429.0,271.0){\rule[-0.200pt]{2.409pt}{0.400pt}}
\put(201.0,295.0){\rule[-0.200pt]{2.409pt}{0.400pt}}
\put(1429.0,295.0){\rule[-0.200pt]{2.409pt}{0.400pt}}
\put(201.0,314.0){\rule[-0.200pt]{2.409pt}{0.400pt}}
\put(1429.0,314.0){\rule[-0.200pt]{2.409pt}{0.400pt}}
\put(201.0,331.0){\rule[-0.200pt]{2.409pt}{0.400pt}}
\put(1429.0,331.0){\rule[-0.200pt]{2.409pt}{0.400pt}}
\put(201.0,345.0){\rule[-0.200pt]{2.409pt}{0.400pt}}
\put(1429.0,345.0){\rule[-0.200pt]{2.409pt}{0.400pt}}
\put(201.0,357.0){\rule[-0.200pt]{2.409pt}{0.400pt}}
\put(1429.0,357.0){\rule[-0.200pt]{2.409pt}{0.400pt}}
\put(201.0,369.0){\rule[-0.200pt]{4.818pt}{0.400pt}}
\put(181,369){\makebox(0,0)[r]{100}}
\put(1419.0,369.0){\rule[-0.200pt]{4.818pt}{0.400pt}}
\put(201.0,443.0){\rule[-0.200pt]{2.409pt}{0.400pt}}
\put(1429.0,443.0){\rule[-0.200pt]{2.409pt}{0.400pt}}
\put(201.0,486.0){\rule[-0.200pt]{2.409pt}{0.400pt}}
\put(1429.0,486.0){\rule[-0.200pt]{2.409pt}{0.400pt}}
\put(201.0,517.0){\rule[-0.200pt]{2.409pt}{0.400pt}}
\put(1429.0,517.0){\rule[-0.200pt]{2.409pt}{0.400pt}}
\put(201.0,540.0){\rule[-0.200pt]{2.409pt}{0.400pt}}
\put(1429.0,540.0){\rule[-0.200pt]{2.409pt}{0.400pt}}
\put(201.0,560.0){\rule[-0.200pt]{2.409pt}{0.400pt}}
\put(1429.0,560.0){\rule[-0.200pt]{2.409pt}{0.400pt}}
\put(201.0,576.0){\rule[-0.200pt]{2.409pt}{0.400pt}}
\put(1429.0,576.0){\rule[-0.200pt]{2.409pt}{0.400pt}}
\put(201.0,591.0){\rule[-0.200pt]{2.409pt}{0.400pt}}
\put(1429.0,591.0){\rule[-0.200pt]{2.409pt}{0.400pt}}
\put(201.0,603.0){\rule[-0.200pt]{2.409pt}{0.400pt}}
\put(1429.0,603.0){\rule[-0.200pt]{2.409pt}{0.400pt}}
\put(201.0,614.0){\rule[-0.200pt]{4.818pt}{0.400pt}}
\put(181,614){\makebox(0,0)[r]{1000}}
\put(1419.0,614.0){\rule[-0.200pt]{4.818pt}{0.400pt}}
\put(201.0,688.0){\rule[-0.200pt]{2.409pt}{0.400pt}}
\put(1429.0,688.0){\rule[-0.200pt]{2.409pt}{0.400pt}}
\put(201.0,732.0){\rule[-0.200pt]{2.409pt}{0.400pt}}
\put(1429.0,732.0){\rule[-0.200pt]{2.409pt}{0.400pt}}
\put(201.0,762.0){\rule[-0.200pt]{2.409pt}{0.400pt}}
\put(1429.0,762.0){\rule[-0.200pt]{2.409pt}{0.400pt}}
\put(201.0,786.0){\rule[-0.200pt]{2.409pt}{0.400pt}}
\put(1429.0,786.0){\rule[-0.200pt]{2.409pt}{0.400pt}}
\put(201.0,805.0){\rule[-0.200pt]{2.409pt}{0.400pt}}
\put(1429.0,805.0){\rule[-0.200pt]{2.409pt}{0.400pt}}
\put(201.0,822.0){\rule[-0.200pt]{2.409pt}{0.400pt}}
\put(1429.0,822.0){\rule[-0.200pt]{2.409pt}{0.400pt}}
\put(201.0,836.0){\rule[-0.200pt]{2.409pt}{0.400pt}}
\put(1429.0,836.0){\rule[-0.200pt]{2.409pt}{0.400pt}}
\put(201.0,849.0){\rule[-0.200pt]{2.409pt}{0.400pt}}
\put(1429.0,849.0){\rule[-0.200pt]{2.409pt}{0.400pt}}
\put(201.0,860.0){\rule[-0.200pt]{4.818pt}{0.400pt}}
\put(181,860){\makebox(0,0)[r]{10000}}
\put(1419.0,860.0){\rule[-0.200pt]{4.818pt}{0.400pt}}
\put(201.0,123.0){\rule[-0.200pt]{0.400pt}{4.818pt}}
\put(201,82){\makebox(0,0){-8}}
\put(201.0,840.0){\rule[-0.200pt]{0.400pt}{4.818pt}}
\put(331.0,123.0){\rule[-0.200pt]{0.400pt}{4.818pt}}
\put(331,82){\makebox(0,0){-6}}
\put(331.0,840.0){\rule[-0.200pt]{0.400pt}{4.818pt}}
\put(462.0,123.0){\rule[-0.200pt]{0.400pt}{4.818pt}}
\put(462,82){\makebox(0,0){-4}}
\put(462.0,840.0){\rule[-0.200pt]{0.400pt}{4.818pt}}
\put(592.0,123.0){\rule[-0.200pt]{0.400pt}{4.818pt}}
\put(592,82){\makebox(0,0){-2}}
\put(592.0,840.0){\rule[-0.200pt]{0.400pt}{4.818pt}}
\put(722.0,123.0){\rule[-0.200pt]{0.400pt}{4.818pt}}
\put(722,82){\makebox(0,0){0}}
\put(722.0,840.0){\rule[-0.200pt]{0.400pt}{4.818pt}}
\put(853.0,123.0){\rule[-0.200pt]{0.400pt}{4.818pt}}
\put(853,82){\makebox(0,0){2}}
\put(853.0,840.0){\rule[-0.200pt]{0.400pt}{4.818pt}}
\put(983.0,123.0){\rule[-0.200pt]{0.400pt}{4.818pt}}
\put(983,82){\makebox(0,0){4}}
\put(983.0,840.0){\rule[-0.200pt]{0.400pt}{4.818pt}}
\put(1113.0,123.0){\rule[-0.200pt]{0.400pt}{4.818pt}}
\put(1113,82){\makebox(0,0){6}}
\put(1113.0,840.0){\rule[-0.200pt]{0.400pt}{4.818pt}}
\put(1244.0,123.0){\rule[-0.200pt]{0.400pt}{4.818pt}}
\put(1244,82){\makebox(0,0){8}}
\put(1244.0,840.0){\rule[-0.200pt]{0.400pt}{4.818pt}}
\put(1374.0,123.0){\rule[-0.200pt]{0.400pt}{4.818pt}}
\put(1374,82){\makebox(0,0){10}}
\put(1374.0,840.0){\rule[-0.200pt]{0.400pt}{4.818pt}}
\put(201.0,123.0){\rule[-0.200pt]{298.234pt}{0.400pt}}
\put(1439.0,123.0){\rule[-0.200pt]{0.400pt}{177.543pt}}
\put(201.0,860.0){\rule[-0.200pt]{298.234pt}{0.400pt}}
\put(40,491){\makebox(0,0){\rotate[l]{Computational complexity}}}
\put(820,21){\makebox(0,0){$\gamma_b$}}
\put(201.0,123.0){\rule[-0.200pt]{0.400pt}{177.543pt}}
\put(1279,820){\makebox(0,0)[r]{Theoretical bound on the GDA}}
\multiput(1299,820)(20.756,0.000){5}{\usebox{\plotpoint}}
\put(1399,820){\usebox{\plotpoint}} \put(201,838){\usebox{\plotpoint}}
\multiput(201,838)(20.756,0.000){2}{\usebox{\plotpoint}}
\multiput(234,838)(20.745,-0.648){2}{\usebox{\plotpoint}}
\put(284.00,836.45){\usebox{\plotpoint}}
\multiput(299,836)(20.745,-0.648){2}{\usebox{\plotpoint}}
\put(346.18,833.62){\usebox{\plotpoint}}
\multiput(364,832)(20.665,-1.937){2}{\usebox{\plotpoint}}
\multiput(396,829)(20.521,-3.109){2}{\usebox{\plotpoint}}
\put(449.14,820.95){\usebox{\plotpoint}}
\multiput(462,819)(20.276,-4.435){2}{\usebox{\plotpoint}}
\put(509.92,807.66){\usebox{\plotpoint}}
\multiput(527,803)(19.811,-6.191){2}{\usebox{\plotpoint}}
\multiput(559,793)(19.506,-7.093){2}{\usebox{\plotpoint}}
\put(608.05,774.19){\usebox{\plotpoint}}
\multiput(625,767)(18.564,-9.282){2}{\usebox{\plotpoint}}
\multiput(657,751)(17.987,-10.356){2}{\usebox{\plotpoint}}
\multiput(690,732)(17.353,-11.388){2}{\usebox{\plotpoint}}
\multiput(722,711)(17.028,-11.868){2}{\usebox{\plotpoint}}
\multiput(755,688)(16.109,-13.088){2}{\usebox{\plotpoint}}
\multiput(787,662)(15.591,-13.701){2}{\usebox{\plotpoint}}
\multiput(820,633)(15.128,-14.211){2}{\usebox{\plotpoint}}
\multiput(853,602)(14.225,-15.114){2}{\usebox{\plotpoint}}
\multiput(885,568)(14.239,-15.101){3}{\usebox{\plotpoint}}
\multiput(918,533)(13.577,-15.699){2}{\usebox{\plotpoint}}
\multiput(950,496)(13.609,-15.671){2}{\usebox{\plotpoint}}
\multiput(983,458)(13.369,-15.876){3}{\usebox{\plotpoint}}
\multiput(1015,420)(14.025,-15.300){2}{\usebox{\plotpoint}}
\multiput(1048,384)(14.239,-15.101){2}{\usebox{\plotpoint}}
\multiput(1081,349)(14.676,-14.676){3}{\usebox{\plotpoint}}
\multiput(1113,317)(16.064,-13.143){2}{\usebox{\plotpoint}}
\multiput(1146,290)(17.103,-11.759){2}{\usebox{\plotpoint}}
\put(1195.88,258.79){\usebox{\plotpoint}}
\multiput(1211,251)(19.506,-7.093){2}{\usebox{\plotpoint}}
\multiput(1244,239)(19.980,-5.619){2}{\usebox{\plotpoint}}
\put(1294.20,227.24){\usebox{\plotpoint}}
\multiput(1309,225)(20.595,-2.574){2}{\usebox{\plotpoint}}
\put(1356.02,220.09){\usebox{\plotpoint}}
\multiput(1374,219)(20.745,-0.648){2}{\usebox{\plotpoint}}
\multiput(1406,218)(20.746,-0.629){2}{\usebox{\plotpoint}}
\put(1439,217){\usebox{\plotpoint}} \put(201,838){\circle{18}}
\put(234,838){\circle{18}} \put(266,837){\circle{18}}
\put(299,836){\circle{18}} \put(331,835){\circle{18}}
\put(364,832){\circle{18}} \put(396,829){\circle{18}}
\put(429,824){\circle{18}} \put(462,819){\circle{18}}
\put(494,812){\circle{18}} \put(527,803){\circle{18}}
\put(559,793){\circle{18}} \put(592,781){\circle{18}}
\put(625,767){\circle{18}} \put(657,751){\circle{18}}
\put(690,732){\circle{18}} \put(722,711){\circle{18}}
\put(755,688){\circle{18}} \put(787,662){\circle{18}}
\put(820,633){\circle{18}} \put(853,602){\circle{18}}
\put(885,568){\circle{18}} \put(918,533){\circle{18}}
\put(950,496){\circle{18}} \put(983,458){\circle{18}}
\put(1015,420){\circle{18}} \put(1048,384){\circle{18}}
\put(1081,349){\circle{18}} \put(1113,317){\circle{18}}
\put(1146,290){\circle{18}} \put(1178,268){\circle{18}}
\put(1211,251){\circle{18}} \put(1244,239){\circle{18}}
\put(1276,230){\circle{18}} \put(1309,225){\circle{18}}
\put(1341,221){\circle{18}} \put(1374,219){\circle{18}}
\put(1406,218){\circle{18}} \put(1439,217){\circle{18}}
\put(1349,820){\circle{18}} \put(1279,779){\makebox(0,0)[r]{Simulation using
the GDA}} \multiput(1299,779)(20.756,0.000){5}{\usebox{\plotpoint}}
\put(1399,779){\usebox{\plotpoint}} \put(201,591){\usebox{\plotpoint}}
\multiput(201,591)(20.750,-0.479){7}{\usebox{\plotpoint}}
\multiput(331,588)(20.750,-0.475){6}{\usebox{\plotpoint}}
\multiput(462,585)(20.636,-2.222){6}{\usebox{\plotpoint}}
\multiput(592,571)(19.922,-5.823){7}{\usebox{\plotpoint}}
\multiput(722,533)(18.157,-10.056){3}{\usebox{\plotpoint}}
\multiput(787,497)(16.665,-12.372){4}{\usebox{\plotpoint}}
\multiput(853,448)(15.605,-13.685){5}{\usebox{\plotpoint}}
\multiput(918,391)(15.019,-14.326){4}{\usebox{\plotpoint}}
\multiput(983,329)(16.451,-12.655){4}{\usebox{\plotpoint}}
\multiput(1048,279)(18.621,-9.167){3}{\usebox{\plotpoint}}
\multiput(1113,247)(20.003,-5.539){4}{\usebox{\plotpoint}}
\multiput(1178,229)(20.605,-2.498){3}{\usebox{\plotpoint}}
\multiput(1244,221)(20.694,-1.592){3}{\usebox{\plotpoint}}
\multiput(1309,216)(20.756,0.000){3}{\usebox{\plotpoint}}
\multiput(1374,216)(20.756,0.000){3}{\usebox{\plotpoint}}
\put(1439,216){\usebox{\plotpoint}} \put(201,591){\makebox(0,0){$+$}}
\put(331,588){\makebox(0,0){$+$}} \put(462,585){\makebox(0,0){$+$}}
\put(592,571){\makebox(0,0){$+$}} \put(722,533){\makebox(0,0){$+$}}
\put(787,497){\makebox(0,0){$+$}} \put(853,448){\makebox(0,0){$+$}}
\put(918,391){\makebox(0,0){$+$}} \put(983,329){\makebox(0,0){$+$}}
\put(1048,279){\makebox(0,0){$+$}} \put(1113,247){\makebox(0,0){$+$}}
\put(1178,229){\makebox(0,0){$+$}} \put(1244,221){\makebox(0,0){$+$}}
\put(1309,216){\makebox(0,0){$+$}} \put(1374,216){\makebox(0,0){$+$}}
\put(1439,216){\makebox(0,0){$+$}} \put(1349,779){\makebox(0,0){$+$}}
\end{picture}

{\caption{Average computational complexity of the simplified GDA
for $(24,12)$ binary extended Golay code.}\label{gdafig1}}
\end{center}
\end{figure}

\begin{figure}[hbt]
\begin{center}
\setlength{\unitlength}{0.240900pt}
\ifx\plotpoint\undefined\newsavebox{\plotpoint}\fi
\sbox{\plotpoint}{\rule[-0.200pt]{0.400pt}{0.400pt}}
\begin{picture}(1500,900)(0,0)
\font\gnuplot=cmr10 at 10pt \gnuplot
\sbox{\plotpoint}{\rule[-0.200pt]{0.400pt}{0.400pt}}
\put(221.0,123.0){\rule[-0.200pt]{4.818pt}{0.400pt}}
\put(201,123){\makebox(0,0)[r]{10}}
\put(1419.0,123.0){\rule[-0.200pt]{4.818pt}{0.400pt}}
\put(221.0,167.0){\rule[-0.200pt]{2.409pt}{0.400pt}}
\put(1429.0,167.0){\rule[-0.200pt]{2.409pt}{0.400pt}}
\put(221.0,226.0){\rule[-0.200pt]{2.409pt}{0.400pt}}
\put(1429.0,226.0){\rule[-0.200pt]{2.409pt}{0.400pt}}
\put(221.0,256.0){\rule[-0.200pt]{2.409pt}{0.400pt}}
\put(1429.0,256.0){\rule[-0.200pt]{2.409pt}{0.400pt}}
\put(221.0,270.0){\rule[-0.200pt]{4.818pt}{0.400pt}}
\put(201,270){\makebox(0,0)[r]{100}}
\put(1419.0,270.0){\rule[-0.200pt]{4.818pt}{0.400pt}}
\put(221.0,315.0){\rule[-0.200pt]{2.409pt}{0.400pt}}
\put(1429.0,315.0){\rule[-0.200pt]{2.409pt}{0.400pt}}
\put(221.0,373.0){\rule[-0.200pt]{2.409pt}{0.400pt}}
\put(1429.0,373.0){\rule[-0.200pt]{2.409pt}{0.400pt}}
\put(221.0,404.0){\rule[-0.200pt]{2.409pt}{0.400pt}}
\put(1429.0,404.0){\rule[-0.200pt]{2.409pt}{0.400pt}}
\put(221.0,418.0){\rule[-0.200pt]{4.818pt}{0.400pt}}
\put(201,418){\makebox(0,0)[r]{1000}}
\put(1419.0,418.0){\rule[-0.200pt]{4.818pt}{0.400pt}}
\put(221.0,462.0){\rule[-0.200pt]{2.409pt}{0.400pt}}
\put(1429.0,462.0){\rule[-0.200pt]{2.409pt}{0.400pt}}
\put(221.0,521.0){\rule[-0.200pt]{2.409pt}{0.400pt}}
\put(1429.0,521.0){\rule[-0.200pt]{2.409pt}{0.400pt}}
\put(221.0,551.0){\rule[-0.200pt]{2.409pt}{0.400pt}}
\put(1429.0,551.0){\rule[-0.200pt]{2.409pt}{0.400pt}}
\put(221.0,565.0){\rule[-0.200pt]{4.818pt}{0.400pt}}
\put(201,565){\makebox(0,0)[r]{10000}}
\put(1419.0,565.0){\rule[-0.200pt]{4.818pt}{0.400pt}}
\put(221.0,610.0){\rule[-0.200pt]{2.409pt}{0.400pt}}
\put(1429.0,610.0){\rule[-0.200pt]{2.409pt}{0.400pt}}
\put(221.0,668.0){\rule[-0.200pt]{2.409pt}{0.400pt}}
\put(1429.0,668.0){\rule[-0.200pt]{2.409pt}{0.400pt}}
\put(221.0,698.0){\rule[-0.200pt]{2.409pt}{0.400pt}}
\put(1429.0,698.0){\rule[-0.200pt]{2.409pt}{0.400pt}}
\put(221.0,713.0){\rule[-0.200pt]{4.818pt}{0.400pt}}
\put(201,713){\makebox(0,0)[r]{100000}}
\put(1419.0,713.0){\rule[-0.200pt]{4.818pt}{0.400pt}}
\put(221.0,757.0){\rule[-0.200pt]{2.409pt}{0.400pt}}
\put(1429.0,757.0){\rule[-0.200pt]{2.409pt}{0.400pt}}
\put(221.0,816.0){\rule[-0.200pt]{2.409pt}{0.400pt}}
\put(1429.0,816.0){\rule[-0.200pt]{2.409pt}{0.400pt}}
\put(221.0,846.0){\rule[-0.200pt]{2.409pt}{0.400pt}}
\put(1429.0,846.0){\rule[-0.200pt]{2.409pt}{0.400pt}}
\put(221.0,860.0){\rule[-0.200pt]{4.818pt}{0.400pt}}
\put(201,860){\makebox(0,0)[r]{1e+006}}
\put(1419.0,860.0){\rule[-0.200pt]{4.818pt}{0.400pt}}
\put(221.0,123.0){\rule[-0.200pt]{0.400pt}{4.818pt}}
\put(221,82){\makebox(0,0){1}}
\put(221.0,840.0){\rule[-0.200pt]{0.400pt}{4.818pt}}
\put(343.0,123.0){\rule[-0.200pt]{0.400pt}{4.818pt}}
\put(343,82){\makebox(0,0){2}}
\put(343.0,840.0){\rule[-0.200pt]{0.400pt}{4.818pt}}
\put(465.0,123.0){\rule[-0.200pt]{0.400pt}{4.818pt}}
\put(465,82){\makebox(0,0){3}}
\put(465.0,840.0){\rule[-0.200pt]{0.400pt}{4.818pt}}
\put(586.0,123.0){\rule[-0.200pt]{0.400pt}{4.818pt}}
\put(586,82){\makebox(0,0){4}}
\put(586.0,840.0){\rule[-0.200pt]{0.400pt}{4.818pt}}
\put(708.0,123.0){\rule[-0.200pt]{0.400pt}{4.818pt}}
\put(708,82){\makebox(0,0){5}}
\put(708.0,840.0){\rule[-0.200pt]{0.400pt}{4.818pt}}
\put(830.0,123.0){\rule[-0.200pt]{0.400pt}{4.818pt}}
\put(830,82){\makebox(0,0){6}}
\put(830.0,840.0){\rule[-0.200pt]{0.400pt}{4.818pt}}
\put(952.0,123.0){\rule[-0.200pt]{0.400pt}{4.818pt}}
\put(952,82){\makebox(0,0){7}}
\put(952.0,840.0){\rule[-0.200pt]{0.400pt}{4.818pt}}
\put(1074.0,123.0){\rule[-0.200pt]{0.400pt}{4.818pt}}
\put(1074,82){\makebox(0,0){8}}
\put(1074.0,840.0){\rule[-0.200pt]{0.400pt}{4.818pt}}
\put(1195.0,123.0){\rule[-0.200pt]{0.400pt}{4.818pt}}
\put(1195,82){\makebox(0,0){9}}
\put(1195.0,840.0){\rule[-0.200pt]{0.400pt}{4.818pt}}
\put(1317.0,123.0){\rule[-0.200pt]{0.400pt}{4.818pt}}
\put(1317,82){\makebox(0,0){10}}
\put(1317.0,840.0){\rule[-0.200pt]{0.400pt}{4.818pt}}
\put(1439.0,123.0){\rule[-0.200pt]{0.400pt}{4.818pt}}
\put(1439,82){\makebox(0,0){11}}
\put(1439.0,840.0){\rule[-0.200pt]{0.400pt}{4.818pt}}
\put(221.0,123.0){\rule[-0.200pt]{293.416pt}{0.400pt}}
\put(1439.0,123.0){\rule[-0.200pt]{0.400pt}{177.543pt}}
\put(221.0,860.0){\rule[-0.200pt]{293.416pt}{0.400pt}}
\put(40,491){\makebox(0,0){\rotate[l]{Computational complexity}}}
\put(830,21){\makebox(0,0){$\gamma_b$}}
\put(221.0,123.0){\rule[-0.200pt]{0.400pt}{177.543pt}}
\put(1279,820){\makebox(0,0)[r]{Theoretical bound on the GDA}}
\multiput(1299,820)(20.756,0.000){5}{\usebox{\plotpoint}}
\put(1399,820){\usebox{\plotpoint}} \put(221,849){\usebox{\plotpoint}}
\multiput(221,849)(17.487,-11.180){4}{\usebox{\plotpoint}}
\multiput(282,810)(17.095,-11.770){4}{\usebox{\plotpoint}}
\multiput(343,768)(16.702,-12.321){3}{\usebox{\plotpoint}}
\multiput(404,723)(16.311,-12.835){4}{\usebox{\plotpoint}}
\multiput(465,675)(16.181,-12.998){4}{\usebox{\plotpoint}}
\multiput(526,626)(15.814,-13.442){4}{\usebox{\plotpoint}}
\multiput(586,575)(15.923,-13.313){3}{\usebox{\plotpoint}}
\multiput(647,524)(16.052,-13.157){4}{\usebox{\plotpoint}}
\multiput(708,474)(16.311,-12.835){4}{\usebox{\plotpoint}}
\multiput(769,426)(16.702,-12.321){4}{\usebox{\plotpoint}}
\multiput(830,381)(17.357,-11.381){3}{\usebox{\plotpoint}}
\multiput(891,341)(18.130,-10.105){3}{\usebox{\plotpoint}}
\multiput(952,307)(18.863,-8.659){4}{\usebox{\plotpoint}}
\multiput(1013,279)(19.625,-6.756){3}{\usebox{\plotpoint}}
\multiput(1074,258)(20.230,-4.643){3}{\usebox{\plotpoint}}
\multiput(1135,244)(20.526,-3.079){3}{\usebox{\plotpoint}}
\multiput(1195,235)(20.686,-1.696){3}{\usebox{\plotpoint}}
\multiput(1256,230)(20.730,-1.020){3}{\usebox{\plotpoint}}
\multiput(1317,227)(20.744,-0.680){3}{\usebox{\plotpoint}}
\multiput(1378,225)(20.753,-0.340){2}{\usebox{\plotpoint}}
\put(1439,224){\usebox{\plotpoint}} \put(221,849){\circle{18}}
\put(282,810){\circle{18}} \put(343,768){\circle{18}}
\put(404,723){\circle{18}} \put(465,675){\circle{18}}
\put(526,626){\circle{18}} \put(586,575){\circle{18}}
\put(647,524){\circle{18}} \put(708,474){\circle{18}}
\put(769,426){\circle{18}} \put(830,381){\circle{18}}
\put(891,341){\circle{18}} \put(952,307){\circle{18}}
\put(1013,279){\circle{18}} \put(1074,258){\circle{18}}
\put(1135,244){\circle{18}} \put(1195,235){\circle{18}}
\put(1256,230){\circle{18}} \put(1317,227){\circle{18}}
\put(1378,225){\circle{18}} \put(1439,224){\circle{18}}
\put(1349,820){\circle{18}} \put(1279,779){\makebox(0,0)[r]{Simulation using
the GDA}} \multiput(1299,779)(20.756,0.000){5}{\usebox{\plotpoint}}
\put(1399,779){\usebox{\plotpoint}} \put(221,690){\usebox{\plotpoint}}
\multiput(221,690)(19.150,-8.005){7}{\usebox{\plotpoint}}
\multiput(343,639)(17.161,-11.675){7}{\usebox{\plotpoint}}
\multiput(465,556)(16.260,-12.900){7}{\usebox{\plotpoint}}
\multiput(586,460)(16.572,-12.497){8}{\usebox{\plotpoint}}
\multiput(708,368)(17.617,-10.974){7}{\usebox{\plotpoint}}
\multiput(830,292)(19.525,-7.042){6}{\usebox{\plotpoint}}
\multiput(952,248)(20.557,-2.864){6}{\usebox{\plotpoint}}
\multiput(1074,231)(20.738,-0.857){6}{\usebox{\plotpoint}}
\multiput(1195,226)(20.755,-0.170){6}{\usebox{\plotpoint}}
\multiput(1317,225)(20.753,-0.340){5}{\usebox{\plotpoint}}
\put(1439,223){\usebox{\plotpoint}} \put(221,690){\makebox(0,0){$+$}}
\put(343,639){\makebox(0,0){$+$}} \put(465,556){\makebox(0,0){$+$}}
\put(586,460){\makebox(0,0){$+$}} \put(708,368){\makebox(0,0){$+$}}
\put(830,292){\makebox(0,0){$+$}} \put(952,248){\makebox(0,0){$+$}}
\put(1074,231){\makebox(0,0){$+$}} \put(1195,226){\makebox(0,0){$+$}}
\put(1317,225){\makebox(0,0){$+$}} \put(1439,223){\makebox(0,0){$+$}}
\put(1349,779){\makebox(0,0){$+$}}
\end{picture}

{\caption{Average computational complexity of the simplified GDA
for $(48,24)$ binary extended quadratic residue
code.}\label{gdafig2}}
\end{center}
\end{figure}
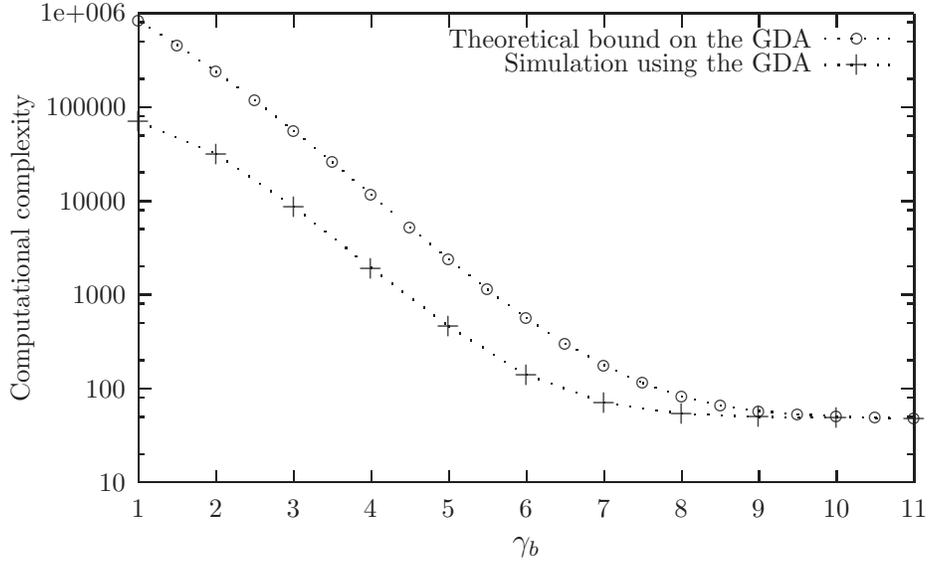

 Figures \ref{gdafig1} and \ref{gdafig2} illustrate the deviation between
the simulated results and the theoretical upper bound in
Theorem~\ref{THM:MAINGDA}. Only one theoretical curve (rather than one enhanced
by Berry-Esseen analysis and the other with simple Chernoff-based analysis) is
plotted in the two figures because no improvement in function ${\cal
B}(\cdot,\cdot,\cdot)$ can be obtained by the introduction of the Berry-Esseen
analysis. According to these figures, the theoretical upper bound is quite
close to the simulation results for high $\gamma_b$ (above $8$ dB). In such a
case, the computational complexity of the simplified GDA reduces to its minimum
possible values, 24 and 48, for $(24,12)$ and $(48,24)$ codes, respectively. As
$\gamma_b$ reaches 1 dB, the theoretical bound for $(48,24)$ code is around 12
times higher than the simulated average complexity. However, for the $(24,12)$
code, the theoretical bound and the simulation results differ only by
$0.671638$ on a log$_{10}$ scale at $\gamma_b=1$ dB, and it is when
$\gamma_b\leq -8$ dB that the upper bound becomes ten times larger than the
simulated complexity. The conclusion section will address the possible cause of
the inaccuracy of the theoretical bounds at low SNR.

\section{Analysis of the Computational Effort of the MLSDA}
\label{SEC:ANA}

Based on
the probability bound established in Lemma~\ref{coro},
the computational complexity of
the maximum-likelihood sequential
decoding algorithm (MLSDA) proposed
in \cite{HANCHEN982} is analyzed for convolutional codes
antipodally transmitted via the AWGN channel.

\subsection{Notation and definitions}
\label{SEC:ANAA}

Let \C\ be an $(n,k,m)$ binary convolutional code, where $k$ is the
number of encoder inputs, $n$ is the number of encoder outputs, and
$m$ is its {\it memory order} defined as the maximum number of shift
register stages from an encoder input to an encoder output. Let $R\Bydef
k/n$ and $N\Bydef n(L+m)$ be the {\it code rate} and
the {\it code length} of \C, respectively, where $L$ represents the length of
applied information sequence. Denote the codeword of \C\ by
$\xx\Bydef(x_0, x_1, . . . , x_{N-1})$. Also denote the left portion of
codeword $\xx$ by $\xx_{(b)}\Bydef(x_0,x_1,\ldots,x_b)$. Assume
that antipodal signaling is used in the codeword transmission such that the
relationship between binary channel codeword $\xx$ and received
vector {\boldmath $r$}$\Bydef(r_0,r_1,\ldots,r_{N-1})$ is
\beq
\mem{awgn1}
r_j=(-1)^{x_j}\sqrt{E}+e_j,\quad0\leq j\leq N-1,
\eeq
where $E$ is the signal energy per channel bit, and $e_j$ is
a noise sample of a Gaussian process with single-sided noise power per hertz
$N_0$.
The signal-to-noise ratio
per information bit
$\gamma_b=(EN)/(N_0kL)$ is again used
to account for the code redundancy
for various code rates.

\begin{figure}[hbt]
\begin{center}
\setlength{\unitlength}{1pt}
\begin{picture}(420,220)(20,0)
\thinlines
\multiput(50,50)(50,0){8}{\circle{15}}
\multiput(50,50)(50,0){8}{\makebox(0,0){$s_0$}}
\multiput(100,150)(50,0){5}{\circle{15}}
\multiput(100,150)(50,0){5}{\makebox(0,0){$s_1$}}
\multiput(150,100)(50,0){5}{\circle{15}}
\multiput(150,100)(50,0){5}{\makebox(0,0){$s_2$}}
\multiput(150,200)(50,0){4}{\circle{15}}
\multiput(150,200)(50,0){4}{\makebox(0,0){$s_3$}}

\multiput(53.75,56.5)(50,0){5}{\line(1,2){43}}
\multiput(105.3,155.3)(50,0){4}{\line(1,1){39.4}}
\multiput(105.3,144.7)(50,0){5}{\line(1,-1){39.4}}
\multiput(155.3,94.7)(50,0){5}{\line(1,-1){39.4}}
\multiput(153.75,193.5)(50,0){4}{\line(1,-2){43}}
\multiput(155.3,105.3)(50,0){3}{\line(1,1){39.4}}
\multiput(57.5,50)(50,0){7}{\line(1,0){35}}
\multiput(157.5,200)(50,0){3}{\line(1,0){35}}

\put(30,50){\makebox(0,30){\shortstack{{\it Start}\\{\it node}}}}
\put(410,50){\makebox(0,40){\shortstack{{\it Goal}\\{\it node}}}}
\put(20,30){\makebox(0,0){level $\ell$}}
\put(50,30){\makebox(0,0){$0$}}
\put(100,30){\makebox(0,0){$1$}}
\put(150,30){\makebox(0,0){$2$}}
\put(200,30){\makebox(0,0){$3$}}
\put(250,30){\makebox(0,0){$4$}}
\put(300,30){\makebox(0,0){$5$}}
\put(350,30){\makebox(0,0){$6$}}
\put(400,30){\makebox(0,0){$7$}}

\multiput(50,55)(50,0){7}{\makebox(50,0){\scriptsize $000$}}
\put(50,50){\makebox(10,60){\scriptsize$111$}}
\multiput(100,50)(50,0){4}{\makebox(2,30){\scriptsize$111$}}
\multiput(100,150)(50,0){5}{\makebox(35,-20){\scriptsize$101$}}
\multiput(100,150)(50,0){4}{\makebox(15,30){\scriptsize$010$}}
\multiput(150,205)(50,0){3}{\makebox(50,0){\scriptsize$001$}}
\multiput(150,200)(50,0){4}{\makebox(30,-25){\scriptsize$110$}}
\multiput(150,100)(50,0){3}{\makebox(15,30){\scriptsize$100$}}
\multiput(150,100)(50,0){5}{\makebox(30,-10){\scriptsize$011$}}

\thicklines
\put(53.75,56.5){\line(1,2){43}}
\put(105.3,155.3){\line(1,1){39.4}}
\put(203.75,193.5){\line(1,-2){43}}
\put(157.5,200){\line(1,0){35}}
\put(255.3,105.3){\line(1,1){39.4}}
\put(305.3,144.7){\line(1,-1){39.4}}
\put(355.3,94.7){\line(1,-1){39.4}}
\end{picture}
\parbox{5in}{\caption{Trellis for a $(3,1,2)$ binary convolutional
code with information length $L=5$. In this case, the code rate $R=1/3$ and
the codeword length $N=3(5+2)=21$.
The code path indicated by the thick line is labelled by
$111$, $010$, $001$, $110$, $100$, $101$ and $011$, thus
its corresponding codeword is
$\xx=(111010001110100101011)$.
}\label{trellis}}
\end{center}
\end{figure}
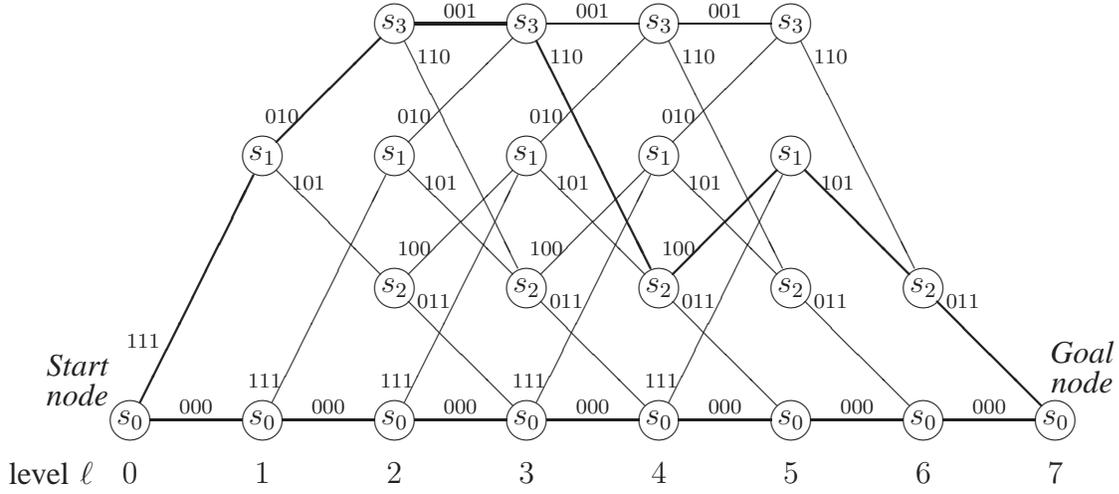

A {\it trellis}, as depicted in Fig.~\ref{trellis} in terms of a
specific example, can be obtained from a code tree by combining
nodes with the same {\it state}. States are characterized by the
content of the shift-register stages in a convolutional encoder.
 For convenience, the leftmost node (at level $0$)
and the rightmost node (at level $L+m$) of
a trellis are named the {\it start node}
and the {\it goal node}, respectively.
A path on a trellis from the single start node to
the single goal node is called a {\it code path}.
Each branch in the trellis is labelled by an appropriate encoder output
of length $n$.

\subsection{Maximum-likelihood soft-decision sequential decoding algorithm
(MLSDA)}
\label{SEC:ANAB}

In \cite{HANCHEN982}, a trellis-based sequential
decoding algorithm specifically for binary convolutional codes
is proposed.
The same paper proves that the algorithm performs
maximum-likelihood decoding, and is thus
named the {\it maximum-likelihood
sequential decoding algorithm} (MLSDA).
Unlike the conventional sequential decoding
algorithm \cite{FOR742,JEL69,LIN83,ZIG66}
which requires only a single stack, the trellis-based
MLSDA maintains two stacks ---
an {\it Open Stack}\ \ and a {\it Closed Stack}.
 For completeness, the algorithm is quoted below.

\begin{list}{Step~\arabic{step}.}
    {\usecounter{step}
    \setlength{\labelwidth}{1cm}
    \setlength{\leftmargin}{1.6cm}\slshape}
\item Put the path that contains only the start node
      into the {\em Open Stack}, and assign its path metric as zero.
\item Compute the path metric for each of the successor paths of the top
      path in the {\em Open Stack} by adding the branch metric of
      the extended branch to the path metric of the top path.
      Put into the {\em Closed Stack}
      both the state and level of the end node of the top path in the {\em Open Stack}.
      Delete the top path from the {\em Open Stack}.
\item Discard any successor path that ends at a node that
      has the same state and level
      as any entry in the {\em Closed Stack}.
      If any successor path
      merges\footnote{``Merging'' of two paths means that the two paths end
      at the same node.}
       with a path already in the {\em Open Stack},
      eliminate the path with higher path metric.
\item Insert the remaining successor paths into the {\em Open Stack} in order of
      ascending path metrics.
\item If the top path in the {\em Open Stack} ends
      at the single goal node, the algorithm stops and output the codeword corresponding to the top path; otherwise go to Step 2.
\end{list}

We remark after the presentation of the MLSDA that the {\em Open Stack}
contains all paths having been visited thus far, but excludes all prefixes of
the paths in it. Hence, the {\em Open Stack} functions in a similar way as the
stack in the conventional sequential decoding algorithm. The {\em Closed Stack}
keeps the information of ending states and ending levels of those paths that
had been the top paths of the {\em Open Stack} at some previous time.
In addition, the path metric for a path labelled by $\xx_{(\ell n-1)}$
 $=$
 $(x_0$, $x_1$, $\ldots$, $x_{\ell n-1})$, upon receipt of $\phiphi_{(\ell n-1)}$, is given by
\beq \mem{pathmetric} \zeta\left(\left.\xx_{(\ell
n-1)}\right|\phiphi_{(\ell n-1)}\right) ~\bydef~\sum_{j=0}^{\ell n-1}(y_j\oplus
x_j)\times|\phi_j|, \eeq where $\phi_j=\log[\Pr(r_j|0)/\Pr(r_j|1)]$ is the
$j$th received log-likelihood ratio, $r_j$ is the $j$th received scalar, and
$y_j=1$ if $\phi_j<0$ and $y_j=0$, otherwise.

\subsection{Analysis of the computational efforts of the MLSDA}

Since the nodes at levels $L$ through $(L+m-1)$
have only one branch leaving them, and $L$ is
typically much larger than $m$,
the contribution of these nodes to the computational complexity
due to path extensions can be reasonably neglected.
Hence, the analysis in the following theorem
only considers those branch metric computations applied up to
level $L$ of the trellis.

Notations that will be used in the next theorem are first introduced.
Denote by $s_j(\ell)$ the node that is located at level $\ell$
and corresponds to state index $j$.
Let ${\cal S}_j(\ell)$ be the set of paths that end
at node $s_j(\ell)$.
Also let ${\cal H}_j(\ell)$ be the set of the Hamming weights
of the paths in ${\cal S}_j(\ell)$.
Denote the minimum
Hamming weight in ${\cal H}_j(\ell)$ by $d_j^\ast(\ell)$.
As an example, ${\cal S}_3(3)$ equals $\{111010001,000111010\}$ in Fig.~\ref{trellis},
which results in ${\cal H}_3(3)=\{5,4\}$ and $d_3^\ast(3)=4$.

\begin{theorem}[Complexity of the MLSDA]
\label{THM:MAIN2}
Consider an $(n,k,m)$ binary convolutional code
transmitted via an AWGN channel.
The average number of branch metric computations
evaluated by the MLSDA,
denoted by ${L_{\rm MLSDA}}(\gamma_b)$, is upper-bounded by
$$
{L_{\rm MLSDA}}(\gamma_b)\leq
2^k\sum_{\ell=0}^{L-1}\sum_{j=0}^{2^m-1}
{\cal B}\left(d_j^\ast(\ell),N-\ell n,\frac{kL}N\gamma_b\right),
$$
where if ${\cal H}_j(\ell)$ is empty, implying
the non-existence of state $j$ at level $\ell$,
then ${\cal B}(d_j^\ast(\ell),N-\ell n,kL\gamma_b/N)=0$.
\end{theorem}
\begin{proof}
Assume without loss of generality that the all-zero
codeword \0 is transmitted.

First, observe that for any two paths that end
at a common node, only one of them will survive in the
Open Stack. In other words, one of the two paths will be discarded either
due to a larger path metric or because
its end node has the same state and level as an entry
in the Closed Stack. In the latter case, the surviving
path has clearly reached the common end node earlier, and has already been extended by
the MLSDA at some previous time (so that the state and level of its end node has
already been stored in the Closed Stack).
Accordingly, unlike the code tree
search in the GDA, the branch metric computations that follow
these two paths will only be performed once.
 It therefore suffices to derive the computational
complexity of the MLSDA based on the nodes that have been extended rather than
the paths that have been extended.

Let $\xx^\ast$ label the minimum-metric code path for a given log-likelihood
ratio $\phiphi$. Then we claim that if a node $s_j(\ell)$ is extended by the
MLSDA, given that $\xx_{(\ell n-1)}$ is the only surviving path (in the Open
Stack) that ends at this node at the time this node is extended, then \beq
\mem{mmm} \zeta(\xx_{(\ell n-1)}|\phiphi_{(\ell n-1)})\leq
\zeta(\xx^\ast|\phiphi) \eeq The validity of the above claim can be simply
proved by contradiction. Suppose $\zeta(\xx_{(\ell n-1)}|\phiphi_{(\ell n-1)})$
$>$ $\zeta(\xx^\ast|\phiphi)$. Then the non-negativity of the individual metric
$(y_j\oplus x_j)|\phi_j|$, which implies $\zeta(\xx^\ast|\phiphi)$ $\geq$
$\zeta(\xx^\ast_{(b)}|\phiphi_{(b)})$ for every $0\leq b\leq N-1$, immediately
gives $\zeta(\xx_{(\ell n-1)}|\phiphi_{(\ell n-1)})$ $>$
$\zeta(\xx^\ast_{(b)}|\phiphi_{(b)})$ for every $0\leq b\leq N-1$. Therefore,
$\xx_{(\ell n-1)}$ cannot be on top of the Open Stack (because some
$\xx^\ast_{(b)}$ always exists in the Open Stack), and hence violates the
assumption that $s_j(\ell)$ is extended by the MLSDA.

For notational convenience, denote by ${\cal A}(s_j(\ell),\xx_{(\ell n-1)})$
the {\em event} that ``{\em $\xx_{(\ell n-1)}$ is the only path
in the intersection of ${\cal S}_j(\ell)$ and the Open Stack at
the time node $s_j(\ell)$ is extended}.'' Notably,
$$\{{\cal A}(s_j(\ell),\xx_{(\ell n-1)})\}_{\xxb_{(\ell n-1)}\in{\cal
S}_j(\ell)}$$
are disjoint, and
$$\sum_{\xxb_{(\ell n-1)}\in{\cal S}_j(\ell)}\Pr\left\{{\cal A}\left(s_j(\ell),\xx_{(\ell n-1)}\right)\right\}=1.$$
Then according to the above claim,
\begin{eqnarray}
&&\Pr\left\{\mbox{node }s_j(\ell)\mbox{ is extended by the MLSDA}\right\}\nono\\
&=&\sum_{\xxb_{(\ell n-1)}\in{\cal S}_j(\ell)}\Pr\left\{{\cal A}\left(s_j(\ell),\xx_{(\ell n-1)}\right)
\right\}
\Pr\left\{\left.\begin{array}{l}
\mbox{node }s_j(\ell)\mbox{ is extended}\\
\mbox{by the MLSDA}
\end{array}\right|{\cal A}\left(s_j(\ell),\xx_{(\ell n-1)}\right)
\right\}\nono\\
&\leq&\max_{\xxb_{(\ell n-1)}\in{\cal S}_j(\ell)}\Pr\left\{\left.\begin{array}{l}
\mbox{node }s_j(\ell)\mbox{ is extended}\\
\mbox{by the MLSDA}
\end{array}\right|{\cal A}\left(s_j(\ell),\xx_{(\ell n-1)}\right)
\right\}\label{mlsda1}\\
&\leq&\max_{\xxb_{(\ell n-1)}\in{\cal S}_j(\ell)} \Pr\left\{\zeta(\xx_{(\ell
n-1)}|\phiphi_{(\ell n-1)})\leq
\zeta(\xx^\ast|\phiphi)\right\}\nono\\
&\leq&\max_{\xxb_{(\ell n-1)}\in{\cal S}_j(\ell)}\Pr\left\{ \zeta(\xx_{(\ell
n-1)}|\phiphi_{(\ell n-1)})\leq
\zeta(\0|\phiphi)\right\}\nono\\
&=&\max_{\xxb_{_{(\ell n-1)}}\in{\cal S}_j(\ell)}\Pr\left\{\sum_{j=0}^{\ell n-1}
(y_j\oplus x_j)|\phi_j|\leq
\sum_{j=0}^{N-1}(y_j\oplus 0)|\phi_j|\right\},\nono
\end{eqnarray}
where the replacement of $\xx^\ast$ by the all-zero codeword $\0$ follows from
$\zeta(\xx^\ast|\phiphi)\leq \zeta(\0|\phiphi)$.
 We then observe that
 for the AWGN channel defined through \rec{awgn1}, $\phi_j=4\sqrt{E}r_j/N_0$; hence,
 $y_j$ can be determined by
 $$y_j=\left\{\begin{array}{ll}
1,&{\rm if}\ r_j<0;\\
0,&{\rm otherwise}.
\end{array}\right.$$
This observation, together with the fact that $2(y_j\oplus x_j)|r_j|
=r_j[(-1)^{y_j}-(-1)^{x_j}]$, gives
\begin{eqnarray*}
&&\Pr\left\{\mbox{node }s_j(\ell)\mbox{ is extended by the MLSDA}\right\}\nono\\
&\leq&\max_{\xxb_{_{(\ell n-1)}}\in{\cal S}_j(\ell)}\Pr\left\{\sum_{j=0}^{\ell n-1}
(y_j\oplus x_j)|r_j|\leq
\sum_{j=0}^{N-1}(y_j\oplus 0)|r_j|\right\},\nono\\
&=&\max_{\xxb_{_{(\ell n-1)}}\in{\cal S}_j(\ell)}\Pr\left\{\sum_{j=0}^{\ell n-1}
r_j\left[(-1)^{y_j}-(-1)^{x_j}\right]
\leq\sum_{j=0}^{N-1}r_j\left[(-1)^{y_j}-(-1)^0\right]\right\}\\
&=&\max_{\xxb_{_{(\ell n-1)}}\in{\cal S}_j(\ell)}\Pr\left\{\sum_{j\in{\cal J}(\xxb_{(\ell n-1)})}r_j
+\sum_{j=\ell n}^{N-1}\min(r_j,0)\leq 0\right\},
\end{eqnarray*}
where ${\cal J}(\xx_{(\ell n-1)})$ is the set of index $j$, where $0\leq j\leq\ell n-1$,
for which $x_j=1$.
As $r_j$ is Gaussian distributed with mean $\sqrt{E}$
and variance $N_0/2$ due to the transmission
of the all-zero codeword, Proposition \ref{claim1} (in
the Appendix) and
Lemma~\ref{coro} can be applied to obtain
\begin{eqnarray*}
&&\Pr\left\{{\rm node}\ s_j(\ell)\ {\rm is\
extended\ by\ the\ MLSDA}\right\}\\
&\leq&
\max_{d\in{\cal H}_j(\ell)}\Pr\left\{r_1+\cdots+r_d
+\sum_{j=\ell n}^{N-1}\min(r_j,0)\leq 0\right\}\\
&=&\Pr\left\{r_1+\cdots+r_{d_j^\ast(\ell)}
+\sum_{j=\ell n}^{N-1}\min(r_j,0)\leq 0\right\}\\
&\leq&
{\cal B}\left(d_j^\ast(\ell),N-\ell n,\frac{kL}N\gamma_b\right).
\end{eqnarray*}
Consequently,
\begin{eqnarray*}
{L_{\rm MLSDA}}(\gamma_b)&\leq&
2^k\sum_{\ell=0}^{L-1}\sum_{j=0}^{2^m-1}
{\cal B}\left(d_j^\ast(\ell),N-\ell n,\frac{kL}N\gamma_b\right),
\end{eqnarray*}
where the multiplication of $2^k$ is due to the fact that
whenever a node is extended, $2^k$ branch metric computations
will follow.
\end{proof}

\subsection{Numerical and simulation results}
\label{NSR}

The accuracy of the previously
derived theoretical upper bound for the computational effort of the MLSDA
is now empirically examined using
two types of convolutional codes.
One is a $(2,1,6)$ code with generators $634,564$ (octal);
the other is a $(2,1,16)$ code with generators $1632044$, $1145734$ (octal).
The lengths of the applied information bits are $60$
and $100$.

\begin{figure}[hbt]
\begin{center}
\setlength{\unitlength}{0.240900pt}
\ifx\plotpoint\undefined\newsavebox{\plotpoint}\fi
\sbox{\plotpoint}{\rule[-0.200pt]{0.400pt}{0.400pt}}
\begin{picture}(1500,900)(0,0)
\font\gnuplot=cmr10 at 10pt \gnuplot
\sbox{\plotpoint}{\rule[-0.200pt]{0.400pt}{0.400pt}}
\put(201.0,123.0){\rule[-0.200pt]{4.818pt}{0.400pt}}
\put(181,123){\makebox(0,0)[r]{100}}
\put(1419.0,123.0){\rule[-0.200pt]{4.818pt}{0.400pt}}
\put(201.0,234.0){\rule[-0.200pt]{2.409pt}{0.400pt}}
\put(1429.0,234.0){\rule[-0.200pt]{2.409pt}{0.400pt}}
\put(201.0,299.0){\rule[-0.200pt]{2.409pt}{0.400pt}}
\put(1429.0,299.0){\rule[-0.200pt]{2.409pt}{0.400pt}}
\put(201.0,345.0){\rule[-0.200pt]{2.409pt}{0.400pt}}
\put(1429.0,345.0){\rule[-0.200pt]{2.409pt}{0.400pt}}
\put(201.0,381.0){\rule[-0.200pt]{2.409pt}{0.400pt}}
\put(1429.0,381.0){\rule[-0.200pt]{2.409pt}{0.400pt}}
\put(201.0,410.0){\rule[-0.200pt]{2.409pt}{0.400pt}}
\put(1429.0,410.0){\rule[-0.200pt]{2.409pt}{0.400pt}}
\put(201.0,434.0){\rule[-0.200pt]{2.409pt}{0.400pt}}
\put(1429.0,434.0){\rule[-0.200pt]{2.409pt}{0.400pt}}
\put(201.0,456.0){\rule[-0.200pt]{2.409pt}{0.400pt}}
\put(1429.0,456.0){\rule[-0.200pt]{2.409pt}{0.400pt}}
\put(201.0,475.0){\rule[-0.200pt]{2.409pt}{0.400pt}}
\put(1429.0,475.0){\rule[-0.200pt]{2.409pt}{0.400pt}}
\put(201.0,492.0){\rule[-0.200pt]{4.818pt}{0.400pt}}
\put(181,492){\makebox(0,0)[r]{1000}}
\put(1419.0,492.0){\rule[-0.200pt]{4.818pt}{0.400pt}}
\put(201.0,602.0){\rule[-0.200pt]{2.409pt}{0.400pt}}
\put(1429.0,602.0){\rule[-0.200pt]{2.409pt}{0.400pt}}
\put(201.0,667.0){\rule[-0.200pt]{2.409pt}{0.400pt}}
\put(1429.0,667.0){\rule[-0.200pt]{2.409pt}{0.400pt}}
\put(201.0,713.0){\rule[-0.200pt]{2.409pt}{0.400pt}}
\put(1429.0,713.0){\rule[-0.200pt]{2.409pt}{0.400pt}}
\put(201.0,749.0){\rule[-0.200pt]{2.409pt}{0.400pt}}
\put(1429.0,749.0){\rule[-0.200pt]{2.409pt}{0.400pt}}
\put(201.0,778.0){\rule[-0.200pt]{2.409pt}{0.400pt}}
\put(1429.0,778.0){\rule[-0.200pt]{2.409pt}{0.400pt}}
\put(201.0,803.0){\rule[-0.200pt]{2.409pt}{0.400pt}}
\put(1429.0,803.0){\rule[-0.200pt]{2.409pt}{0.400pt}}
\put(201.0,824.0){\rule[-0.200pt]{2.409pt}{0.400pt}}
\put(1429.0,824.0){\rule[-0.200pt]{2.409pt}{0.400pt}}
\put(201.0,843.0){\rule[-0.200pt]{2.409pt}{0.400pt}}
\put(1429.0,843.0){\rule[-0.200pt]{2.409pt}{0.400pt}}
\put(201.0,860.0){\rule[-0.200pt]{4.818pt}{0.400pt}}
\put(181,860){\makebox(0,0)[r]{10000}}
\put(1419.0,860.0){\rule[-0.200pt]{4.818pt}{0.400pt}}
\put(201.0,123.0){\rule[-0.200pt]{0.400pt}{4.818pt}}
\put(201,82){\makebox(0,0){1}}
\put(201.0,840.0){\rule[-0.200pt]{0.400pt}{4.818pt}}
\put(325.0,123.0){\rule[-0.200pt]{0.400pt}{4.818pt}}
\put(325,82){\makebox(0,0){2}}
\put(325.0,840.0){\rule[-0.200pt]{0.400pt}{4.818pt}}
\put(449.0,123.0){\rule[-0.200pt]{0.400pt}{4.818pt}}
\put(449,82){\makebox(0,0){3}}
\put(449.0,840.0){\rule[-0.200pt]{0.400pt}{4.818pt}}
\put(572.0,123.0){\rule[-0.200pt]{0.400pt}{4.818pt}}
\put(572,82){\makebox(0,0){4}}
\put(572.0,840.0){\rule[-0.200pt]{0.400pt}{4.818pt}}
\put(696.0,123.0){\rule[-0.200pt]{0.400pt}{4.818pt}}
\put(696,82){\makebox(0,0){5}}
\put(696.0,840.0){\rule[-0.200pt]{0.400pt}{4.818pt}}
\put(820.0,123.0){\rule[-0.200pt]{0.400pt}{4.818pt}}
\put(820,82){\makebox(0,0){6}}
\put(820.0,840.0){\rule[-0.200pt]{0.400pt}{4.818pt}}
\put(944.0,123.0){\rule[-0.200pt]{0.400pt}{4.818pt}}
\put(944,82){\makebox(0,0){7}}
\put(944.0,840.0){\rule[-0.200pt]{0.400pt}{4.818pt}}
\put(1068.0,123.0){\rule[-0.200pt]{0.400pt}{4.818pt}}
\put(1068,82){\makebox(0,0){8}}
\put(1068.0,840.0){\rule[-0.200pt]{0.400pt}{4.818pt}}
\put(1191.0,123.0){\rule[-0.200pt]{0.400pt}{4.818pt}}
\put(1191,82){\makebox(0,0){9}}
\put(1191.0,840.0){\rule[-0.200pt]{0.400pt}{4.818pt}}
\put(1315.0,123.0){\rule[-0.200pt]{0.400pt}{4.818pt}}
\put(1315,82){\makebox(0,0){10}}
\put(1315.0,840.0){\rule[-0.200pt]{0.400pt}{4.818pt}}
\put(1439.0,123.0){\rule[-0.200pt]{0.400pt}{4.818pt}}
\put(1439,82){\makebox(0,0){11}}
\put(1439.0,840.0){\rule[-0.200pt]{0.400pt}{4.818pt}}
\put(201.0,123.0){\rule[-0.200pt]{298.234pt}{0.400pt}}
\put(1439.0,123.0){\rule[-0.200pt]{0.400pt}{177.543pt}}
\put(201.0,860.0){\rule[-0.200pt]{298.234pt}{0.400pt}}
\put(40,491){\makebox(0,0){\rotate[l]{Computational complexity}}}
\put(820,21){\makebox(0,0){$\gamma_b$}}
\put(201.0,123.0){\rule[-0.200pt]{0.400pt}{177.543pt}}
\put(1279,820){\makebox(0,0)[r]{Theoretical bound}}
\multiput(1299,820)(20.756,0.000){5}{\usebox{\plotpoint}}
\put(1399,820){\usebox{\plotpoint}} \put(201,860){\usebox{\plotpoint}}
\multiput(201,860)(20.436,-3.626){4}{\usebox{\plotpoint}}
\multiput(263,849)(20.097,-5.186){3}{\usebox{\plotpoint}}
\multiput(325,833)(19.250,-7.762){3}{\usebox{\plotpoint}}
\multiput(387,808)(17.823,-10.636){3}{\usebox{\plotpoint}}
\multiput(449,771)(15.777,-13.487){4}{\usebox{\plotpoint}}
\multiput(511,718)(13.636,-15.648){5}{\usebox{\plotpoint}}
\multiput(572,648)(12.231,-16.769){5}{\usebox{\plotpoint}}
\multiput(634,563)(11.599,-17.212){5}{\usebox{\plotpoint}}
\multiput(696,471)(12.046,-16.903){5}{\usebox{\plotpoint}}
\multiput(758,384)(13.872,-15.439){5}{\usebox{\plotpoint}}
\multiput(820,315)(16.926,-12.012){3}{\usebox{\plotpoint}}
\multiput(882,271)(19.561,-6.941){4}{\usebox{\plotpoint}}
\multiput(944,249)(20.491,-3.305){3}{\usebox{\plotpoint}}
\multiput(1006,239)(20.731,-1.003){3}{\usebox{\plotpoint}}
\multiput(1068,236)(20.745,-0.669){3}{\usebox{\plotpoint}}
\multiput(1130,234)(20.756,0.000){3}{\usebox{\plotpoint}}
\multiput(1191,234)(20.756,0.000){3}{\usebox{\plotpoint}}
\multiput(1253,234)(20.756,0.000){3}{\usebox{\plotpoint}}
\multiput(1315,234)(20.756,0.000){2}{\usebox{\plotpoint}}
\multiput(1377,234)(20.756,0.000){3}{\usebox{\plotpoint}}
\put(1439,234){\usebox{\plotpoint}} \put(201,860){\circle{18}}
\put(263,849){\circle{18}} \put(325,833){\circle{18}}
\put(387,808){\circle{18}} \put(449,771){\circle{18}}
\put(511,718){\circle{18}} \put(572,648){\circle{18}}
\put(634,563){\circle{18}} \put(696,471){\circle{18}}
\put(758,384){\circle{18}} \put(820,315){\circle{18}}
\put(882,271){\circle{18}} \put(944,249){\circle{18}}
\put(1006,239){\circle{18}} \put(1068,236){\circle{18}}
\put(1130,234){\circle{18}} \put(1191,234){\circle{18}}
\put(1253,234){\circle{18}} \put(1315,234){\circle{18}}
\put(1377,234){\circle{18}} \put(1439,234){\circle{18}}
\put(1349,820){\circle{18}} \put(1279,779){\makebox(0,0)[r]{Simulation}}
\multiput(1299,779)(20.756,0.000){5}{\usebox{\plotpoint}}
\put(1399,779){\usebox{\plotpoint}} \put(201,850){\usebox{\plotpoint}}
\multiput(201,850)(19.408,-7.356){7}{\usebox{\plotpoint}}
\multiput(325,803)(16.093,-13.108){8}{\usebox{\plotpoint}}
\multiput(449,702)(11.981,-16.948){10}{\usebox{\plotpoint}}
\multiput(572,528)(11.428,-17.326){11}{\usebox{\plotpoint}}
\multiput(696,340)(16.991,-11.921){7}{\usebox{\plotpoint}}
\multiput(820,253)(20.585,-2.656){6}{\usebox{\plotpoint}}
\multiput(944,237)(20.749,-0.502){6}{\usebox{\plotpoint}}
\multiput(1068,234)(20.756,0.000){6}{\usebox{\plotpoint}}
\multiput(1191,234)(20.756,0.000){6}{\usebox{\plotpoint}}
\multiput(1315,234)(20.756,0.000){6}{\usebox{\plotpoint}}
\put(1439,234){\usebox{\plotpoint}} \put(201,850){\makebox(0,0){$+$}}
\put(325,803){\makebox(0,0){$+$}} \put(449,702){\makebox(0,0){$+$}}
\put(572,528){\makebox(0,0){$+$}} \put(696,340){\makebox(0,0){$+$}}
\put(820,253){\makebox(0,0){$+$}} \put(944,237){\makebox(0,0){$+$}}
\put(1068,234){\makebox(0,0){$+$}} \put(1191,234){\makebox(0,0){$+$}}
\put(1315,234){\makebox(0,0){$+$}} \put(1439,234){\makebox(0,0){$+$}}
\put(1349,779){\makebox(0,0){$+$}}
\sbox{\plotpoint}{\rule[-0.400pt]{0.800pt}{0.800pt}}
\put(1279,738){\makebox(0,0)[r]{Theoretical bound (BE)}}
\put(1299.0,738.0){\rule[-0.400pt]{24.090pt}{0.800pt}}
\put(201,860){\usebox{\plotpoint}}
\multiput(201.00,858.08)(3.032,-0.512){15}{\rule{4.709pt}{0.123pt}}
\multiput(201.00,858.34)(52.226,-11.000){2}{\rule{2.355pt}{0.800pt}}
\multiput(263.00,847.09)(2.013,-0.507){25}{\rule{3.300pt}{0.122pt}}
\multiput(263.00,847.34)(55.151,-16.000){2}{\rule{1.650pt}{0.800pt}}
\multiput(325.00,831.09)(1.260,-0.504){43}{\rule{2.184pt}{0.121pt}}
\multiput(325.00,831.34)(57.467,-25.000){2}{\rule{1.092pt}{0.800pt}}
\multiput(387.00,806.09)(0.843,-0.503){67}{\rule{1.541pt}{0.121pt}}
\multiput(387.00,806.34)(58.803,-37.000){2}{\rule{0.770pt}{0.800pt}}
\multiput(449.00,769.09)(0.585,-0.502){99}{\rule{1.136pt}{0.121pt}}
\multiput(449.00,769.34)(59.642,-53.000){2}{\rule{0.568pt}{0.800pt}}
\multiput(512.41,713.36)(0.502,-0.573){115}{\rule{0.121pt}{1.118pt}}
\multiput(509.34,715.68)(61.000,-67.679){2}{\rule{0.800pt}{0.559pt}}
\multiput(573.41,642.62)(0.502,-0.686){117}{\rule{0.121pt}{1.297pt}}
\multiput(570.34,645.31)(62.000,-82.308){2}{\rule{0.800pt}{0.648pt}}
\multiput(635.41,557.24)(0.502,-0.743){117}{\rule{0.121pt}{1.387pt}}
\multiput(632.34,560.12)(62.000,-89.121){2}{\rule{0.800pt}{0.694pt}}
\multiput(697.41,465.51)(0.502,-0.703){117}{\rule{0.121pt}{1.323pt}}
\multiput(694.34,468.25)(62.000,-84.255){2}{\rule{0.800pt}{0.661pt}}
\multiput(759.41,379.47)(0.502,-0.556){117}{\rule{0.121pt}{1.090pt}}
\multiput(756.34,381.74)(62.000,-66.737){2}{\rule{0.800pt}{0.545pt}}
\multiput(820.00,313.09)(0.706,-0.502){81}{\rule{1.327pt}{0.121pt}}
\multiput(820.00,313.34)(59.245,-44.000){2}{\rule{0.664pt}{0.800pt}}
\multiput(882.00,269.09)(1.439,-0.505){37}{\rule{2.455pt}{0.122pt}}
\multiput(882.00,269.34)(56.905,-22.000){2}{\rule{1.227pt}{0.800pt}}
\multiput(944.00,247.08)(3.382,-0.514){13}{\rule{5.160pt}{0.124pt}}
\multiput(944.00,247.34)(51.290,-10.000){2}{\rule{2.580pt}{0.800pt}}
\put(1006,235.84){\rule{14.936pt}{0.800pt}}
\multiput(1006.00,237.34)(31.000,-3.000){2}{\rule{7.468pt}{0.800pt}}
\put(1068,233.34){\rule{14.936pt}{0.800pt}}
\multiput(1068.00,234.34)(31.000,-2.000){2}{\rule{7.468pt}{0.800pt}}
\put(201,860){\raisebox{-.8pt}{\makebox(0,0){$\Box$}}}
\put(263,849){\raisebox{-.8pt}{\makebox(0,0){$\Box$}}}
\put(325,833){\raisebox{-.8pt}{\makebox(0,0){$\Box$}}}
\put(387,808){\raisebox{-.8pt}{\makebox(0,0){$\Box$}}}
\put(449,771){\raisebox{-.8pt}{\makebox(0,0){$\Box$}}}
\put(511,718){\raisebox{-.8pt}{\makebox(0,0){$\Box$}}}
\put(572,648){\raisebox{-.8pt}{\makebox(0,0){$\Box$}}}
\put(634,563){\raisebox{-.8pt}{\makebox(0,0){$\Box$}}}
\put(696,471){\raisebox{-.8pt}{\makebox(0,0){$\Box$}}}
\put(758,384){\raisebox{-.8pt}{\makebox(0,0){$\Box$}}}
\put(820,315){\raisebox{-.8pt}{\makebox(0,0){$\Box$}}}
\put(882,271){\raisebox{-.8pt}{\makebox(0,0){$\Box$}}}
\put(944,249){\raisebox{-.8pt}{\makebox(0,0){$\Box$}}}
\put(1006,239){\raisebox{-.8pt}{\makebox(0,0){$\Box$}}}
\put(1068,236){\raisebox{-.8pt}{\makebox(0,0){$\Box$}}}
\put(1130,234){\raisebox{-.8pt}{\makebox(0,0){$\Box$}}}
\put(1191,234){\raisebox{-.8pt}{\makebox(0,0){$\Box$}}}
\put(1253,234){\raisebox{-.8pt}{\makebox(0,0){$\Box$}}}
\put(1315,234){\raisebox{-.8pt}{\makebox(0,0){$\Box$}}}
\put(1377,234){\raisebox{-.8pt}{\makebox(0,0){$\Box$}}}
\put(1439,234){\raisebox{-.8pt}{\makebox(0,0){$\Box$}}}
\put(1349,738){\raisebox{-.8pt}{\makebox(0,0){$\Box$}}}
\put(1130.0,234.0){\rule[-0.400pt]{74.438pt}{0.800pt}}
\end{picture}
\parbox{5in}
{\caption{Average computational complexity
of the MLSDA for $(2,1,6)$ convolutional code
with generators $634,564$ (octal) and
information length $L=100$.
}\label{figure1}}
\end{center}
\end{figure}
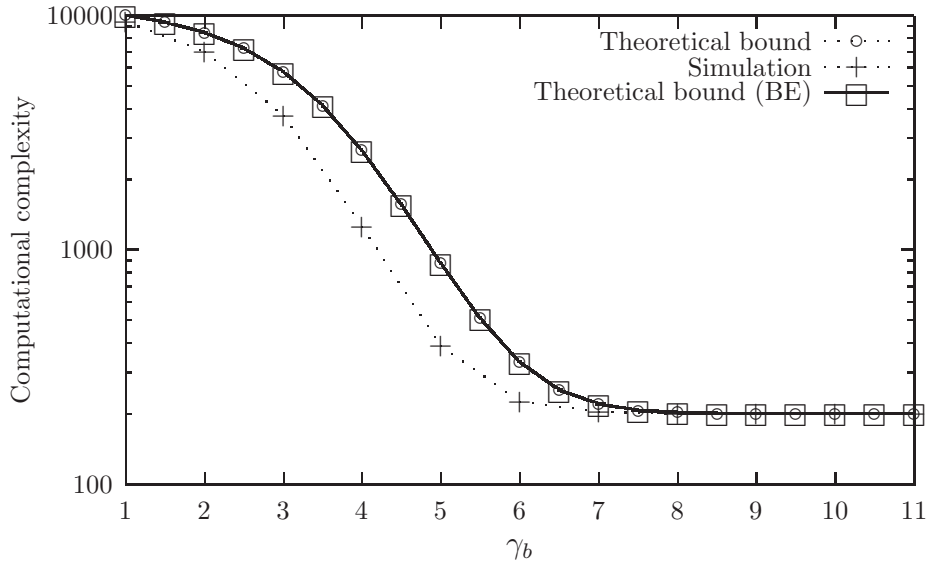

\begin{figure}[hbt]
\begin{center}
\setlength{\unitlength}{0.240900pt}
\ifx\plotpoint\undefined\newsavebox{\plotpoint}\fi
\sbox{\plotpoint}{\rule[-0.200pt]{0.400pt}{0.400pt}}
\begin{picture}(1500,900)(0,0)
\font\gnuplot=cmr10 at 10pt \gnuplot
\sbox{\plotpoint}{\rule[-0.200pt]{0.400pt}{0.400pt}}
\put(221.0,123.0){\rule[-0.200pt]{4.818pt}{0.400pt}}
\put(201,123){\makebox(0,0)[r]{100}}
\put(1419.0,123.0){\rule[-0.200pt]{4.818pt}{0.400pt}}
\put(221.0,197.0){\rule[-0.200pt]{2.409pt}{0.400pt}}
\put(1429.0,197.0){\rule[-0.200pt]{2.409pt}{0.400pt}}
\put(221.0,240.0){\rule[-0.200pt]{2.409pt}{0.400pt}}
\put(1429.0,240.0){\rule[-0.200pt]{2.409pt}{0.400pt}}
\put(221.0,271.0){\rule[-0.200pt]{2.409pt}{0.400pt}}
\put(1429.0,271.0){\rule[-0.200pt]{2.409pt}{0.400pt}}
\put(221.0,295.0){\rule[-0.200pt]{2.409pt}{0.400pt}}
\put(1429.0,295.0){\rule[-0.200pt]{2.409pt}{0.400pt}}
\put(221.0,314.0){\rule[-0.200pt]{2.409pt}{0.400pt}}
\put(1429.0,314.0){\rule[-0.200pt]{2.409pt}{0.400pt}}
\put(221.0,331.0){\rule[-0.200pt]{2.409pt}{0.400pt}}
\put(1429.0,331.0){\rule[-0.200pt]{2.409pt}{0.400pt}}
\put(221.0,345.0){\rule[-0.200pt]{2.409pt}{0.400pt}}
\put(1429.0,345.0){\rule[-0.200pt]{2.409pt}{0.400pt}}
\put(221.0,357.0){\rule[-0.200pt]{2.409pt}{0.400pt}}
\put(1429.0,357.0){\rule[-0.200pt]{2.409pt}{0.400pt}}
\put(221.0,369.0){\rule[-0.200pt]{4.818pt}{0.400pt}}
\put(201,369){\makebox(0,0)[r]{1000}}
\put(1419.0,369.0){\rule[-0.200pt]{4.818pt}{0.400pt}}
\put(221.0,443.0){\rule[-0.200pt]{2.409pt}{0.400pt}}
\put(1429.0,443.0){\rule[-0.200pt]{2.409pt}{0.400pt}}
\put(221.0,486.0){\rule[-0.200pt]{2.409pt}{0.400pt}}
\put(1429.0,486.0){\rule[-0.200pt]{2.409pt}{0.400pt}}
\put(221.0,517.0){\rule[-0.200pt]{2.409pt}{0.400pt}}
\put(1429.0,517.0){\rule[-0.200pt]{2.409pt}{0.400pt}}
\put(221.0,540.0){\rule[-0.200pt]{2.409pt}{0.400pt}}
\put(1429.0,540.0){\rule[-0.200pt]{2.409pt}{0.400pt}}
\put(221.0,560.0){\rule[-0.200pt]{2.409pt}{0.400pt}}
\put(1429.0,560.0){\rule[-0.200pt]{2.409pt}{0.400pt}}
\put(221.0,576.0){\rule[-0.200pt]{2.409pt}{0.400pt}}
\put(1429.0,576.0){\rule[-0.200pt]{2.409pt}{0.400pt}}
\put(221.0,591.0){\rule[-0.200pt]{2.409pt}{0.400pt}}
\put(1429.0,591.0){\rule[-0.200pt]{2.409pt}{0.400pt}}
\put(221.0,603.0){\rule[-0.200pt]{2.409pt}{0.400pt}}
\put(1429.0,603.0){\rule[-0.200pt]{2.409pt}{0.400pt}}
\put(221.0,614.0){\rule[-0.200pt]{4.818pt}{0.400pt}}
\put(201,614){\makebox(0,0)[r]{10000}}
\put(1419.0,614.0){\rule[-0.200pt]{4.818pt}{0.400pt}}
\put(221.0,688.0){\rule[-0.200pt]{2.409pt}{0.400pt}}
\put(1429.0,688.0){\rule[-0.200pt]{2.409pt}{0.400pt}}
\put(221.0,732.0){\rule[-0.200pt]{2.409pt}{0.400pt}}
\put(1429.0,732.0){\rule[-0.200pt]{2.409pt}{0.400pt}}
\put(221.0,762.0){\rule[-0.200pt]{2.409pt}{0.400pt}}
\put(1429.0,762.0){\rule[-0.200pt]{2.409pt}{0.400pt}}
\put(221.0,786.0){\rule[-0.200pt]{2.409pt}{0.400pt}}
\put(1429.0,786.0){\rule[-0.200pt]{2.409pt}{0.400pt}}
\put(221.0,805.0){\rule[-0.200pt]{2.409pt}{0.400pt}}
\put(1429.0,805.0){\rule[-0.200pt]{2.409pt}{0.400pt}}
\put(221.0,822.0){\rule[-0.200pt]{2.409pt}{0.400pt}}
\put(1429.0,822.0){\rule[-0.200pt]{2.409pt}{0.400pt}}
\put(221.0,836.0){\rule[-0.200pt]{2.409pt}{0.400pt}}
\put(1429.0,836.0){\rule[-0.200pt]{2.409pt}{0.400pt}}
\put(221.0,849.0){\rule[-0.200pt]{2.409pt}{0.400pt}}
\put(1429.0,849.0){\rule[-0.200pt]{2.409pt}{0.400pt}}
\put(221.0,860.0){\rule[-0.200pt]{4.818pt}{0.400pt}}
\put(201,860){\makebox(0,0)[r]{100000}}
\put(1419.0,860.0){\rule[-0.200pt]{4.818pt}{0.400pt}}
\put(221.0,123.0){\rule[-0.200pt]{0.400pt}{4.818pt}}
\put(221,82){\makebox(0,0){1}}
\put(221.0,840.0){\rule[-0.200pt]{0.400pt}{4.818pt}}
\put(343.0,123.0){\rule[-0.200pt]{0.400pt}{4.818pt}}
\put(343,82){\makebox(0,0){2}}
\put(343.0,840.0){\rule[-0.200pt]{0.400pt}{4.818pt}}
\put(465.0,123.0){\rule[-0.200pt]{0.400pt}{4.818pt}}
\put(465,82){\makebox(0,0){3}}
\put(465.0,840.0){\rule[-0.200pt]{0.400pt}{4.818pt}}
\put(586.0,123.0){\rule[-0.200pt]{0.400pt}{4.818pt}}
\put(586,82){\makebox(0,0){4}}
\put(586.0,840.0){\rule[-0.200pt]{0.400pt}{4.818pt}}
\put(708.0,123.0){\rule[-0.200pt]{0.400pt}{4.818pt}}
\put(708,82){\makebox(0,0){5}}
\put(708.0,840.0){\rule[-0.200pt]{0.400pt}{4.818pt}}
\put(830.0,123.0){\rule[-0.200pt]{0.400pt}{4.818pt}}
\put(830,82){\makebox(0,0){6}}
\put(830.0,840.0){\rule[-0.200pt]{0.400pt}{4.818pt}}
\put(952.0,123.0){\rule[-0.200pt]{0.400pt}{4.818pt}}
\put(952,82){\makebox(0,0){7}}
\put(952.0,840.0){\rule[-0.200pt]{0.400pt}{4.818pt}}
\put(1074.0,123.0){\rule[-0.200pt]{0.400pt}{4.818pt}}
\put(1074,82){\makebox(0,0){8}}
\put(1074.0,840.0){\rule[-0.200pt]{0.400pt}{4.818pt}}
\put(1195.0,123.0){\rule[-0.200pt]{0.400pt}{4.818pt}}
\put(1195,82){\makebox(0,0){9}}
\put(1195.0,840.0){\rule[-0.200pt]{0.400pt}{4.818pt}}
\put(1317.0,123.0){\rule[-0.200pt]{0.400pt}{4.818pt}}
\put(1317,82){\makebox(0,0){10}}
\put(1317.0,840.0){\rule[-0.200pt]{0.400pt}{4.818pt}}
\put(1439.0,123.0){\rule[-0.200pt]{0.400pt}{4.818pt}}
\put(1439,82){\makebox(0,0){11}}
\put(1439.0,840.0){\rule[-0.200pt]{0.400pt}{4.818pt}}
\put(221.0,123.0){\rule[-0.200pt]{293.416pt}{0.400pt}}
\put(1439.0,123.0){\rule[-0.200pt]{0.400pt}{177.543pt}}
\put(221.0,860.0){\rule[-0.200pt]{293.416pt}{0.400pt}}
\put(40,491){\makebox(0,0){\rotate[l]{Computational complexity}}}
\put(830,21){\makebox(0,0){$\gamma_b$}}
\put(221.0,123.0){\rule[-0.200pt]{0.400pt}{177.543pt}}
\put(1279,820){\makebox(0,0)[r]{Theoretical bound}}
\multiput(1299,820)(20.756,0.000){5}{\usebox{\plotpoint}}
\put(1399,820){\usebox{\plotpoint}} \put(221,541){\usebox{\plotpoint}}
\multiput(221,541)(20.230,-4.643){4}{\usebox{\plotpoint}}
\multiput(282,527)(19.722,-6.466){3}{\usebox{\plotpoint}}
\multiput(343,507)(18.979,-8.401){3}{\usebox{\plotpoint}}
\multiput(404,480)(17.875,-10.549){3}{\usebox{\plotpoint}}
\multiput(465,444)(16.964,-11.958){4}{\usebox{\plotpoint}}
\multiput(526,401)(16.076,-13.129){4}{\usebox{\plotpoint}}
\multiput(586,352)(15.668,-13.613){3}{\usebox{\plotpoint}}
\multiput(647,299)(15.923,-13.313){4}{\usebox{\plotpoint}}
\multiput(708,248)(17.095,-11.770){4}{\usebox{\plotpoint}}
\multiput(769,206)(18.625,-9.160){3}{\usebox{\plotpoint}}
\multiput(830,176)(19.907,-5.874){3}{\usebox{\plotpoint}}
\multiput(891,158)(20.533,-3.029){3}{\usebox{\plotpoint}}
\multiput(952,149)(20.711,-1.358){3}{\usebox{\plotpoint}}
\multiput(1013,145)(20.744,-0.680){3}{\usebox{\plotpoint}}
\multiput(1074,143)(20.756,0.000){3}{\usebox{\plotpoint}}
\multiput(1135,143)(20.756,0.000){3}{\usebox{\plotpoint}}
\multiput(1195,143)(20.753,-0.340){3}{\usebox{\plotpoint}}
\multiput(1256,142)(20.756,0.000){3}{\usebox{\plotpoint}}
\multiput(1317,142)(20.756,0.000){3}{\usebox{\plotpoint}}
\multiput(1378,142)(20.756,0.000){3}{\usebox{\plotpoint}}
\put(1439,142){\usebox{\plotpoint}} \put(221,541){\circle{18}}
\put(282,527){\circle{18}} \put(343,507){\circle{18}}
\put(404,480){\circle{18}} \put(465,444){\circle{18}}
\put(526,401){\circle{18}} \put(586,352){\circle{18}}
\put(647,299){\circle{18}} \put(708,248){\circle{18}}
\put(769,206){\circle{18}} \put(830,176){\circle{18}}
\put(891,158){\circle{18}} \put(952,149){\circle{18}}
\put(1013,145){\circle{18}} \put(1074,143){\circle{18}}
\put(1135,143){\circle{18}} \put(1195,143){\circle{18}}
\put(1256,142){\circle{18}} \put(1317,142){\circle{18}}
\put(1378,142){\circle{18}} \put(1439,142){\circle{18}}
\put(1349,820){\circle{18}} \put(1279,779){\makebox(0,0)[r]{Simulation}}
\multiput(1299,779)(20.756,0.000){5}{\usebox{\plotpoint}}
\put(1399,779){\usebox{\plotpoint}} \put(221,526){\usebox{\plotpoint}}
\multiput(221,526)(18.979,-8.401){7}{\usebox{\plotpoint}}
\multiput(343,472)(16.506,-12.583){7}{\usebox{\plotpoint}}
\multiput(465,379)(14.983,-14.364){8}{\usebox{\plotpoint}}
\multiput(586,263)(17.095,-11.770){8}{\usebox{\plotpoint}}
\multiput(708,179)(20.155,-4.956){6}{\usebox{\plotpoint}}
\multiput(830,149)(20.730,-1.020){5}{\usebox{\plotpoint}}
\multiput(952,143)(20.755,-0.170){6}{\usebox{\plotpoint}}
\multiput(1074,142)(20.756,0.000){6}{\usebox{\plotpoint}}
\multiput(1195,142)(20.756,0.000){6}{\usebox{\plotpoint}}
\multiput(1317,142)(20.756,0.000){6}{\usebox{\plotpoint}}
\put(1439,142){\usebox{\plotpoint}} \put(221,526){\makebox(0,0){$+$}}
\put(343,472){\makebox(0,0){$+$}} \put(465,379){\makebox(0,0){$+$}}
\put(586,263){\makebox(0,0){$+$}} \put(708,179){\makebox(0,0){$+$}}
\put(830,149){\makebox(0,0){$+$}} \put(952,143){\makebox(0,0){$+$}}
\put(1074,142){\makebox(0,0){$+$}} \put(1195,142){\makebox(0,0){$+$}}
\put(1317,142){\makebox(0,0){$+$}} \put(1439,142){\makebox(0,0){$+$}}
\put(1349,779){\makebox(0,0){$+$}}
\sbox{\plotpoint}{\rule[-0.400pt]{0.800pt}{0.800pt}}
\put(1279,738){\makebox(0,0)[r]{Theoretical bound (BE)}}
\put(1299.0,738.0){\rule[-0.400pt]{24.090pt}{0.800pt}}
\put(221,541){\usebox{\plotpoint}}
\multiput(221.00,539.09)(2.285,-0.509){21}{\rule{3.686pt}{0.123pt}}
\multiput(221.00,539.34)(53.350,-14.000){2}{\rule{1.843pt}{0.800pt}}
\multiput(282.00,525.09)(1.564,-0.505){33}{\rule{2.640pt}{0.122pt}}
\multiput(282.00,525.34)(55.521,-20.000){2}{\rule{1.320pt}{0.800pt}}
\multiput(343.00,505.09)(1.145,-0.504){47}{\rule{2.007pt}{0.121pt}}
\multiput(343.00,505.34)(56.834,-27.000){2}{\rule{1.004pt}{0.800pt}}
\multiput(404.00,478.09)(0.852,-0.503){65}{\rule{1.556pt}{0.121pt}}
\multiput(404.00,478.34)(57.771,-36.000){2}{\rule{0.778pt}{0.800pt}}
\multiput(465.00,442.09)(0.711,-0.502){79}{\rule{1.335pt}{0.121pt}}
\multiput(465.00,442.34)(58.229,-43.000){2}{\rule{0.667pt}{0.800pt}}
\multiput(526.00,399.09)(0.612,-0.502){91}{\rule{1.180pt}{0.121pt}}
\multiput(526.00,399.34)(57.552,-49.000){2}{\rule{0.590pt}{0.800pt}}
\multiput(586.00,350.09)(0.575,-0.502){99}{\rule{1.121pt}{0.121pt}}
\multiput(586.00,350.34)(58.674,-53.000){2}{\rule{0.560pt}{0.800pt}}
\multiput(647.00,297.09)(0.598,-0.502){95}{\rule{1.157pt}{0.121pt}}
\multiput(647.00,297.34)(58.599,-51.000){2}{\rule{0.578pt}{0.800pt}}
\multiput(708.00,246.09)(0.728,-0.502){77}{\rule{1.362pt}{0.121pt}}
\multiput(708.00,246.34)(58.173,-42.000){2}{\rule{0.681pt}{0.800pt}}
\multiput(769.00,204.09)(1.027,-0.503){53}{\rule{1.827pt}{0.121pt}}
\multiput(769.00,204.34)(57.209,-30.000){2}{\rule{0.913pt}{0.800pt}}
\multiput(830.00,174.09)(1.747,-0.506){29}{\rule{2.911pt}{0.122pt}}
\multiput(830.00,174.34)(54.958,-18.000){2}{\rule{1.456pt}{0.800pt}}
\multiput(891.00,156.08)(3.766,-0.516){11}{\rule{5.622pt}{0.124pt}}
\multiput(891.00,156.34)(49.331,-9.000){2}{\rule{2.811pt}{0.800pt}}
\put(952,145.34){\rule{12.400pt}{0.800pt}}
\multiput(952.00,147.34)(35.263,-4.000){2}{\rule{6.200pt}{0.800pt}}
\put(1013,142.34){\rule{14.695pt}{0.800pt}}
\multiput(1013.00,143.34)(30.500,-2.000){2}{\rule{7.347pt}{0.800pt}}
\put(1195,140.84){\rule{14.695pt}{0.800pt}}
\multiput(1195.00,141.34)(30.500,-1.000){2}{\rule{7.347pt}{0.800pt}}
\put(1074.0,143.0){\rule[-0.400pt]{29.149pt}{0.800pt}}
\put(221,541){\raisebox{-.8pt}{\makebox(0,0){$\Box$}}}
\put(282,527){\raisebox{-.8pt}{\makebox(0,0){$\Box$}}}
\put(343,507){\raisebox{-.8pt}{\makebox(0,0){$\Box$}}}
\put(404,480){\raisebox{-.8pt}{\makebox(0,0){$\Box$}}}
\put(465,444){\raisebox{-.8pt}{\makebox(0,0){$\Box$}}}
\put(526,401){\raisebox{-.8pt}{\makebox(0,0){$\Box$}}}
\put(586,352){\raisebox{-.8pt}{\makebox(0,0){$\Box$}}}
\put(647,299){\raisebox{-.8pt}{\makebox(0,0){$\Box$}}}
\put(708,248){\raisebox{-.8pt}{\makebox(0,0){$\Box$}}}
\put(769,206){\raisebox{-.8pt}{\makebox(0,0){$\Box$}}}
\put(830,176){\raisebox{-.8pt}{\makebox(0,0){$\Box$}}}
\put(891,158){\raisebox{-.8pt}{\makebox(0,0){$\Box$}}}
\put(952,149){\raisebox{-.8pt}{\makebox(0,0){$\Box$}}}
\put(1013,145){\raisebox{-.8pt}{\makebox(0,0){$\Box$}}}
\put(1074,143){\raisebox{-.8pt}{\makebox(0,0){$\Box$}}}
\put(1135,143){\raisebox{-.8pt}{\makebox(0,0){$\Box$}}}
\put(1195,143){\raisebox{-.8pt}{\makebox(0,0){$\Box$}}}
\put(1256,142){\raisebox{-.8pt}{\makebox(0,0){$\Box$}}}
\put(1317,142){\raisebox{-.8pt}{\makebox(0,0){$\Box$}}}
\put(1378,142){\raisebox{-.8pt}{\makebox(0,0){$\Box$}}}
\put(1439,142){\raisebox{-.8pt}{\makebox(0,0){$\Box$}}}
\put(1349,738){\raisebox{-.8pt}{\makebox(0,0){$\Box$}}}
\put(1256.0,142.0){\rule[-0.400pt]{44.085pt}{0.800pt}}
\end{picture}

\parbox{5in}
{\caption{Average computational complexity of the MLSDA for
$(2,1,6)$ convolutional code with generators $634,564$ (octal)
and information length $L=60$.}\label{figure2}}
\end{center}
\end{figure}
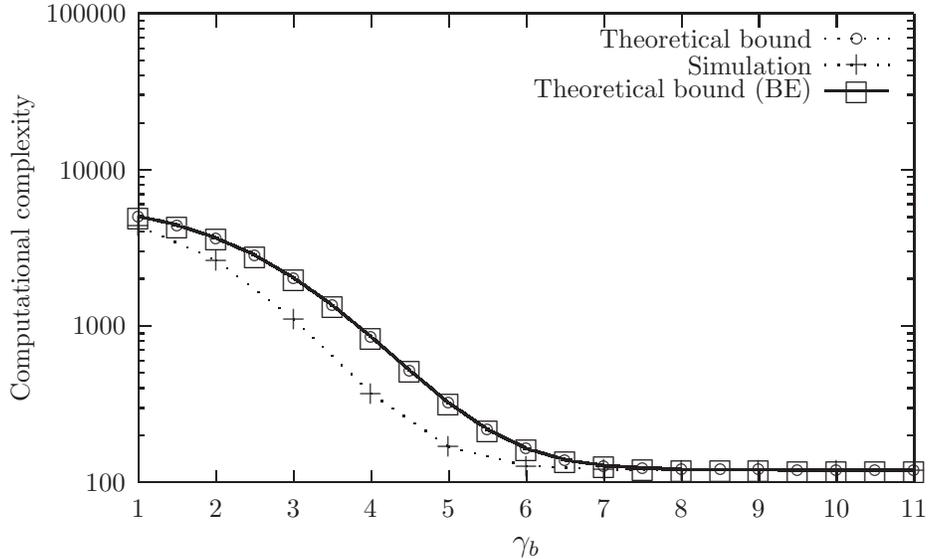

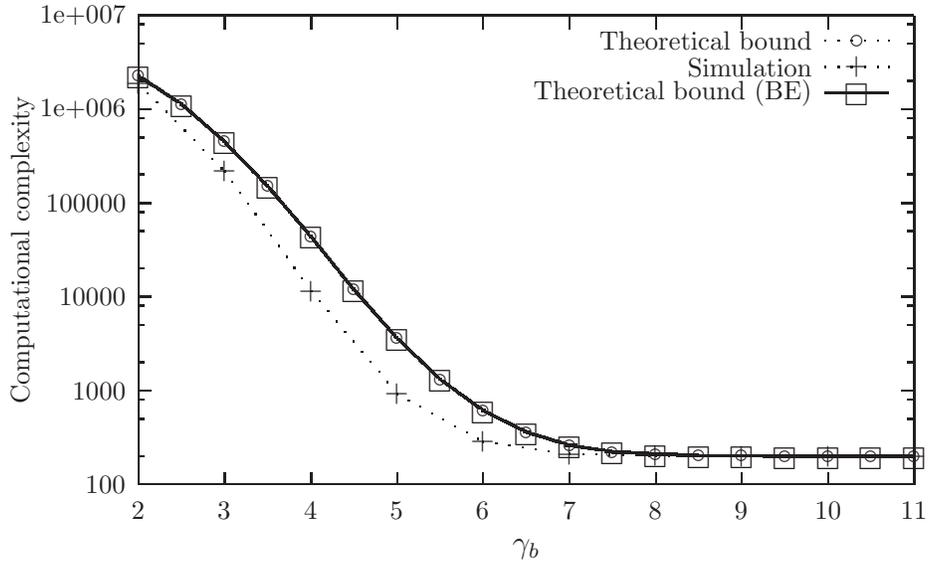
\begin{figure}[hbt]
\begin{center}
\setlength{\unitlength}{0.240900pt}
\ifx\plotpoint\undefined\newsavebox{\plotpoint}\fi
\sbox{\plotpoint}{\rule[-0.200pt]{0.400pt}{0.400pt}}
\begin{picture}(1500,900)(0,0)
\font\gnuplot=cmr10 at 10pt \gnuplot
\sbox{\plotpoint}{\rule[-0.200pt]{0.400pt}{0.400pt}}
\put(221.0,123.0){\rule[-0.200pt]{4.818pt}{0.400pt}}
\put(201,123){\makebox(0,0)[r]{100}}
\put(1419.0,123.0){\rule[-0.200pt]{4.818pt}{0.400pt}}
\put(221.0,167.0){\rule[-0.200pt]{2.409pt}{0.400pt}}
\put(1429.0,167.0){\rule[-0.200pt]{2.409pt}{0.400pt}}
\put(221.0,226.0){\rule[-0.200pt]{2.409pt}{0.400pt}}
\put(1429.0,226.0){\rule[-0.200pt]{2.409pt}{0.400pt}}
\put(221.0,256.0){\rule[-0.200pt]{2.409pt}{0.400pt}}
\put(1429.0,256.0){\rule[-0.200pt]{2.409pt}{0.400pt}}
\put(221.0,270.0){\rule[-0.200pt]{4.818pt}{0.400pt}}
\put(201,270){\makebox(0,0)[r]{1000}}
\put(1419.0,270.0){\rule[-0.200pt]{4.818pt}{0.400pt}}
\put(221.0,315.0){\rule[-0.200pt]{2.409pt}{0.400pt}}
\put(1429.0,315.0){\rule[-0.200pt]{2.409pt}{0.400pt}}
\put(221.0,373.0){\rule[-0.200pt]{2.409pt}{0.400pt}}
\put(1429.0,373.0){\rule[-0.200pt]{2.409pt}{0.400pt}}
\put(221.0,404.0){\rule[-0.200pt]{2.409pt}{0.400pt}}
\put(1429.0,404.0){\rule[-0.200pt]{2.409pt}{0.400pt}}
\put(221.0,418.0){\rule[-0.200pt]{4.818pt}{0.400pt}}
\put(201,418){\makebox(0,0)[r]{10000}}
\put(1419.0,418.0){\rule[-0.200pt]{4.818pt}{0.400pt}}
\put(221.0,462.0){\rule[-0.200pt]{2.409pt}{0.400pt}}
\put(1429.0,462.0){\rule[-0.200pt]{2.409pt}{0.400pt}}
\put(221.0,521.0){\rule[-0.200pt]{2.409pt}{0.400pt}}
\put(1429.0,521.0){\rule[-0.200pt]{2.409pt}{0.400pt}}
\put(221.0,551.0){\rule[-0.200pt]{2.409pt}{0.400pt}}
\put(1429.0,551.0){\rule[-0.200pt]{2.409pt}{0.400pt}}
\put(221.0,565.0){\rule[-0.200pt]{4.818pt}{0.400pt}}
\put(201,565){\makebox(0,0)[r]{100000}}
\put(1419.0,565.0){\rule[-0.200pt]{4.818pt}{0.400pt}}
\put(221.0,610.0){\rule[-0.200pt]{2.409pt}{0.400pt}}
\put(1429.0,610.0){\rule[-0.200pt]{2.409pt}{0.400pt}}
\put(221.0,668.0){\rule[-0.200pt]{2.409pt}{0.400pt}}
\put(1429.0,668.0){\rule[-0.200pt]{2.409pt}{0.400pt}}
\put(221.0,698.0){\rule[-0.200pt]{2.409pt}{0.400pt}}
\put(1429.0,698.0){\rule[-0.200pt]{2.409pt}{0.400pt}}
\put(221.0,713.0){\rule[-0.200pt]{4.818pt}{0.400pt}}
\put(201,713){\makebox(0,0)[r]{1e+006}}
\put(1419.0,713.0){\rule[-0.200pt]{4.818pt}{0.400pt}}
\put(221.0,757.0){\rule[-0.200pt]{2.409pt}{0.400pt}}
\put(1429.0,757.0){\rule[-0.200pt]{2.409pt}{0.400pt}}
\put(221.0,816.0){\rule[-0.200pt]{2.409pt}{0.400pt}}
\put(1429.0,816.0){\rule[-0.200pt]{2.409pt}{0.400pt}}
\put(221.0,846.0){\rule[-0.200pt]{2.409pt}{0.400pt}}
\put(1429.0,846.0){\rule[-0.200pt]{2.409pt}{0.400pt}}
\put(221.0,860.0){\rule[-0.200pt]{4.818pt}{0.400pt}}
\put(201,860){\makebox(0,0)[r]{1e+007}}
\put(1419.0,860.0){\rule[-0.200pt]{4.818pt}{0.400pt}}
\put(221.0,123.0){\rule[-0.200pt]{0.400pt}{4.818pt}}
\put(221,82){\makebox(0,0){2}}
\put(221.0,840.0){\rule[-0.200pt]{0.400pt}{4.818pt}}
\put(356.0,123.0){\rule[-0.200pt]{0.400pt}{4.818pt}}
\put(356,82){\makebox(0,0){3}}
\put(356.0,840.0){\rule[-0.200pt]{0.400pt}{4.818pt}}
\put(492.0,123.0){\rule[-0.200pt]{0.400pt}{4.818pt}}
\put(492,82){\makebox(0,0){4}}
\put(492.0,840.0){\rule[-0.200pt]{0.400pt}{4.818pt}}
\put(627.0,123.0){\rule[-0.200pt]{0.400pt}{4.818pt}}
\put(627,82){\makebox(0,0){5}}
\put(627.0,840.0){\rule[-0.200pt]{0.400pt}{4.818pt}}
\put(762.0,123.0){\rule[-0.200pt]{0.400pt}{4.818pt}}
\put(762,82){\makebox(0,0){6}}
\put(762.0,840.0){\rule[-0.200pt]{0.400pt}{4.818pt}}
\put(898.0,123.0){\rule[-0.200pt]{0.400pt}{4.818pt}}
\put(898,82){\makebox(0,0){7}}
\put(898.0,840.0){\rule[-0.200pt]{0.400pt}{4.818pt}}
\put(1033.0,123.0){\rule[-0.200pt]{0.400pt}{4.818pt}}
\put(1033,82){\makebox(0,0){8}}
\put(1033.0,840.0){\rule[-0.200pt]{0.400pt}{4.818pt}}
\put(1168.0,123.0){\rule[-0.200pt]{0.400pt}{4.818pt}}
\put(1168,82){\makebox(0,0){9}}
\put(1168.0,840.0){\rule[-0.200pt]{0.400pt}{4.818pt}}
\put(1304.0,123.0){\rule[-0.200pt]{0.400pt}{4.818pt}}
\put(1304,82){\makebox(0,0){10}}
\put(1304.0,840.0){\rule[-0.200pt]{0.400pt}{4.818pt}}
\put(1439.0,123.0){\rule[-0.200pt]{0.400pt}{4.818pt}}
\put(1439,82){\makebox(0,0){11}}
\put(1439.0,840.0){\rule[-0.200pt]{0.400pt}{4.818pt}}
\put(221.0,123.0){\rule[-0.200pt]{293.416pt}{0.400pt}}
\put(1439.0,123.0){\rule[-0.200pt]{0.400pt}{177.543pt}}
\put(221.0,860.0){\rule[-0.200pt]{293.416pt}{0.400pt}}
\put(40,491){\makebox(0,0){\rotate[l]{Computational complexity}}}
\put(830,21){\makebox(0,0){$\gamma_b$}}
\put(221.0,123.0){\rule[-0.200pt]{0.400pt}{177.543pt}}
\put(1279,820){\makebox(0,0)[r]{Theoretical bound}}
\multiput(1299,820)(20.756,0.000){5}{\usebox{\plotpoint}}
\put(1399,820){\usebox{\plotpoint}} \put(221,765){\usebox{\plotpoint}}
\multiput(221,765)(17.309,-11.454){4}{\usebox{\plotpoint}}
\multiput(289,720)(15.692,-13.584){5}{\usebox{\plotpoint}}
\multiput(356,662)(14.462,-14.887){4}{\usebox{\plotpoint}}
\multiput(424,592)(13.540,-15.731){5}{\usebox{\plotpoint}}
\multiput(492,513)(13.037,-16.150){6}{\usebox{\plotpoint}}
\multiput(559,430)(13.739,-15.557){5}{\usebox{\plotpoint}}
\multiput(627,353)(15.004,-14.342){4}{\usebox{\plotpoint}}
\multiput(695,288)(16.753,-12.252){4}{\usebox{\plotpoint}}
\multiput(762,239)(18.564,-9.282){4}{\usebox{\plotpoint}}
\multiput(830,205)(19.831,-6.124){3}{\usebox{\plotpoint}}
\multiput(898,184)(20.528,-3.064){3}{\usebox{\plotpoint}}
\multiput(965,174)(20.720,-1.219){4}{\usebox{\plotpoint}}
\multiput(1033,170)(20.747,-0.610){3}{\usebox{\plotpoint}}
\multiput(1101,168)(20.756,0.000){3}{\usebox{\plotpoint}}
\multiput(1168,168)(20.753,-0.305){3}{\usebox{\plotpoint}}
\multiput(1236,167)(20.756,0.000){4}{\usebox{\plotpoint}}
\multiput(1304,167)(20.756,0.000){3}{\usebox{\plotpoint}}
\multiput(1371,167)(20.756,0.000){3}{\usebox{\plotpoint}}
\put(1439,167){\usebox{\plotpoint}} \put(221,765){\circle{18}}
\put(289,720){\circle{18}} \put(356,662){\circle{18}}
\put(424,592){\circle{18}} \put(492,513){\circle{18}}
\put(559,430){\circle{18}} \put(627,353){\circle{18}}
\put(695,288){\circle{18}} \put(762,239){\circle{18}}
\put(830,205){\circle{18}} \put(898,184){\circle{18}}
\put(965,174){\circle{18}} \put(1033,170){\circle{18}}
\put(1101,168){\circle{18}} \put(1168,168){\circle{18}}
\put(1236,167){\circle{18}} \put(1304,167){\circle{18}}
\put(1371,167){\circle{18}} \put(1439,167){\circle{18}}
\put(1349,820){\circle{18}} \put(1279,779){\makebox(0,0)[r]{Simulation}}
\multiput(1299,779)(20.756,0.000){5}{\usebox{\plotpoint}}
\put(1399,779){\usebox{\plotpoint}} \put(221,753){\usebox{\plotpoint}}
\multiput(221,753)(14.514,-14.837){10}{\usebox{\plotpoint}}
\multiput(356,615)(12.165,-16.817){11}{\usebox{\plotpoint}}
\multiput(492,427)(13.287,-15.945){10}{\usebox{\plotpoint}}
\multiput(627,265)(18.144,-10.080){8}{\usebox{\plotpoint}}
\multiput(762,190)(20.556,-2.872){6}{\usebox{\plotpoint}}
\multiput(898,171)(20.750,-0.461){7}{\usebox{\plotpoint}}
\multiput(1033,168)(20.755,-0.154){6}{\usebox{\plotpoint}}
\multiput(1168,167)(20.756,0.000){7}{\usebox{\plotpoint}}
\multiput(1304,167)(20.756,0.000){6}{\usebox{\plotpoint}}
\put(1439,167){\usebox{\plotpoint}} \put(221,753){\makebox(0,0){$+$}}
\put(356,615){\makebox(0,0){$+$}} \put(492,427){\makebox(0,0){$+$}}
\put(627,265){\makebox(0,0){$+$}} \put(762,190){\makebox(0,0){$+$}}
\put(898,171){\makebox(0,0){$+$}} \put(1033,168){\makebox(0,0){$+$}}
\put(1168,167){\makebox(0,0){$+$}} \put(1304,167){\makebox(0,0){$+$}}
\put(1439,167){\makebox(0,0){$+$}} \put(1349,779){\makebox(0,0){$+$}}
\sbox{\plotpoint}{\rule[-0.400pt]{0.800pt}{0.800pt}}
\put(1279,738){\makebox(0,0)[r]{Theoretical bound (BE)}}
\put(1299.0,738.0){\rule[-0.400pt]{24.090pt}{0.800pt}}
\put(221,765){\usebox{\plotpoint}}
\multiput(221.00,763.09)(0.758,-0.502){83}{\rule{1.409pt}{0.121pt}}
\multiput(221.00,763.34)(65.076,-45.000){2}{\rule{0.704pt}{0.800pt}}
\multiput(289.00,718.09)(0.577,-0.502){109}{\rule{1.124pt}{0.121pt}}
\multiput(289.00,718.34)(64.667,-58.000){2}{\rule{0.562pt}{0.800pt}}
\multiput(357.41,657.75)(0.501,-0.514){129}{\rule{0.121pt}{1.024pt}}
\multiput(354.34,659.88)(68.000,-67.876){2}{\rule{0.800pt}{0.512pt}}
\multiput(425.41,587.31)(0.501,-0.581){129}{\rule{0.121pt}{1.129pt}}
\multiput(422.34,589.66)(68.000,-76.656){2}{\rule{0.800pt}{0.565pt}}
\multiput(493.41,508.06)(0.501,-0.619){127}{\rule{0.121pt}{1.191pt}}
\multiput(490.34,510.53)(67.000,-80.528){2}{\rule{0.800pt}{0.596pt}}
\multiput(560.41,425.41)(0.501,-0.566){129}{\rule{0.121pt}{1.106pt}}
\multiput(557.34,427.70)(68.000,-74.705){2}{\rule{0.800pt}{0.553pt}}
\multiput(627.00,351.09)(0.522,-0.501){123}{\rule{1.037pt}{0.121pt}}
\multiput(627.00,351.34)(65.848,-65.000){2}{\rule{0.518pt}{0.800pt}}
\multiput(695.00,286.09)(0.685,-0.502){91}{\rule{1.294pt}{0.121pt}}
\multiput(695.00,286.34)(64.314,-49.000){2}{\rule{0.647pt}{0.800pt}}
\multiput(762.00,237.09)(1.009,-0.503){61}{\rule{1.800pt}{0.121pt}}
\multiput(762.00,237.34)(64.264,-34.000){2}{\rule{0.900pt}{0.800pt}}
\multiput(830.00,203.09)(1.659,-0.505){35}{\rule{2.790pt}{0.122pt}}
\multiput(830.00,203.34)(62.208,-21.000){2}{\rule{1.395pt}{0.800pt}}
\multiput(898.00,182.08)(3.660,-0.514){13}{\rule{5.560pt}{0.124pt}}
\multiput(898.00,182.34)(55.460,-10.000){2}{\rule{2.780pt}{0.800pt}}
\put(965,170.34){\rule{13.800pt}{0.800pt}}
\multiput(965.00,172.34)(39.357,-4.000){2}{\rule{6.900pt}{0.800pt}}
\put(1033,167.34){\rule{16.381pt}{0.800pt}}
\multiput(1033.00,168.34)(34.000,-2.000){2}{\rule{8.191pt}{0.800pt}}
\put(1168,165.84){\rule{16.381pt}{0.800pt}}
\multiput(1168.00,166.34)(34.000,-1.000){2}{\rule{8.191pt}{0.800pt}}
\put(1101.0,168.0){\rule[-0.400pt]{16.140pt}{0.800pt}}
\put(221,765){\raisebox{-.8pt}{\makebox(0,0){$\Box$}}}
\put(289,720){\raisebox{-.8pt}{\makebox(0,0){$\Box$}}}
\put(356,662){\raisebox{-.8pt}{\makebox(0,0){$\Box$}}}
\put(424,592){\raisebox{-.8pt}{\makebox(0,0){$\Box$}}}
\put(492,513){\raisebox{-.8pt}{\makebox(0,0){$\Box$}}}
\put(559,430){\raisebox{-.8pt}{\makebox(0,0){$\Box$}}}
\put(627,353){\raisebox{-.8pt}{\makebox(0,0){$\Box$}}}
\put(695,288){\raisebox{-.8pt}{\makebox(0,0){$\Box$}}}
\put(762,239){\raisebox{-.8pt}{\makebox(0,0){$\Box$}}}
\put(830,205){\raisebox{-.8pt}{\makebox(0,0){$\Box$}}}
\put(898,184){\raisebox{-.8pt}{\makebox(0,0){$\Box$}}}
\put(965,174){\raisebox{-.8pt}{\makebox(0,0){$\Box$}}}
\put(1033,170){\raisebox{-.8pt}{\makebox(0,0){$\Box$}}}
\put(1101,168){\raisebox{-.8pt}{\makebox(0,0){$\Box$}}}
\put(1168,168){\raisebox{-.8pt}{\makebox(0,0){$\Box$}}}
\put(1236,167){\raisebox{-.8pt}{\makebox(0,0){$\Box$}}}
\put(1304,167){\raisebox{-.8pt}{\makebox(0,0){$\Box$}}}
\put(1371,167){\raisebox{-.8pt}{\makebox(0,0){$\Box$}}}
\put(1439,167){\raisebox{-.8pt}{\makebox(0,0){$\Box$}}}
\put(1349,738){\raisebox{-.8pt}{\makebox(0,0){$\Box$}}}
\put(1236.0,167.0){\rule[-0.400pt]{48.903pt}{0.800pt}}
\end{picture}

\parbox{5in}
{\caption{Average computational complexity of the MLSDA for
$(2,1,16)$ convolutional code with generators $1632044,1145734$ (octal)
and information length $L=100$.}\label{figure3}}
\end{center}
\end{figure}

\begin{figure}[hbt]
\begin{center}
\setlength{\unitlength}{0.240900pt}
\ifx\plotpoint\undefined\newsavebox{\plotpoint}\fi
\sbox{\plotpoint}{\rule[-0.200pt]{0.400pt}{0.400pt}}
\begin{picture}(1500,900)(0,0)
\font\gnuplot=cmr10 at 10pt \gnuplot
\sbox{\plotpoint}{\rule[-0.200pt]{0.400pt}{0.400pt}}
\put(221.0,123.0){\rule[-0.200pt]{4.818pt}{0.400pt}}
\put(201,123){\makebox(0,0)[r]{100}}
\put(1419.0,123.0){\rule[-0.200pt]{4.818pt}{0.400pt}}
\put(221.0,167.0){\rule[-0.200pt]{2.409pt}{0.400pt}}
\put(1429.0,167.0){\rule[-0.200pt]{2.409pt}{0.400pt}}
\put(221.0,226.0){\rule[-0.200pt]{2.409pt}{0.400pt}}
\put(1429.0,226.0){\rule[-0.200pt]{2.409pt}{0.400pt}}
\put(221.0,256.0){\rule[-0.200pt]{2.409pt}{0.400pt}}
\put(1429.0,256.0){\rule[-0.200pt]{2.409pt}{0.400pt}}
\put(221.0,270.0){\rule[-0.200pt]{4.818pt}{0.400pt}}
\put(201,270){\makebox(0,0)[r]{1000}}
\put(1419.0,270.0){\rule[-0.200pt]{4.818pt}{0.400pt}}
\put(221.0,315.0){\rule[-0.200pt]{2.409pt}{0.400pt}}
\put(1429.0,315.0){\rule[-0.200pt]{2.409pt}{0.400pt}}
\put(221.0,373.0){\rule[-0.200pt]{2.409pt}{0.400pt}}
\put(1429.0,373.0){\rule[-0.200pt]{2.409pt}{0.400pt}}
\put(221.0,404.0){\rule[-0.200pt]{2.409pt}{0.400pt}}
\put(1429.0,404.0){\rule[-0.200pt]{2.409pt}{0.400pt}}
\put(221.0,418.0){\rule[-0.200pt]{4.818pt}{0.400pt}}
\put(201,418){\makebox(0,0)[r]{10000}}
\put(1419.0,418.0){\rule[-0.200pt]{4.818pt}{0.400pt}}
\put(221.0,462.0){\rule[-0.200pt]{2.409pt}{0.400pt}}
\put(1429.0,462.0){\rule[-0.200pt]{2.409pt}{0.400pt}}
\put(221.0,521.0){\rule[-0.200pt]{2.409pt}{0.400pt}}
\put(1429.0,521.0){\rule[-0.200pt]{2.409pt}{0.400pt}}
\put(221.0,551.0){\rule[-0.200pt]{2.409pt}{0.400pt}}
\put(1429.0,551.0){\rule[-0.200pt]{2.409pt}{0.400pt}}
\put(221.0,565.0){\rule[-0.200pt]{4.818pt}{0.400pt}}
\put(201,565){\makebox(0,0)[r]{100000}}
\put(1419.0,565.0){\rule[-0.200pt]{4.818pt}{0.400pt}}
\put(221.0,610.0){\rule[-0.200pt]{2.409pt}{0.400pt}}
\put(1429.0,610.0){\rule[-0.200pt]{2.409pt}{0.400pt}}
\put(221.0,668.0){\rule[-0.200pt]{2.409pt}{0.400pt}}
\put(1429.0,668.0){\rule[-0.200pt]{2.409pt}{0.400pt}}
\put(221.0,698.0){\rule[-0.200pt]{2.409pt}{0.400pt}}
\put(1429.0,698.0){\rule[-0.200pt]{2.409pt}{0.400pt}}
\put(221.0,713.0){\rule[-0.200pt]{4.818pt}{0.400pt}}
\put(201,713){\makebox(0,0)[r]{1e+006}}
\put(1419.0,713.0){\rule[-0.200pt]{4.818pt}{0.400pt}}
\put(221.0,757.0){\rule[-0.200pt]{2.409pt}{0.400pt}}
\put(1429.0,757.0){\rule[-0.200pt]{2.409pt}{0.400pt}}
\put(221.0,816.0){\rule[-0.200pt]{2.409pt}{0.400pt}}
\put(1429.0,816.0){\rule[-0.200pt]{2.409pt}{0.400pt}}
\put(221.0,846.0){\rule[-0.200pt]{2.409pt}{0.400pt}}
\put(1429.0,846.0){\rule[-0.200pt]{2.409pt}{0.400pt}}
\put(221.0,860.0){\rule[-0.200pt]{4.818pt}{0.400pt}}
\put(201,860){\makebox(0,0)[r]{1e+007}}
\put(1419.0,860.0){\rule[-0.200pt]{4.818pt}{0.400pt}}
\put(221.0,123.0){\rule[-0.200pt]{0.400pt}{4.818pt}}
\put(221,82){\makebox(0,0){2}}
\put(221.0,840.0){\rule[-0.200pt]{0.400pt}{4.818pt}}
\put(356.0,123.0){\rule[-0.200pt]{0.400pt}{4.818pt}}
\put(356,82){\makebox(0,0){3}}
\put(356.0,840.0){\rule[-0.200pt]{0.400pt}{4.818pt}}
\put(492.0,123.0){\rule[-0.200pt]{0.400pt}{4.818pt}}
\put(492,82){\makebox(0,0){4}}
\put(492.0,840.0){\rule[-0.200pt]{0.400pt}{4.818pt}}
\put(627.0,123.0){\rule[-0.200pt]{0.400pt}{4.818pt}}
\put(627,82){\makebox(0,0){5}}
\put(627.0,840.0){\rule[-0.200pt]{0.400pt}{4.818pt}}
\put(762.0,123.0){\rule[-0.200pt]{0.400pt}{4.818pt}}
\put(762,82){\makebox(0,0){6}}
\put(762.0,840.0){\rule[-0.200pt]{0.400pt}{4.818pt}}
\put(898.0,123.0){\rule[-0.200pt]{0.400pt}{4.818pt}}
\put(898,82){\makebox(0,0){7}}
\put(898.0,840.0){\rule[-0.200pt]{0.400pt}{4.818pt}}
\put(1033.0,123.0){\rule[-0.200pt]{0.400pt}{4.818pt}}
\put(1033,82){\makebox(0,0){8}}
\put(1033.0,840.0){\rule[-0.200pt]{0.400pt}{4.818pt}}
\put(1168.0,123.0){\rule[-0.200pt]{0.400pt}{4.818pt}}
\put(1168,82){\makebox(0,0){9}}
\put(1168.0,840.0){\rule[-0.200pt]{0.400pt}{4.818pt}}
\put(1304.0,123.0){\rule[-0.200pt]{0.400pt}{4.818pt}}
\put(1304,82){\makebox(0,0){10}}
\put(1304.0,840.0){\rule[-0.200pt]{0.400pt}{4.818pt}}
\put(1439.0,123.0){\rule[-0.200pt]{0.400pt}{4.818pt}}
\put(1439,82){\makebox(0,0){11}}
\put(1439.0,840.0){\rule[-0.200pt]{0.400pt}{4.818pt}}
\put(221.0,123.0){\rule[-0.200pt]{293.416pt}{0.400pt}}
\put(1439.0,123.0){\rule[-0.200pt]{0.400pt}{177.543pt}}
\put(221.0,860.0){\rule[-0.200pt]{293.416pt}{0.400pt}}
\put(40,491){\makebox(0,0){\rotate[l]{Computational complexity}}}
\put(830,21){\makebox(0,0){$\gamma_b$}}
\put(221.0,123.0){\rule[-0.200pt]{0.400pt}{177.543pt}}
\put(1279,820){\makebox(0,0)[r]{Theoretical bound}}
\multiput(1299,820)(20.756,0.000){5}{\usebox{\plotpoint}}
\put(1399,820){\usebox{\plotpoint}} \put(221,669){\usebox{\plotpoint}}
\multiput(221,669)(16.722,-12.295){5}{\usebox{\plotpoint}}
\multiput(289,619)(15.692,-13.584){4}{\usebox{\plotpoint}}
\multiput(356,561)(15.004,-14.342){4}{\usebox{\plotpoint}}
\multiput(424,496)(14.462,-14.887){5}{\usebox{\plotpoint}}
\multiput(492,426)(14.352,-14.994){5}{\usebox{\plotpoint}}
\multiput(559,356)(15.225,-14.106){4}{\usebox{\plotpoint}}
\multiput(627,293)(16.370,-12.759){4}{\usebox{\plotpoint}}
\multiput(695,240)(17.704,-10.834){4}{\usebox{\plotpoint}}
\multiput(762,199)(18.990,-8.378){4}{\usebox{\plotpoint}}
\multiput(830,169)(20.064,-5.311){3}{\usebox{\plotpoint}}
\multiput(898,151)(20.571,-2.763){3}{\usebox{\plotpoint}}
\multiput(965,142)(20.700,-1.522){4}{\usebox{\plotpoint}}
\multiput(1033,137)(20.753,-0.305){3}{\usebox{\plotpoint}}
\multiput(1101,136)(20.753,-0.310){3}{\usebox{\plotpoint}}
\multiput(1168,135)(20.756,0.000){3}{\usebox{\plotpoint}}
\multiput(1236,135)(20.756,0.000){4}{\usebox{\plotpoint}}
\multiput(1304,135)(20.756,0.000){3}{\usebox{\plotpoint}}
\multiput(1371,135)(20.756,0.000){3}{\usebox{\plotpoint}}
\put(1439,135){\usebox{\plotpoint}} \put(221,669){\circle{18}}
\put(289,619){\circle{18}} \put(356,561){\circle{18}}
\put(424,496){\circle{18}} \put(492,426){\circle{18}}
\put(559,356){\circle{18}} \put(627,293){\circle{18}}
\put(695,240){\circle{18}} \put(762,199){\circle{18}}
\put(830,169){\circle{18}} \put(898,151){\circle{18}}
\put(965,142){\circle{18}} \put(1033,137){\circle{18}}
\put(1101,136){\circle{18}} \put(1168,135){\circle{18}}
\put(1236,135){\circle{18}} \put(1304,135){\circle{18}}
\put(1371,135){\circle{18}} \put(1439,135){\circle{18}}
\put(1349,820){\circle{18}} \put(1279,779){\makebox(0,0)[r]{Simulation}}
\multiput(1299,779)(20.756,0.000){5}{\usebox{\plotpoint}}
\put(1399,779){\usebox{\plotpoint}} \put(221,652){\usebox{\plotpoint}}
\multiput(221,652)(13.936,-15.381){10}{\usebox{\plotpoint}}
\multiput(356,503)(13.106,-16.094){11}{\usebox{\plotpoint}}
\multiput(492,336)(15.173,-14.162){8}{\usebox{\plotpoint}}
\multiput(627,210)(19.172,-7.953){8}{\usebox{\plotpoint}}
\multiput(762,154)(20.613,-2.425){6}{\usebox{\plotpoint}}
\multiput(898,138)(20.750,-0.461){7}{\usebox{\plotpoint}}
\multiput(1033,135)(20.756,0.000){6}{\usebox{\plotpoint}}
\multiput(1168,135)(20.756,0.000){7}{\usebox{\plotpoint}}
\multiput(1304,135)(20.756,0.000){6}{\usebox{\plotpoint}}
\put(1439,135){\usebox{\plotpoint}} \put(221,652){\makebox(0,0){$+$}}
\put(356,503){\makebox(0,0){$+$}} \put(492,336){\makebox(0,0){$+$}}
\put(627,210){\makebox(0,0){$+$}} \put(762,154){\makebox(0,0){$+$}}
\put(898,138){\makebox(0,0){$+$}} \put(1033,135){\makebox(0,0){$+$}}
\put(1168,135){\makebox(0,0){$+$}} \put(1304,135){\makebox(0,0){$+$}}
\put(1439,135){\makebox(0,0){$+$}} \put(1349,779){\makebox(0,0){$+$}}
\sbox{\plotpoint}{\rule[-0.400pt]{0.800pt}{0.800pt}}
\put(1279,738){\makebox(0,0)[r]{Theoretical bound (BE)}}
\put(1299.0,738.0){\rule[-0.400pt]{24.090pt}{0.800pt}}
\put(221,669){\usebox{\plotpoint}}
\multiput(221.00,667.09)(0.681,-0.502){93}{\rule{1.288pt}{0.121pt}}
\multiput(221.00,667.34)(65.327,-50.000){2}{\rule{0.644pt}{0.800pt}}
\multiput(289.00,617.09)(0.577,-0.502){109}{\rule{1.124pt}{0.121pt}}
\multiput(289.00,617.34)(64.667,-58.000){2}{\rule{0.562pt}{0.800pt}}
\multiput(356.00,559.09)(0.522,-0.501){123}{\rule{1.037pt}{0.121pt}}
\multiput(356.00,559.34)(65.848,-65.000){2}{\rule{0.518pt}{0.800pt}}
\multiput(425.41,491.75)(0.501,-0.514){129}{\rule{0.121pt}{1.024pt}}
\multiput(422.34,493.88)(68.000,-67.876){2}{\rule{0.800pt}{0.512pt}}
\multiput(493.41,421.70)(0.501,-0.521){127}{\rule{0.121pt}{1.036pt}}
\multiput(490.34,423.85)(67.000,-67.850){2}{\rule{0.800pt}{0.518pt}}
\multiput(559.00,354.09)(0.539,-0.502){119}{\rule{1.063pt}{0.121pt}}
\multiput(559.00,354.34)(65.793,-63.000){2}{\rule{0.532pt}{0.800pt}}
\multiput(627.00,291.09)(0.642,-0.502){99}{\rule{1.226pt}{0.121pt}}
\multiput(627.00,291.34)(65.455,-53.000){2}{\rule{0.613pt}{0.800pt}}
\multiput(695.00,238.09)(0.821,-0.502){75}{\rule{1.507pt}{0.121pt}}
\multiput(695.00,238.34)(63.871,-41.000){2}{\rule{0.754pt}{0.800pt}}
\multiput(762.00,197.09)(1.147,-0.503){53}{\rule{2.013pt}{0.121pt}}
\multiput(762.00,197.34)(63.821,-30.000){2}{\rule{1.007pt}{0.800pt}}
\multiput(830.00,167.09)(1.951,-0.506){29}{\rule{3.222pt}{0.122pt}}
\multiput(830.00,167.34)(61.312,-18.000){2}{\rule{1.611pt}{0.800pt}}
\multiput(898.00,149.08)(4.145,-0.516){11}{\rule{6.156pt}{0.124pt}}
\multiput(898.00,149.34)(54.224,-9.000){2}{\rule{3.078pt}{0.800pt}}
\multiput(965.00,140.06)(11.003,-0.560){3}{\rule{11.080pt}{0.135pt}}
\multiput(965.00,140.34)(45.003,-5.000){2}{\rule{5.540pt}{0.800pt}}
\put(1033,134.84){\rule{16.381pt}{0.800pt}}
\multiput(1033.00,135.34)(34.000,-1.000){2}{\rule{8.191pt}{0.800pt}}
\put(1101,133.84){\rule{16.140pt}{0.800pt}}
\multiput(1101.00,134.34)(33.500,-1.000){2}{\rule{8.070pt}{0.800pt}}
\put(221,669){\raisebox{-.8pt}{\makebox(0,0){$\Box$}}}
\put(289,619){\raisebox{-.8pt}{\makebox(0,0){$\Box$}}}
\put(356,561){\raisebox{-.8pt}{\makebox(0,0){$\Box$}}}
\put(424,496){\raisebox{-.8pt}{\makebox(0,0){$\Box$}}}
\put(492,426){\raisebox{-.8pt}{\makebox(0,0){$\Box$}}}
\put(559,356){\raisebox{-.8pt}{\makebox(0,0){$\Box$}}}
\put(627,293){\raisebox{-.8pt}{\makebox(0,0){$\Box$}}}
\put(695,240){\raisebox{-.8pt}{\makebox(0,0){$\Box$}}}
\put(762,199){\raisebox{-.8pt}{\makebox(0,0){$\Box$}}}
\put(830,169){\raisebox{-.8pt}{\makebox(0,0){$\Box$}}}
\put(898,151){\raisebox{-.8pt}{\makebox(0,0){$\Box$}}}
\put(965,142){\raisebox{-.8pt}{\makebox(0,0){$\Box$}}}
\put(1033,137){\raisebox{-.8pt}{\makebox(0,0){$\Box$}}}
\put(1101,136){\raisebox{-.8pt}{\makebox(0,0){$\Box$}}}
\put(1168,135){\raisebox{-.8pt}{\makebox(0,0){$\Box$}}}
\put(1236,135){\raisebox{-.8pt}{\makebox(0,0){$\Box$}}}
\put(1304,135){\raisebox{-.8pt}{\makebox(0,0){$\Box$}}}
\put(1371,135){\raisebox{-.8pt}{\makebox(0,0){$\Box$}}}
\put(1439,135){\raisebox{-.8pt}{\makebox(0,0){$\Box$}}}
\put(1349,738){\raisebox{-.8pt}{\makebox(0,0){$\Box$}}}
\put(1168.0,135.0){\rule[-0.400pt]{65.284pt}{0.800pt}}
\end{picture}

\parbox{5in}
{\caption{Average computational complexity of the MLSDA
for $(2,1,16)$ convolutional code with generators $1632044,1145734$ (octal)
and information length $L=60$.}\label{figure4}}
\end{center}
\end{figure}
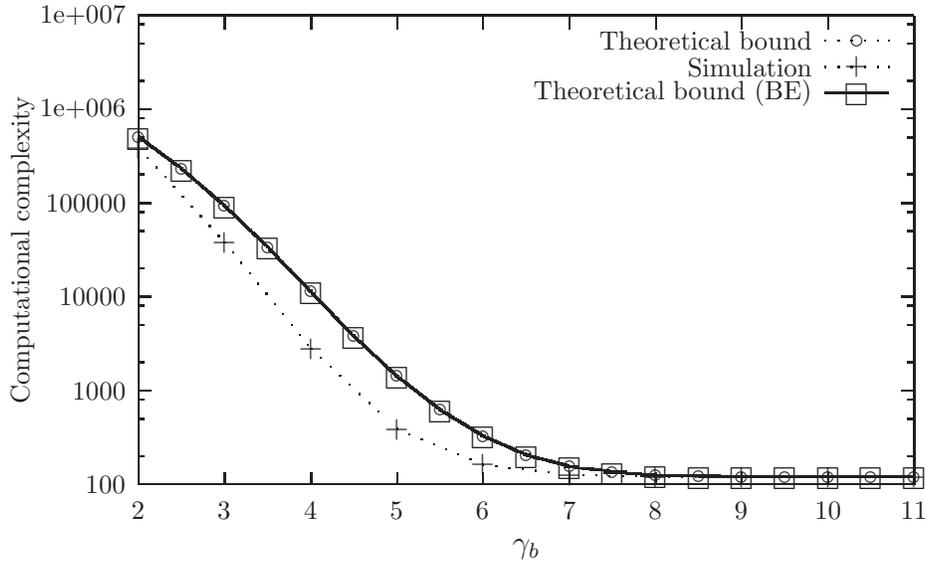

 Figures \ref{figure1}--\ref{figure4} present the deviation between
the simulated results and the two theoretical upper bounds on the computational
complexity of the MLSDA. According to these figures, the Berry-Esseen-enhanced
theoretical upper bound is fairly close to the simulation results for both high
$\gamma_b$ (above $6$ dB) and low $\gamma_b$ (below $2$ dB). Even for moderate
$\gamma_b$, they only differ by no more than $0.8$ for
Figs.~\ref{figure1}--\ref{figure4} on a $\log_{10}$
scale. The differences between the two theoretical upper bounds with and
without Berry-Esseen analysis are now visible in these figures. For example,
the ratios of the two theoretical bounds are respectively 0.86, 0.90 and 0.95
at 4.0 dB, 4.5 dB and 5.0 dB in Fig.~\ref{figure3}.

A side observation from these figures
is that the codes with longer constraint
length, although having a lower bit error rate,
require more computations. However,
such a tradeoff on constraint length and bit error rate
 can be moderately eased at high SNR. Notably,
when $\gamma_b>6$ dB,
the average computational effort of the MLSDA in
all four figures is reduced to approximately $2^kL$
in spite of the constraint length.

\section{Conclusions}
\label{SEC:CON}

In terms of the large deviations technique and Berry-Esseen theorem, this study
established theoretical upper bounds on the computational effort of the
simplified GDA and the MLSDA for AWGN channels.

There may be two factors determining the accuracy of the complexity upper
bound. The first factor is the accuracy of the large deviations probability
bound for sum of independent samples in Lemma 2, and the second one is the
accuracy of the estimate of the node extension probability for sequential-type
decoding. We however found that the main inaccuracy may not come from the
latter. Taking the GDA algorithm as an example, \rec{last0} is actually the
exact event for path $\xx_{[\ell]}$ to be extended by the simplified GDA, and
\rec{last1} becomes equality when the maximum-likelihood decision is exactly
the transmitted all-zero codewords. Notably, as long as the node expanding
distribution for each node is known, the average decoding complexity can be
exactly obtained (specifically, if $Z_j=1$ when node $j$ is visited and
expanded, and $Z_j=0$, otherwise, then the average number of computations is
exactly $2\sum_jE[Z_j]=2\sum_j\Pr[Z_j=1]$ since the extension of each path
causes two branch metric computations). Hence, the main inaccuracy is due to
the overestimate of the large deviations probability bound for sum of
independent variables (and, of course, accumulating such overestimate by
summing for all nodes may make worse the situation). Since the codes simulated
for the GDA algorithm are of lengths $24$ and $48$ under which the large
deviations probability bound is very inaccurate, the resultant complexity bound
is also inaccurate, and Berry-Esseen inequality does not provide much help in
decreasing such inaccuracy. As for the MLSDA algorithm, a looser estimate is
used to bound the node expanding probability by replacing ``summation'' by
``maximization'' as shown in \eqref{mlsda1}. However, the resultant complexity
bound is much more accurate simply because the large deviations probability
bound is more exact at larger block length.

\section*{Appendix}

\begin{proposition}\label{claim1}For a fixed non-negative integer $k$,
the probability mass of
$$\Pr\left\{r_1+\cdots+r_d+\min(w_{1},0)+\ldots+\min(w_{k},0)\leq 0\right\}$$
is a decreasing function for non-negative integer $d$, provided that
$r_1$, $r_2$, $\ldots$, $r_d$, $w_1$, $w_2$, $\ldots$, $w_k$ are i.i.d.~with a Gaussian marginal
distribution of positive mean $\mu$ and variance $\sigma^2$.
\end{proposition}
\begin{proof}
Assume without loss of generality that
$\sigma^2=1$. Also, assume $k\geq 1$ since the proposition
is trivially valid for $k=0$.

Let $\Omega_d\Bydef r_1+\cdots+r_d$. Denote the probability density function
of $w_1$ by $f(\cdot)$.
Then putting $\nu\Bydef\Pr\{w_{j}=0\}$ yields
\begin{eqnarray*}
&&\Pr\left\{\Omega_d+w_{1}+w_{2}+\cdots+w_{k}\leq 0\right\}\\
&=&\sum_{j=0}^k\Pr\left\{\mbox{exactly }(k-j)\mbox{ zeros in }(w_{1},w_{2},\ldots,w_{k})\right\}\\
&&\quad\quad\Pr\left\{\left.\Omega_d+w_{1}+w_{2}+\cdots+w_{k}\leq 0\right|
\mbox{exactly }(k-j)\mbox{ zeros in }(w_{1},w_{2},\ldots,w_{k})\right\}\\
&=&
{k\choose 0}\nu^k\Pr\{\Omega_d\leq 0\}
+{k\choose 1}\nu^{k-1}(1-\nu)
\int_{-\infty}^0f(x)\Pr\{\Omega_d\leq -x\}dx\\
&&+{k\choose 2}\nu^{k-2}(1-\nu)^2
\int_{-\infty}^0\int_{-\infty}^0
f(x_1)f(x_2)\Pr\{\Omega_d\leq-(x_1+x_2)\}dx_1dx_2\\
&&+\cdots\\
&&+{k\choose k}(1-\nu)^k
\int_{-\infty}^0\cdots\int_{-\infty}^0
f(x_1)\cdots f(x_k)\Pr\{\Omega_d\leq-(x_1+\cdots+x_k)\}dx_1\cdots dx_k.
\end{eqnarray*}
Accordingly, if each of the above $(k+1)$ terms is
non-increasing in $d$,
so is their sum.
Let
\begin{eqnarray*}
q_j(d)&\bydef&
\int_{-\infty}^0\cdots\int_{-\infty}^0
f(x_1)\cdots f(x_j)\Pr\{\Omega_d\leq -(x_1+\cdots+x_j)\}dx_1\cdots dx_j\\
&=&
\int_{-\infty}^0\cdots\int_{-\infty}^0
f(x_1)\cdots f(x_j)
\Phi\left(-\frac{x_1+\cdots+x_j}{\sqrt{d}}-\sqrt{d}\mu\right)dx_1\cdots dx_j.
\end{eqnarray*}
Then
\begin{eqnarray}
\frac{\partial q_j(d)}{\partial\left(\sqrt{d}\right)}
&=&\int_{-\infty}^0\cdots\int_{-\infty}^0
f(x_1)\cdots f(x_j)\nono\\
&&\times\left(
\frac{x_1+\cdots+x_j}d-\mu\right)
\frac 1{\sqrt{2\pi}}e^{-(x_1+\cdots+x_j+d\cdot\mu)^2/(2d)}dx_1\cdots dx_j\nono\\
&\leq&-\frac{\mu}{\sqrt{2\pi}}\int_{-\infty}^0
\cdots\int_{-\infty}^0
f(x_1)\cdots f(x_j)
e^{-(x_1+\cdots+x_j+d\cdot\mu)^2/(2d)}dx_1\cdots dx_j\mem{last}\\
&<&0,\nono
\end{eqnarray}
where \rec{last} follows from $x_i\leq 0$ (according to
the range of integration) for
$1\leq i\leq j$. Consequently, $q_j(d)$ is decreasing
in $d$ for $d$ positive and every $1\leq j\leq k$.
The proof is completed by noting that the first term,
$\Pr\{\Omega_d\leq 0\}=\Phi(-\sqrt{d}\mu)$, is also decreasing in $d$.
\end{proof}

\end{document}